\documentclass[aps,prb,twocolumn,floatfix,showpacs,citeautoscript,superscriptaddress,longbibliography,hyperlinks]{revtex4-2}
\usepackage{amsmath}
\usepackage{amssymb}
\usepackage{amsfonts}
\usepackage[colorlinks=true,citecolor=blue]{hyperref}
\usepackage{graphicx}
\usepackage{enumitem}
\setlist[enumerate]{leftmargin=6mm}
\usepackage[caption = false]{subfig}
\usepackage{braket}
\usepackage{siunitx}
\usepackage{tikz}
\usepackage{bbm}
\usepackage{circuitikz}
\usepackage[capitalise]{cleveref}
\usepackage{booktabs}
\usepackage[normalem]{ulem}

\usepackage{glossaries}
\glsdisablehyper
\newacronym{jj}{JJ}{Josephson junction}
\newacronym{ldos}{LDoS}{local density of states}
\newacronym{dos}{DoS}{density of states}
\newacronym{mbs}{MBS}{Majorana bound state}
\newacronym{abs}{ABS}{Andreev bound state}

\newcommand{\up}{\uparrow}
\newcommand{\dw}{\downarrow}

\newcommand{\sii}{\hat{\sigma}_{0}}
\newcommand{\sx}{\hat{\sigma}_{x}}
\newcommand{\sy}{\hat{\sigma}_{y}}
\newcommand{\sz}{\hat{\sigma}_{z}}

\newcommand{\ti}{\hat{\tau}_{0}}

\newcommand{\tz}{\hat{\tau}_{z}}

\newcommand{\Tr}{\textrm{Tr}}

\newcommand{\md}{\mathrm{d}}

\newcommand{\me}{\mathrm{e}}

\definecolor{mahogany}{RGB}{192,64,0}
\definecolor{magenta2}{RGB}{255,102,255}
\definecolor{crimson}{RGB}{220,20,60}

\usepackage{sidecap,tikz}
\definecolor{lime}{HTML}{A6CE39}
\DeclareRobustCommand{\orcidicon}{\hspace{-1mm}
	\begin{tikzpicture}
		\draw[lime, fill=lime] (0,0) 
		circle [radius=0.16] 
		node[white] {{\fontfamily{qag}\selectfont \tiny \,ID}};
		\draw[white, fill=white] (-0.0525,0.095) 
		circle [radius=0.007];
	\end{tikzpicture}
	\hspace{-3mm}
}
\foreach \x in {A, ..., Z}{\expandafter\xdef\csname orcid\x\endcsname{\noexpand\href{https://orcid.org/\csname orcidauthor\x\endcsname}
		{\noexpand\orcidicon}}
}
\begin{document}

\title{Odd-frequency Pairing in Josephson Junctions Coupled by Magnetic Textures} 

\author{Ignacio Sardinero\orcidA{}}
\affiliation{Department of Theoretical Condensed Matter Physics\char`,~Universidad Aut\'onoma de Madrid, 28049 Madrid, Spain}
\affiliation{Condensed Matter Physics Center (IFIMAC), Universidad Aut\'onoma de Madrid, 28049 Madrid, Spain}

\author{Jorge Cayao\orcidB{}}
\affiliation{Department of Physics and Astronomy, Uppsala University, Box 516, S-751 20 Uppsala, Sweden}

\author{Rubén Seoane Souto\orcidC{}}
\affiliation{Consejo Superior de Investigaciones Científicas (CSIC), Sor Juana Inés de la Cruz 3, 28049 Madrid, Spain}

\author{Pablo Burset\orcidD{}}
\affiliation{Department of Theoretical Condensed Matter Physics\char`,~Universidad Aut\'onoma de Madrid, 28049 Madrid, Spain}
\affiliation{Condensed Matter Physics Center (IFIMAC), Universidad Aut\'onoma de Madrid, 28049 Madrid, Spain}
\affiliation{Instituto Nicol\'as Cabrera, Universidad Aut\'onoma de Madrid, 28049 Madrid, Spain}

\date{\today}

\begin{abstract}
Josephson junctions coupled through magnetic textures provide a controllable platform for odd-frequency superconductivity and Majorana physics. Within a tight-binding Green function framework, induced pair correlations and spectral properties are analyzed under various magnetic and geometric conditions. When the junction is in the topologically trivial regime, even-frequency singlet pairing is dominant, whereas the topological phase is characterized by the coexistence of Majorana bound states and robust odd-frequency equal-spin triplet pairing at the interface edges. The odd-frequency polarized triplets reveal a divergent $1/\omega$ behavior when the Majorana states are decoupled, which is intrinsically connected to their self-conjugation property. The zero-frequency divergence evolves into shifted resonances and linear low-frequency behavior once hybridization occurs. A nonmagnetic interruption in the texture separates the topological superconductor into two topological segments and generates additional inner Majorana modes. 
When the nonmagnetic barrier is comparable to the inner Majorana states localization length, they hybridize and modify their associated odd-frequency triplet pairing, while the outer edge modes preserve their self-conjugated nature. Tuning the superconducting phase difference further controls the onset of the topological regime and the stability of localized Majorana states. The results highlight the central role of odd-frequency triplet correlations as a probe of topological superconductivity in magnetically engineered Josephson junctions. 
\end{abstract}

\maketitle

%%%%%%%%%%%%%%%%%%%%%%%%%%%%%%%
%  INTRO
%%%%%%%%%%%%%%%%%%%%%%%%%%%%%%%

\section{Introduction\label{sec:intro}}

Topological superconductors host \glspl{mbs}~\cite{Tanaka_reviewJPSJ,Sato2016,Sato_2017,Aguado_review2017,LutchynReview,frolov2019quest,Flensberg_NRM2021,Tanaka2024May,Fukaya_2025,Souto_chapter}, emergent quasiparticles that obey non-Abelian statistics and exhibit potential for decoherence-free quantum computation~\cite{Sarma_NPJ2015,Lahtinen_2017, beenakker2019search, aguado2020majorana, Marra_2022, aasen2025roadmap}. 
These exotic modes are associated with localized, zero-energy excitations at interfaces, vortices, or boundaries in one- and two-dimensional superconducting systems~\cite{Sato_2017,Aguado_review2017,Tanaka2024May}. 
Recent efforts in condensed matter physics have focused on engineering physical platforms where \glspl{mbs} can be realized, manipulated, and detected through unambiguous experimental signatures~\cite{prada2020,beenakker2020search}. 
For future applications in topological quantum computation, such platforms must exploit the intrinsic properties of \glspl{mbs} like their self-conjugation \cite{Tanaka2024May}, which is tied to their charge neutrality and spatial nonlocality~\cite{Cayao2017Nov, Clarke2017Nov, Prada2017Aug, Deng2018Aug, Smirnov2022May, Smirnov2024May,  Vimal2024Dec, Dutta2024Mar, Cayao2024Feb, mondal2025JDE}.
%Initial 
Most of the experimental advances on finding signatures of \glspl{mbs} focused on their behavior at zero energy~\cite{frolov2019quest,prada2020,Flensberg_NRM2021}. However, it is now well accepted that topologically trivial states can appear at low energies,  mimicking the behavior of nontrivial \glspl{mbs} at low 
bias~\cite{Kells2012Sep, Prada2012Nov, SanJoseSciRep, Cayao2015Jan, Liu_PRB_2017, Liu_PRB_2018, Awoga2019Sep, Vuik_SPP_2019,Cayao2021Jul, Burset_2021, deMendonca2023May, baldo2023zero, Awoga2023May, hes23, Cayao2024Aug, ahmed2025}, which reduces the effectiveness of spectroscopy to unambiguously demonstrate Majorana states.
The field is thus moving towards probing the nonlocal behavior of \glspl{mbs} through coupling of several of them, a crucial step towards applications. 
%To achieve these goals, both new material platforms and different signatures must be explored. \red{[RS: I would remove this sentence (or turn it down). It may sound that we say that what people did is fundamentally insufficient. Some people may disagree]}

\begin{figure}[b]
    \centering
    \includegraphics[width=0.99\linewidth]{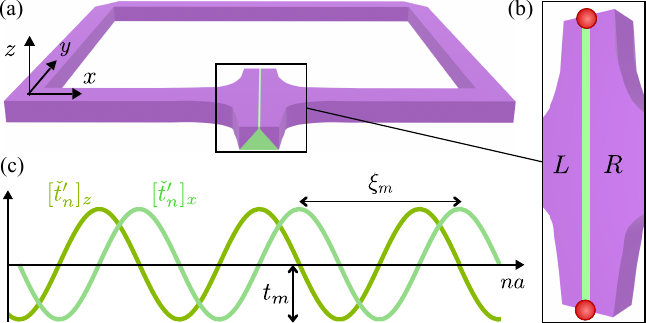}
    \caption{Josephson junction coupled by a magnetic texture. 
    (a) Superconducting loop closed around a magnetic texture (green). 
    (b) Two superconductors, $L$ and $R$, form the junction. Majorana states (red circles) emerge in the nontrivial regime at the interface edges. (c) Helical magnetic texture along the interface with period $\xi_m$ and amplitude $t_m$. 
    }
    \label{fig:1_setup}
\end{figure}

%Intro paragraph 2. Magnetic textures with ferro or antiferro. Challenge: our research question (emergence of MZMs and interplay between them?)
Topological superconductivity relies on the combination of conventional superconductivity and different spin fields, especially spin-orbit and magnetism~\cite{Braunecker2010, Beenakker_2011, Egger_2012_PRB, Braunecker_2013, Klinovaja_2013_PRL, Franz_2013, Perge_PRB2013, Pientka_PRB2013, Ojanen_2014, Heimes_2014, Xiao2015, Paaske_2016, Paaske_2016b, Cuoco_2017, Kim2018, cayao2018andreev, Kornich_2020, escribano2021, Liu_PRB2021, Langbehn_PRB2021, Awoga2022Apr, Escribano_NPJ2022, Frolov_2021, Mondal_PRB2023, mizushima2025}. Magnetic textures such as domain walls, helical patterns, and antiferromagnetic modulations, provide a controllable mechanism for engineering the spin-dependent properties of superconducting junctions~\cite{Mourik_Science2012,Nadj-Perge2014,Ruby_PRL_2015,Jeon_Science_2017,manna2020,maiani2021,Woods_PRB2021,Khindanov_PRB2021,vaitiekenas2021,vaitiekenas2022,Razmadze_PRB2023, Desjardins_NatMater_2019,Steffensen_2022_PRR}. 
When incorporated into Josephson junctions~\cite{Pientka2017,Hell_PRL2017,Laeven_PRL2020,Paudel_PRB2021,Lesser_JoPD2022,Kuiri_PRB2023}, these textures can serve as tunable barriers with nontrivial symmetry properties~\cite{Ai2021, Idzuchi2021, kang_van_2022, wu_field-free_2022, hu_long-range-skin_2023, Spuri2023, Sardinero2024, Gonzalez-Sanchez2025May}, see \Cref{fig:1_setup}. A key question is how different magnetic textures affect the emergence, symmetry, and coupling of Majorana modes, and how such effects manifest in measurable quantities. 

%Intro paragraph 2*. Odd-w pairing and multiple majoranas: 
In addition to the spatial localization and spectral signatures of Majorana modes, their presence is also reflected in the symmetry of the induced superconducting correlations~\cite{Tanaka_reviewJPSJ,Tanaka2024May}. 
Indeed, odd-frequency pairing, a type of superconducting order that is nonlocal and odd in time (or frequency)~\cite{Cayao2020odd, Linder_2019}, plays a central role in the characterization of \glspl{mbs}. 
Majorana states are defined by a self-conjugation property which yields anomalous correlations that are odd under the exchange of coordinates, but such antisymmetry only reflects on the time dependence since Majoranas are spinless, spatially localized zero-energy modes~\cite{Asano2013Mar, Huang2015Sep, Crepin2015Sep, Kashuba2017May, Cayao2017Oct, Tamura2020Jun, Lee2017May, Keidel_PRB2018, Fleckenstein2018Apr, Linder_2019, Tsintzis2019Sep, Ziani2020May, Takagi2020Jan, Cayao2020odd, Kuzmanovski2020Mar, Lu2022Dec, Cayao2022Sep, Yang_2023, Ahmed2025Jan, Cayao2024Sep, Ahmed2025Jun}. 
Consequently, the simplest form of odd-frequency pairing with a characteristic $1/\omega$ behavior at low frequencies is a natural fingerprint of the self-conjugation of Majorana states~\cite{Tamura2019May,Tamura2021Oct, Cayao2024Sep, Ahmed2025Jan, Ahmed2025Jun}. 

However, odd-frequency equal-spin triplet correlations with no connection to Majorana physics often emerge in systems with magnetic inhomogeneity or broken spin-rotation symmetry~\cite{BergeretRMP2005, eschrig2007symmetries, Sothmann2014Dec, linder2015strong, Burset2015Nov, Burset2016May, Lu2016Jul, Burset2017Jun, diesch2018, Hwang2018Oct, Bobkova2019, Johnsen2021Feb, hijano2022quasiparticle, Fyhn_2023_PRL, Johnsen2023, Akash2023, Li2024Nov}. In that context, odd-frequency triplet states are of great interest in the field of superconducting spintronics~\cite{Linder_NatPhys_2015, Eschrig_RPP_2015}. Establishing a direct correspondence between Majorana modes and induced odd-frequency pairing provides an alternative route for their detection, and enables a deeper understanding of the symmetries underlying topological superconductivity. The study of topological odd-frequency equal-spin triplet pairing can also unveil novel applications for superconducting spintronics~\cite{Breunig2018Jan, Keidel2020Apr}. 

%Intro paragraph 3. In this work we\ldots
This work explores the emergence and interplay of \glspl{mbs} in Josephson junctions with magnetic textures (\Cref{fig:1_setup}) by investigating their pairing symmetries and spatial structure. 
Within a microscopic tight-binding Green function formalism, we explore the induced pairing and spectral properties of the junction. 
The analysis reveals the presence of odd-frequency equal-spin triplet correlations and their connection to emergent \glspl{mbs} in the topologically nontrivial regime. 
In particular, when \glspl{mbs} are decoupled, therefore being described by local self-conjugate operators, the odd-frequency spin-polarized triplet displays the expected $1/\omega $ behavior at zero energy. By contrast, Majorana edge states that couple in narrow junctions acquire a finite energy from hybridization and lose their self-conjugation property. This effect is manifested by the odd-frequency pairing becoming linear with $\omega$ at low energy, indicating that the \glspl{mbs} are no longer self-conjugated. 

When the magnetic texture is interrupted by a nonmagnetic barrier an extra pair of \glspl{mbs} emerges at the edges of the barrier. These states can also hybridize, and thus loose their self-conjugation property, if the barrier is narrow enough that their wavefunctions overlap. 
The phase difference across the junction is shown to affect the localization of the \glspl{mbs} and can help recover the low-energy $1/\omega $ behavior of the odd-frequency triplet for hybridized Majorana modes. Consequently, our work introduces a tunable platform to probe Majorana physics in magnetically engineered superconducting systems. 

The rest of the article is organized as follows. 
We present our model and microscopic Green function formalism in \Cref{sec:model}. We first analyze the case of an uninterrupted magnetic texture in \Cref{sec:spiral}. Then, we explore in \Cref{sec:barrier} the induced pairing and spectral properties of a magnetic texture interrupted by a nonmagnetic barrier. Finally, \Cref{sec:conc} presents a summary of our results and our conclusions. 

%%%%%%%%%%%%%%%%%%%%%%%%%%%%%%%
%  MODEL
%%%%%%%%%%%%%%%%%%%%%%%%%%%%%%%

\section{Model and formalism\label{sec:model}}

\subsection{Hamiltonian}

We consider a two-dimensional (2D) \gls{jj} consisting of two conventional singlet $s$-wave superconductors coupled by a one-dimensional (1D) magnetic-textured barrier as sketched in \Cref{fig:1_setup}. The junction is described by a tight-binding square lattice with $2N_x$ horizontal and $N_y$ vertical sites given by the Hamiltonian
\begin{equation}\label{eq:H-tot}
    \check{H} = \check{H}_L + \check{H}_R + \check{H}_t ,
\end{equation}
where $\check{H}_{L,R}$ describe the left and right superconductors and $\check{H}_t$ the magnetic tunnel barrier. The superconductors have the same uniform chemical potential $\mu$, local superconducting pairing $\Delta>0$, and hopping parameter $t$, with Hamiltonians
\begin{align}
	\check{H}_{L,R} = 
    & -\sum_{\sigma=\up,\dw} \Bigl( 
    t \sum_{ \langle x, x' \rangle } c_{mn,\sigma}^\dagger c_{m'n',\sigma} 
%    \nonumber \\ {}& 
    + \mu \sum_{m,n} c_{mn,\sigma}^\dagger c_{mn,\sigma} \Bigr)
	\nonumber \\  
	& 
    + \Delta \me^{i \phi_{L,R}} \sum_{j, k} c^\dagger_{mn,\up} c^\dagger_{mn,\dw} + \mathrm{H.c.} 
    \label{eq:hlr}
\end{align}
Here, the operators $c^\dagger_{mn,\sigma}$ ($c_{mn,\sigma}$) create (annihilate) electrons with spin $\sigma=\up,\dw$ on the lattice site $(m,n)$; $\langle x,x' \rangle$ stands for nearest-neighbors combinations of the horizontal and vertical indices $m,m'$ and $n,n'$; and $\phi_{L,R}$ is the superconducting phase at each side ($L$ or $R$) of the junction. 
Each superconducting region has the same length $L_x=N_x a$ and width $L_y=N_y a$, with $a$ being the lattice constant. 

\begin{figure}[b]
    \centering
    \includegraphics[width=\linewidth]{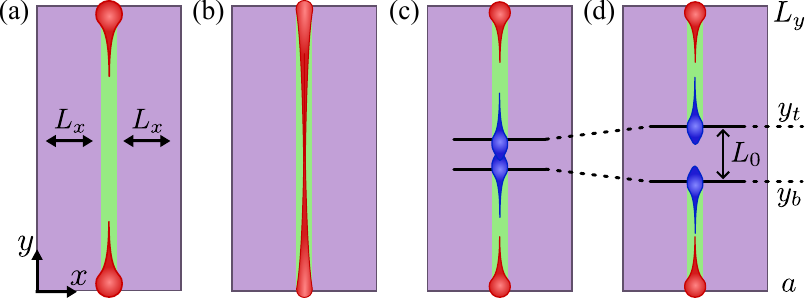}
    \caption{Different setup configurations: 
    (a,b) Uninterrupted magnetic texture (green) of length $L_y$ (a) longer or (b) comparable to the Majorana localization length. 
    (c,d) Magnetic texture interrupted by a nonmagnetic barrier of length $L_0$ that is (c) shorter or (d) longer than the inner edge states wavefunction decay. 
    The wavefunction localization of the outer and inner edge states is respectively shown in red and blue. 
    }
    \label{fig:2_sketch}
\end{figure}

The barrier that connects the superconductors has a magnetization that changes spatially along the $xz$ plane, see \Cref{fig:1_setup}(c). 
Defining the Nambu spinor $\check{\Psi}_{mn}^\dagger=[c_{mn,\up}^\dagger, c_{mn,\dw}^\dagger, c_{mn,\up}, c_{mn,\dw}]$, we have 
\begin{equation} \label{eq:vlr}
    \check{H}_t = -\frac{1}{2} \sum_{\sigma, \sigma'}\sum_{n=1}^{N_y} \check{\Psi}_{N_xn,\sigma}^\dagger \left( \check{t}'_{n} \right)_{\sigma\sigma'} 
 \check{\Psi}_{(N_x+1)n,\sigma'} ~,
\end{equation}
where $\check{t}'_{n}$, with the index $n$ running along the junction interface, is a matrix in spin space containing the magnetic details of the barrier.  
We divide the general barrier into one or three regions as sketched in \Cref{fig:2_sketch}. In the latter case, see \Cref{fig:2_sketch}(c,d), one \textit{inner} barrier can separate two \textit{outer} segments (green). The inner barrier represents possible defects in the magnetic texture and ranges from $y_b$ to $y_t$ with a width $L_0 =y_t - y_b = N_0a $. 
We only consider the case where the inner barrier becomes nonmagnetic while the outer segments maintain the magnetic texture. 
As a result, the magnetic barrier between superconductors is given by
\begin{widetext} 
\begin{equation}\label{eq:barrierModel}
    \check{t}'_n = -t_0 \tz \sii 
    %\\ 
	- \left\{ 
    \begin{array}{cl}
		 0, & y_b/a < n \leq y_t/a, \\
	   t_m \tz \left[ \cos \left( \frac{2\pi a}{\xi_m} n \right) \sz 
		- \sin \left( \frac{2\pi a }{ \xi_m } n \right) \sx \right], &  \text{otherwise} 
	\end{array} 
    \right. . 
    %\notag
\end{equation}
Here, $t_0$ is the spin-independent hopping between superconductors, $t_m$ the amplitude of the spin-texture and $\xi_m$ its period, see \Cref{fig:1_setup}(c). 
The Pauli matrices $\hat{\tau}_{0,x,y,z}$ and $\hat{\sigma}_{0,x,y,z}$ respectively act on Nambu and spin degrees of freedom, with $\ti$ and $\sii$ being identity matrices. 

\end{widetext}

\subsection{Green functions}

The objective of this work is to analyze the pairing amplitudes induced at the \gls{jj} interface. The information about the induced pairings is encoded in the anomalous part of the Green function \cite{mahan2013many,zagoskin}. 
We thus define the retarded (R) and advanced (A) Green's functions associated to the Hamiltonian in \Cref{eq:H-tot} as
\begin{equation}\label{eq:GF-RA}
    \check{G}^\text{R,A}(\omega)= \left[ \left( \omega \pm i 0^+ \right) \check{1} - \check{H} \right]^{-1} ,
\end{equation}
with $\omega$ the energy and $\check{1}$ the identity matrix. 
We define the orthonormal local basis ${\ket{mn,\sigma,\tau}}$, where $(mn)$ labels each lattice site, $\sigma$ denotes spin and $\tau$ Nambu (electron-hole) indices. The projector $P_{mn}=\sum_{\sigma,\tau}\ket{mn,\sigma,\tau}\bra{mn,\sigma,\tau}$ then extracts the spin-Nambu matrix block for lattice site $mn$. The lattice representation of the Green function is thus $
\check{G}_{mn,m'n'}^\text{R,A}(\omega)=P_{mn} \check{G}^\text{R,A}(\omega)P_{m'n'}
$, which can be computed efficiently using sparse solvers~\cite{Quantica}.
In the following, we compute the system spectral properties from the diagonal part of the Green function and the induced pairing from the off-diagonal or anomalous components. 

\subsection{Anomalous Green function and pairing amplitudes}

The anomalous Green function contains the electron-hole Nambu components of \Cref{eq:GF-RA}, $\hat{F}_{mn,m'n'}^\text{R,A}=(\check{G}_{mn,m'n'}^\text{R,A})_{eh}$. In what follows, we only need the retarded Green function since the advanced one is defined as $\check{G}_{mn,m'n'}^\text{A}(\omega)= [\check{G}_{m'n',mn}^\text{R}(\omega)]^\dagger$. While $\check{G}_{mn,m'n'}^\text{R}$ are matrices in spin and Nambu spaces, $\hat{F}_{mn,m'n'}^\text{R}$ are matrices only in spin space, so we can decompose them into one singlet ($\nu=0$) and three triplet components ($\nu=+,-,z$) as~\cite{Keidel_PRB2018}
\begin{equation}\label{eq:anom-GF-spin}
  F^\nu_{mn,m'n'} (\omega)
  \equiv \mathrm{Tr} \{ -i\sy \boldsymbol{ [\hat{\sigma}} ]_{\nu}\,\hat{F}^\text{R}_{mn,m'n'} (\omega) \},
\end{equation}
where we have defined the vector $\boldsymbol{\hat{\sigma}}= [\sii,\hat{\sigma}_+,\hat{\sigma}_-,\sz]^T$, with $\hat{\sigma}_\pm = (\sx\pm i\sy)/2$. The anomalous retarded Green functions in \Cref{eq:anom-GF-spin} correspond to the singlet component $F^0 = (\hat{F}^\text{R})_{\up\dw} - (\hat{F}^\text{R})_{\dw\up}$, the non-polarized triplet one $F^z = (\hat{F}^\text{R})_{\up\dw} + (\hat{F}^\text{R})_{\dw\up}$, and the polarized triplets along the $z$-direction $F^+ = (\hat{F}^\text{R})_{\dw\dw} $ and $F^- = - (\hat{F}^\text{R})_{\up\up} $ (we have omitted the site indices $F_{mn,m'n'}$ for simplicity). 

We can further symmetrize the anomalous correlators into their spatially symmetric and antisymmetric parts,
\begin{equation} \label{eq:Fspatial}
	F^{\nu\pm}_{mn,m'n'} (\omega) = \frac{1}{2} \left[ F^\nu_{mn,m'n'} (\omega) \pm F^\nu_{m'n',mn} (\omega) \right] ,
\end{equation}
where the $+$ ($-$) sign selects the even (odd) parity component in real space. The odd-parity correlators contribute only to nonlocal pairing, since they vanish identically when considering the same lattice sites $mn=m'n'$. 

Finally, to fully symmetrize the anomalous Green function we need to consider the frequency dependence. By choosing even- and odd-frequency combinations we reach
\begin{subequations}\label{eq:Fsym}
    \begin{align}
        F^\text{ESE}_{mn,m'n'} (\omega) = {} & \frac{1}{2} \left[ F^{0+}_{mn,m'n'} (\omega) + F^{0+}_{mn,m'n'} (-\omega) \right] , \\ 
        F^\text{OSO}_{mn,m'n'} (\omega) = {} & \frac{1}{2} \left[ F^{0-}_{mn,m'n'} (\omega) - F^{0-}_{mn,m'n'} (-\omega) \right] , \\ 
        F^{\text{ETO},t}_{mn,m'n'} (\omega) = {} & \frac{1}{2} \left[ F^{t-}_{mn,m'n'} (\omega) + F^{t-}_{mn,m'n'} (-\omega) \right] , \\ 
        F^{\text{OTE},t}_{mn,m'n'} (\omega) = {} & \frac{1}{2} \left[ F^{t+}_{mn,m'n'} (\omega) - F^{t+}_{mn,m'n'} (-\omega) \right] ,
    \end{align}
\end{subequations}
with $t=+,-,z$ labeling the triplet components. 
We have thus obtained the four fermionic pairing channels: even-frequency singlet even-parity (ESE), odd-frequency singlet odd-parity (OSO), even-frequency triplet odd-parity (ETO), and odd-frequency triplet even-parity (OTE)~\cite{Tanaka_reviewJPSJ,Cayao2020odd,Tanaka2024May}. 

In the following sections we only consider Green functions at the \gls{jj} interface, where the \glspl{mbs} emerge, so we always set the horizontal components to be $m=m'=N_x$ and only allow changes in the vertical components along the interface. We thus define the short-hand notation
\begin{equation}\label{eq:short-notation}
    F_{nn'} \equiv F_{N_xn,N_xn'} ,
\end{equation}
which we apply to the fermionic pairing channels in \Cref{eq:Fsym}. 
Additionally, to better visualize the polarized triplet components, we also define
%\begin{subequations}\label{eq:ote_pol}
%    \begin{align}
%    p^{\textrm{ETO}}_{nn'} = {}& \mathrm{sgn}( F^{\text{ETO},+}_{nn'} - F^{\text{ETO},-}_{nn'} ) \sqrt{ |F^{\text{ETO},+}_{nn'}|^2 + |F^{\text{ETO},-}_{nn'}|^2}, \\
%    p^{\textrm{OTE}}_{nn'} = {}& \mathrm{sgn}( F^{\text{OTE},+}_{nn'} - F^{\text{OTE},-}_{nn'} ) \sqrt{ |F^{\text{OTE},+}_{nn'}|^2 + |F^{\text{OTE},-}_{nn'}|^2} .
%\end{align}
%\end{subequations}
\begin{equation}\label{eq:ote_pol}
    p^{\textrm{OTE}}_{nn'} = s_{nn'}^\text{OTE} \sqrt{ |F^{\text{OTE},+}_{nn'}|^2 + |F^{\text{OTE},-}_{nn'}|^2}, 
\end{equation}
with 
\begin{equation}
    s_{nn'}^\text{OTE} = \mathrm{sgn}( F^{\text{OTE},+}_{nn'} - F^{\text{OTE},-}_{nn'} ). 
\end{equation}
This expression contains the total magnitude of the polarized triplets in the square root term, while keeping the information about the local spin polarization in the sign term $s_{nn}^\text{OTE}$. Below, we focus on the local components of \Cref{eq:ote_pol}, $p_{nn}^\text{OTE}$. 

\subsection{Density of states at the interface}

The anomalous Green function is not directly observable, but can be inferred from the density of states via scanning tunneling microscopy spectra~\cite{Perrin_PRL_2020}. 
The \gls{ldos} at site $mn$ is obtained directly from the retarded Green function as
\begin{equation}\label{eq:dos}
    \rho_{mn}(\omega)= -\frac{1}{\pi} \Im \left\{ \Tr \left[ \check{G}_{mn,mn}^\text{R}(\omega) \right] \right\},
\end{equation}
where the trace runs over the spin-Nambu subspace. Since we are focusing on the induced pairing effects at the junction interface, we can again apply a short-hand notation for the \gls{ldos} along the interface as
\begin{equation}\label{eq:dos_interface}
    \rho_n(\omega) \equiv \rho_{N_xn}(\omega) .
\end{equation}
The total \gls{dos} of the interface is
\begin{equation}\label{eq:dos_total}
    \rho(\omega) = \sum_{n=1}^{N_y} \rho_n(\omega).
\end{equation}

As reference to compare our calculations we consider an unbiased junction without magnetic texture, i.e., with $\phi=0$ and $t_m=0$. This trivial and conventional \gls{jj} only displays ESE pairing, which at zero energy takes the value 
\begin{equation}\label{eq:F0}
    F_0\equiv \sum_{n=1}^{2N_x}\sum_{n=1}^{N_y/2} F^\text{ESE}_{mn,mn}(\omega=0,t_m=0,\phi=0) , 
\end{equation}
where we summed over all horizontal sites and half vertical ones, due to symmetry constraints. 
We use $F_0$ as a normalization when exploring the low-energy pairing in the next sections. 

We also define a normalization for the density of states,
\begin{equation}\label{eq:dos-norm}
    \rho_0= \frac{1}{\Delta} \int_{\mu-\Delta/2}^{\mu+\Delta/2}\rho(\omega) \rm{d}\omega,
\end{equation}
which is computed at a range of energies away from the superconducting gap appearing around $\omega\sim0$. 

%%%%%%%%%%%%%%%%%%%%%%%%%%%%%%%
%  PAIRINGS
%%%%%%%%%%%%%%%%%%%%%%%%%%%%%%%

\begin{figure}
	\centering
		%\subfloat{\includegraphics[width=1\columnwidth]{./draft-img/fig-1/setup-conf-sinus_v4.pdf}}\\
	\subfloat{\includegraphics[width=\linewidth]{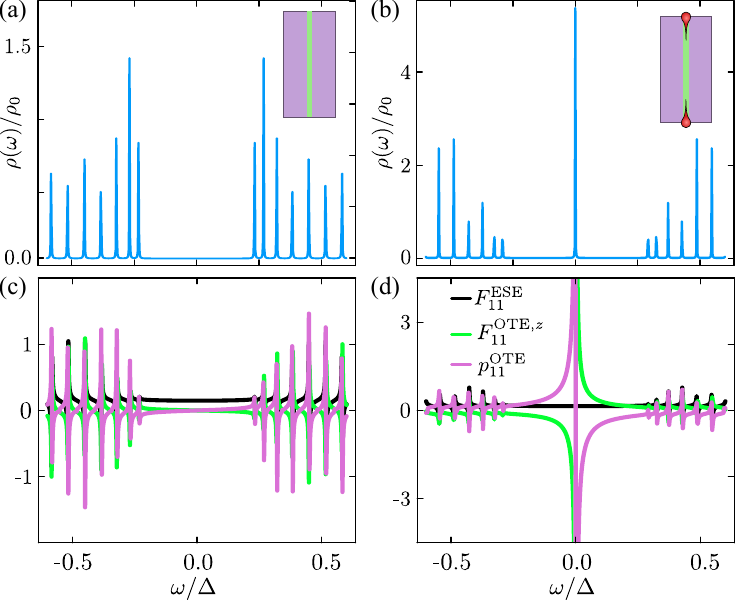}}
	\caption{Interface density of states and pairing. 
    (a,b) Total \gls{dos} as a function of $\omega$ in the (a) trivial ($t_m=0.4t$) and (b) topological ($t_m=0.6t$) phases. (c,d) Local correlators at the junction edge ($L_x, a$) in the (c) trivial and (d) topological phases. In all cases, $\mu=-3.9t$, $t_0=t$, $\phi=0$, $\Delta=0.2t$. 
     }\label{fig:3_dos}
\end{figure}

\section{Pairing induced by the magnetic texture\label{sec:spiral}}

We explore a 2D \gls{jj} coupled through a magnetic-textured barrier with a spatial modulation along the junction interface, see \Cref{fig:1_setup,fig:2_sketch}. At zero phase bias ($\phi=0$), this system enters the topological regime when the superconducting coherence length $\xi_S=\hbar v_F/\Delta$ and the magnetization periodicity $\xi_m$ are comparable~\cite{Sardinero2024}. In the topological regime, a pair of \glspl{mbs} emerge and localize at the edges of the junction interface, i.e., at sites ($N_x,1$) and ($N_x,N_y$) [\Cref{fig:1_setup,fig:2_sketch}]. 
We only consider a harmonic variation of the magnetic texture that maintains the amplitude constant, see \Cref{eq:barrierModel}. Other spiral-like textures also lead to topologically non-trivial phases, see discussion in Ref.~\cite{Sardinero2024}.%The topological phase diagram for this \textit{spiral} texture is representative of other periodic magnetization profiles~\cite{Sardinero2024}. 

\begin{figure*}
	\centering
%\subfloat{\includegraphics[width=1\columnwidth]{./draft-img/fig-1/setup-conf-sinus_v4.pdf}}\\
	\subfloat{\includegraphics[width=1.\linewidth]{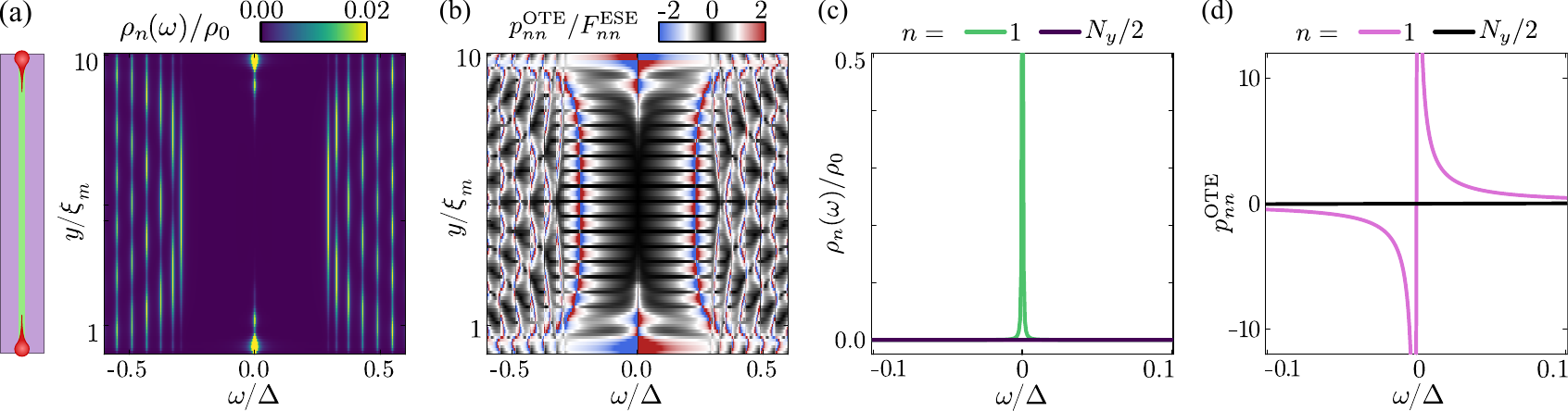}}
	\caption{Wide junction in the topological regime. 
    (a) \gls{ldos} along the interface as a function of the energy. The Majorana edge states are decoupled as shown in the sketch. 
    %Colorbar saturates at $0.02$.
    (b) Ratio between the local polarized odd-frequency triplet and even-frequency singlet, as a function of energy. 
    %Colorbar saturates at $\pm2$.
    (c) \gls{ldos} as a function of the energy computed at the bottom edge $y=a$ and in the middle of the interface $y=L_y/2$. 
    (d) Ratio $p^\text{OTE}/F^\text{ESE}$ at the same points. In all cases, $\mu=-3.9t$, $t_0=t$, $t_m=0.6t$, $\phi=0$, $\xi_m/L_y=0.1$ and $\Delta=0.2t$.
    }\label{fig:4_wide} 
\end{figure*}

\subsection{Trivial and nontrivial regimes}

We first explore the connection between the density of states and the induced pairing in the trivial and topological regimes. 
In \Cref{fig:3_dos}, we show the total \gls{dos}, \Cref{eq:dos_total}, for energies around the superconducting gap, both in the trivial (left panels) and the nontrivial (right  panels) topological phases. 
The resonances over the induced gap $\Delta_0\sim 0.25\Delta$ are \glspl{abs} and emerge for both the trivial and nontrivial phases. In the nontrivial phase, the sharp peak at $\omega=0$ corresponds to the \gls{mbs} [\Cref{fig:3_dos}(b)]. 

Since the \glspl{mbs} emerge at the edges of the interface, which are equivalent, we focus on the anomalous local correlator $F_{11}$ and decompose it into the symmetric pairings defined in \Cref{eq:Fsym}. For local pairings only ESE and OTE states are allowed, but the triplet can still present the non-polarized and the polarized components. These triplet components have similar magnitude and opposite sign since the local magnetization is not usually aligned with the spin quantization axis. 

The local correlators in the topologically trivial phase, \Cref{fig:3_dos}(c), peak around the \glspl{abs} and have a small but finite value inside the induced gap $\Delta_0$~\cite{SeoaneSouto2024Oct}. 
The even-frequency singlet pairing (black line) is usually dominant over the odd-frequency triplets. 
In the nontrivial regime, the correlators for \glspl{abs} resonances have qualitatively the same behavior. 
However, the induced pairings are drastically different below $\Delta_0$, see \Cref{fig:3_dos}(d). 

Ignoring extra degrees of freedom and focusing only on the frequency dependence, the anomalous Green function associated to a Majorana operator $\gamma^\dagger=\gamma$ must adopt the form $F(\omega)\propto \langle \gamma\gamma \rangle\sim1/\omega$~\cite{Kashuba2017May,Tamura2019May,Kuzmanovski2020Mar,Cayao2024Sep,Ahmed2025Jun,Cayao2020odd,Tanaka2024May}. In fact, the Majorana property of self-conjugation yields that the propagating and anomalous Green functions must coincide, $\langle \gamma^\dagger\gamma \rangle = \langle \gamma\gamma \rangle \sim 1/\omega$. Consequently, for \textit{pure} Majorana modes there is a connection between a resonant zero-energy density of states and a $1/\omega$ divergence of the odd-frequency equal-spin triplet~\cite{Asano2013Mar}. 
Indeed, we observe in \Cref{fig:3_dos}(d) that the odd-frequency triplets become dominant and display a typical $1/\omega$ resonant behavior when the \glspl{mbs} emerge and localize~\cite{Tamura2019May,Tamura2021Oct}. 
%This is a consequence of the intrinsic odd-frequency nature of the Majorana modes~\cite{Kashuba2017May}. 

\begin{figure*}
	\centering
    \subfloat{\includegraphics[width=1.\linewidth]{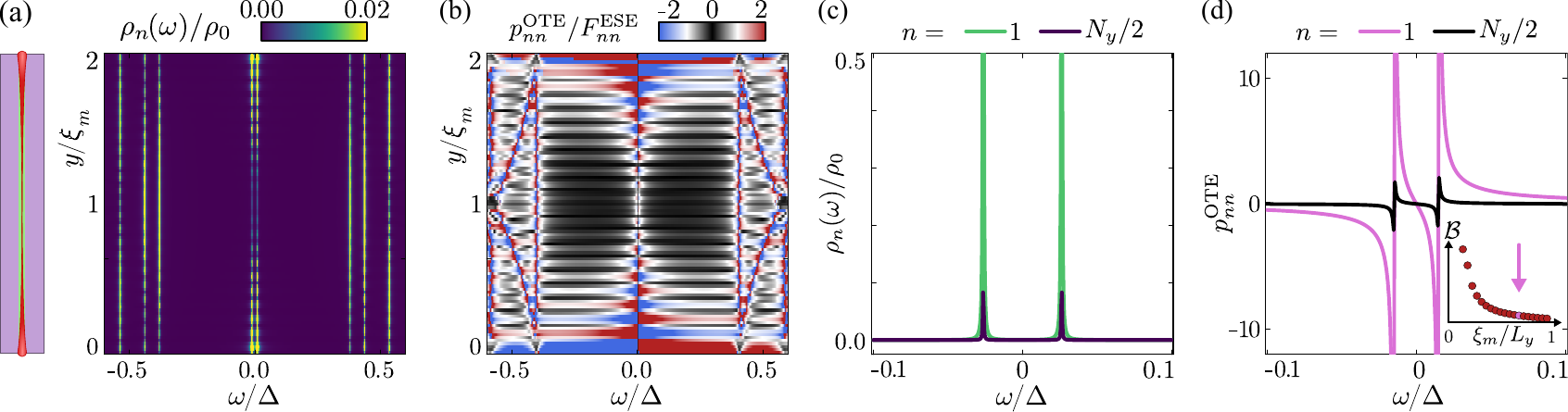}}
	\caption{Narrow junction in the topological regime. 
    (a) \gls{ldos} along the interface as a function of the energy. The wave function of the different Majorana edge states overlaps as indicated in the sketch. 
    %Colorbar saturates at $0.1$.
    (b) Ratio $p^\text{OTE}/F^\text{ESE}$ along the interface as a function of $\omega$. 
    %Colorbar saturates at $\pm2$.
    (c) \gls{ldos} as a function of the energy computed at $y=a$ and $y=L_y/2$. 
    (d) Ratio $p^\text{OTE}/F^\text{ESE}$ at the same points as a function of $\omega$. 
    Inset: Log-scale plot of the zero-energy slope $\mathcal{B}$ as a function of $\xi_m/L_y$, with an arrow indicating the value corresponding to this figure. 
    In all cases, $\mu=-3.9t$, $t_0=t$, $t_m=0.6t$, $\phi=0$, $\xi_m/L_y=0.5$ and $\Delta=0.04t$. 
}\label{fig:5_narrow} 
\end{figure*}

\subsection{Role of Majorana hybridization on odd-frequency pairing}

The previous analysis assumed that the junction width $L_y$ was large enough so that the emerging \glspl{mbs} in the nontrivial phase were decoupled [\Cref{fig:2_sketch}(a)]. Indeed, when $L_y$ is larger than the localization length of the \glspl{mbs}, which depends on the induced gap $\Delta_0$, the wavefunctions of the edge Majorana states do not overlap as they decay exponentially inside the \gls{jj} interface. For narrower junctions, the wavefunctions can overlap and the \glspl{mbs} hybridize. We now compare the regimes where \glspl{mbs} are isolated or overlap. 

%We start with a wide junction where the \glspl{mbs} are decoupled, as shown in the sketch in \Cref{fig:4_wide}(a). This sketch coincides with the computed \gls{ldos} along the interface as a function of the energy for a topologically nontrivial case. 
We start with a wide junction where the \glspl{mbs} are decoupled by computing the \gls{ldos} along the interface as a function of the energy for a topologically nontrivial case in \Cref{fig:4_wide}(a).
We have set the junction width $L_y$ to be larger than the superconducting coherence length $\xi_S$ and magnetic period $\xi_m$ that control the decay of the Majorana edge modes. Consequently, the \glspl{mbs} are not hybridized and emerge at zero energy. 
Moreover, for energies over the induced gap we see the \glspl{abs}, which are modulated according to the solutions of a potential well, with one maximum at $L_y/2$ for the first state, two maxima for the second state, three for the third, and so on. 

\Cref{fig:4_wide}(b) shows the energy dependence of the polarized odd-frequency triplets, \Cref{eq:ote_pol}, along the interface. We have rescaled the magnitude of $p^\text{OTE}_{nn}$ in the figure dividing it by the corresponding value of the even-frequency singlet, $F^\text{ESE}_{nn}$. Black color then indicates vanishing polarized triplet and white colors correspond to a similar contribution from $p^\text{OTE}_{nn}$ and $F^\text{ESE}_{nn}$ (although in some cases both are vanishingly small). The red and blue colors, by contrast, indicate a strong presence of spin-polarized odd-frequency triplet. The red-blue checkered behavior of $p^\text{OTE}_{nn}$ is a consequence of the spin-polarization axis changing with the orientation of the magnetic texture. That is, the dominant polarized triplet component ($F^{\text{OTE,}+}$ or $F^{\text{OTE,}-}$) is determined by the local spin texture. 
As expected, $p^\text{OTE}_{nn}$ is very dominant over the singlet pairing around zero energy and at the edges of the interface, where the \glspl{mbs} localize. There is also a dominant presence of polarized odd-frequency triplets for the trivial \glspl{abs} at higher energies~\cite{Tanaka_reviewJPSJ,Tamura2019May}. 
We focus here on the low-energy topological states, but the strong spin-polarization of the trivial \glspl{abs} could be interesting for exploring applications in superconducting spintronics~\cite{Linder_NatPhys_2015, Eschrig_RPP_2015}. 

The zero-energy peak at the interface edge corresponding to the \glspl{mbs} is clearly visible in \Cref{fig:4_wide}(c) and it vanishes before reaching the middle of the interface at $y=L_y/2$. 
Since the \glspl{mbs} at the interface edges are decoupled, $\rho_1(\omega)$ features a sharp peak at $\omega=0$ and the corresponding polarized triplet $p^\text{OTE}_{11}$ diverges as $1/\omega$, see \Cref{fig:4_wide}(d). 

We now consider in \Cref{fig:5_narrow}(a) the \gls{ldos} in a narrow junction where the \glspl{mbs} have overlapping wavefunctions and, thus, hybridize. 
%Again, the simple sketch coincides with the \gls{ldos} calculation in \Cref{fig:5_narrow}(a), which also shows the Majorana hybridization. 
The energy splitting is visible in the \gls{ldos} shown in  \Cref{fig:5_narrow}(a) along the interface. 
The ratio $p^\text{OTE}_{nn}/F^\text{ESE}_{nn}$ in \Cref{fig:5_narrow}(b) indicates that the triplet pairing extends along the interface, following the decay and overlap of the edge \glspl{mbs}. 
%
%We now study in more detail the zero-energy behavior of the odd-frequency triplet pairings comparing panels \Cref{fig:4_wide}(d) and \Cref{fig:5_narrow}(d). 
%
%When the \glspl{mbs} are decoupled the \gls{ldos} at the site where the Majorana state emerges, e.g., $\rho_1(\omega)$, features a sharp peak at $\omega=0$. At the same time, the corresponding polarized triplet $p^\text{OTE}_{11}$ diverges as $1/\omega$, see \Cref{fig:4_wide}(c,d). 
%
We plot in \Cref{fig:5_narrow}(c) the \gls{ldos} $\rho_1(\omega)$ of a hybridized Majorana mode displaying two resonances at $\omega=\pm \varepsilon$ and a vanishing density at zero energy. 
The resonances of the polarized triplet $p^\text{OTE}_{nn}$ in \Cref{fig:5_narrow}(d) are shifted accordingly to $1/(\omega\pm \varepsilon)$. Moreover, while $p^\text{OTE}_{nn}$ for the decoupled system simply changed sign at $\omega=0$ [\Cref{fig:4_wide}(d)], the odd-frequency triplet pairings now feature an additional sign change at $\omega\sim\varepsilon$, see \Cref{fig:5_narrow}(d). 

Majorana states that hybridize in narrow junctions loose the self-conjugation property and no longer follow a pure $1/\omega$ behavior~\cite{Ahmed2025Jun}. 
The low-energy behavior of odd-frequency polarized triplets can be approximated as 
\begin{equation}\label{eq:low-freq_P}
    p^\text{OTE}_{11} (\omega\ll\Delta_0) \sim \frac{\mathcal{W}}{2} \Bigl( \frac{1}{\omega + \varepsilon} + \frac{1}{\omega - \varepsilon} \Bigr), 
\end{equation}
where the the parameter $\mathcal{W}$ can be associated to the junction topological invariant~\cite{Tamura2019May,Tamura2021Oct,Ahmed2025Jan}, as we discuss in the next section. 
When the \glspl{mbs} are decoupled in wider junctions with $\xi_m\ll L_y$ the hybridization vanishes ($\varepsilon=0$) and $p^\text{OTE}_{11}$ recovers the $1/\omega$ behavior of self-conjugated Majorana modes. 
By contrast, in narrow junctions with $\varepsilon>0$ the behavior at zero frequency is linear with slope $\mathcal{B}=-\mathcal{W}/\varepsilon^2$. 
%In the limit of strong coupling between Majorana states, $\xi_m\sim L_y$, the hybridization is maximal and the slope $\mathcal{B}$ tends to a small constant value. 
We confirm this approximation by plotting the slope $\mathcal{B}$ in the inset of \Cref{fig:5_narrow}(d), where the vertical axis is in log scale. 
Consequently, the zero-frequency behavior of the polarized triplets at low energy is a signature of the \textit{purity} (as in self-conjugation property) of emergent Majorana states. 

\begin{figure*}
	\centering
	\subfloat{\includegraphics[width=\linewidth]{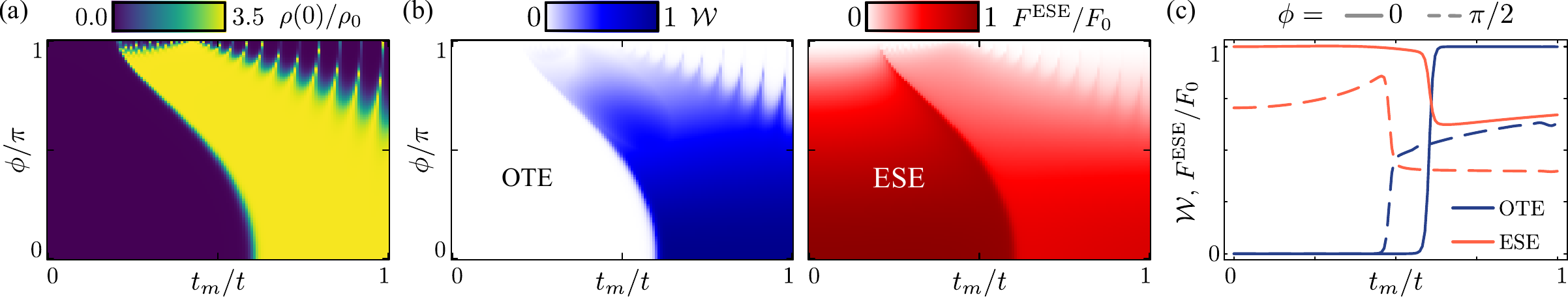}}
	\caption{Topological phase diagram. 
    (a) Zero-energy total \gls{dos} at the interface as a function of $t_m$ and $\phi$. 
    (b) Same maps for $\mathcal{W}$ (left, in blue) and $F^\text{ESE}(0) $ (right, in red). 
    (c) $\mathcal{W}$ and $F^\text{ESE}(0)$ as a function of $t_m$ for $\phi=0$ (solid) and $\phi=\pi/2$ (dashed). 
    In all cases, $\mu=-3.75t$, $\Delta=0.2t$, $t_0=t$, $N_x=10$, $\xi_m=0.1L_y$, $N_y=100$. 
    }\label{fig:6_topodiagram} 
\end{figure*}

\subsection{Majorana hybridization at finite superconducting phase difference}

Thus far we only explored unbiased junctions with $\phi=0$. However, the phase difference affects the topological phase diagram of a \gls{jj} coupled by a magnetic texture and the \glspl{mbs} localization length~\cite{Sardinero2024}. 
Previous works on \glspl{jj} have illustrated that the topologically nontrivial phase can be controlled by the superconducting phase difference~\cite{Pientka2017,Hell_PRL2017,Ren_Nature2019,Ikegaya2020Oct,Lesser_2022,Oshima2022May}. In our setup, a finite phase difference $0<\phi<\pi$ reduces the energy of all subgap states so that the topological phase transition requires smaller amplitudes $t_m$ of the magnetization~\cite{Sardinero2024}. Usually, such a phase-induced topological transition comes at the cost of reducing the gap to the excited states. Consequently, the localization length of the \glspl{mbs} is also affected by the phase difference and, in some cases, the hybridization of Majorana modes is \textit{enhanced} by the phase. 

We explore the topological phase diagram in \Cref{fig:6_topodiagram}(a) by computing the zero-energy total \gls{dos} at the interface as a function of the magnetic texture amplitude $t_m$ and the phase difference $\phi$. In the trivial phase (dark regions) the zero-energy total \gls{dos} is zero, but it becomes finite when \glspl{mbs} emerge (lighter regions). For simplicity, we only consider in this phase diagram a wide junction with decoupled \glspl{mbs}. \Cref{fig:6_topodiagram}(a) shows how the phase difference facilitates a topological phase transition: For example, for $t_m\sim t/2$ the junction is in the trivial regime at $\phi=0$ but becomes nontrivial when $\pi/2<\phi<\pi$. 
Moreover, as the phase approaches $\pi$ the topological gap closes and the zero-energy \gls{dos} displays some oscillations, which is another characteristic behavior of topological edge states~\cite{Prada2012Nov,Cayao2017Nov,Cayao2021Jul}. 

As explained above, the induced pairings are intimately connected to the spectral and topological properties of the junction. 
To compare with the zero-energy \gls{dos}, we compute in \Cref{fig:6_topodiagram}(b) the zero-frequency pairing amplitudes. Since the odd-frequency state is, by definition, zero at $\omega=0$, we define 
\begin{equation}\label{eq:invariant}
        \mathcal{W} = \lim_{z\to 0} z \Bigl[ \sum_{\sigma=\pm} \sum_{m=1}^{2N_x} \sum_{n=1}^{N_y/2} F^{\text{OTE},\sigma}_{mn,mn} (z) \Bigr], 
\end{equation}
with $z=\omega+i0^+$. 
The quantity $\mathcal{W}$, cf. \Cref{eq:low-freq_P}, corresponds to the topological invariant at $\phi=0$~\cite{Tamura2019May,Tamura2021Oct,Ahmed2025Jan}. 
On the other hand, the even frequency singlet is finite at zero energy, so we can compute it as
\begin{equation}\label{eq:ESElocal}
    F^\text{ESE} \equiv \sum_{m=1}^{2N_x} \sum_{n=1}^{N_y/2} F_{mn,mn}^\text{ESE}(\omega=0),
\end{equation}
and normalize it with respect to the singlet pairing in a conventional \gls{jj}, see \Cref{eq:F0}. 

\Cref{fig:6_topodiagram}(b) shows that in the trivial regions the zero-frequency limit of the even-frequency singlet is maximum and the odd-frequency triplet is zero. By contrast, the even-frequency singlet is reduced in the nontrivial regime and the odd-frequency triplet becomes dominant. At zero phase, \Cref{eq:invariant} is quantized as corresponds to the topological invariant. For clarity, \Cref{fig:6_topodiagram}(c) shows these quantities for $\phi=0$ and $\pi/2$.  

\begin{figure}[b]
	\centering
	\subfloat{\includegraphics[width=\linewidth]{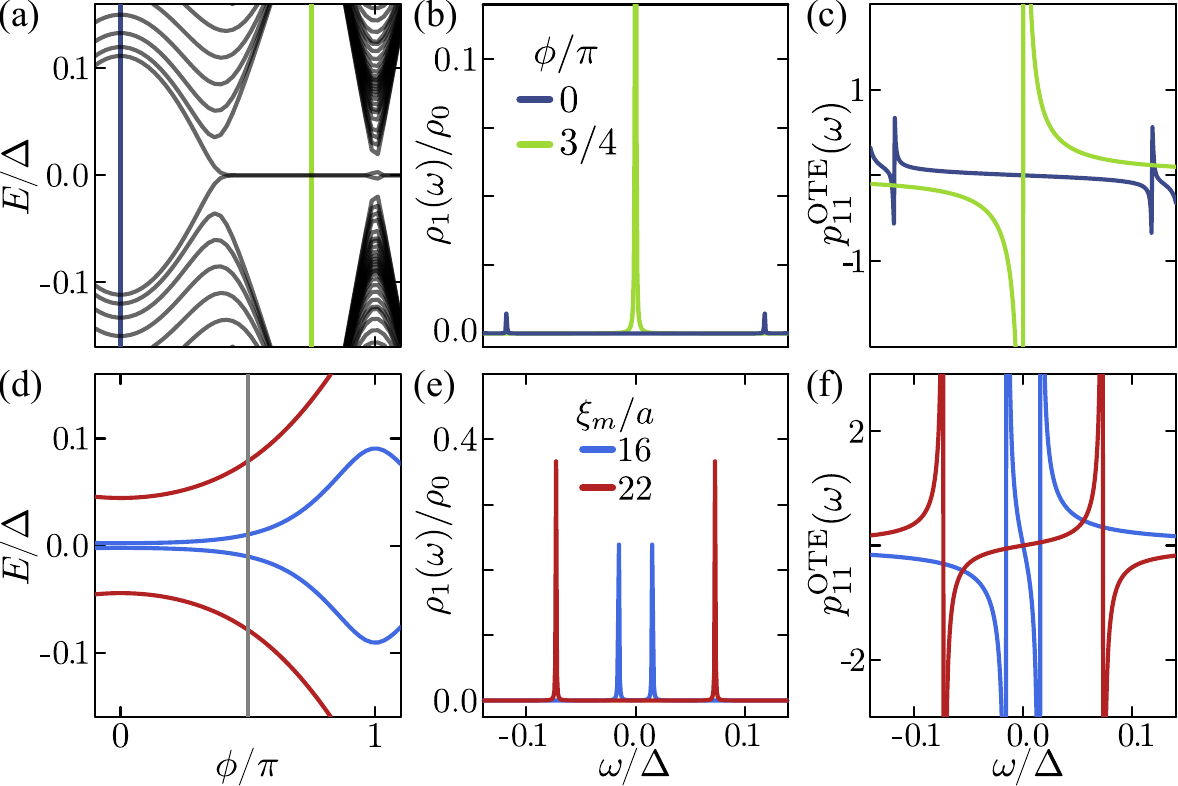}}
	\caption{Phase bias effect on Majorana hybridization.
    (a,b,c) Wide junction in the trivial regime at $\phi=0$, with $t_m=0.5t$ and $\xi_S=0.6\xi_m$. (a) Lowest energy levels (black lines). 
    (b) Edge \gls{ldos} and (c) $p^\text{OTE}$ at $\phi=0$ [blue line in panel (a)] and $3\pi/4$ (green line). 
    (d,e,f) Narrow junction in the nontrivial regime at $\phi=0$, with $t_m=0.8t$ and $\xi_S=1.2\xi_m$. 
    (d) Lowest two energy levels for $\xi_m=16a$ (blue lines) and $22a$ (red lines). (e) Edge \gls{ldos} and (f) $p^\text{OTE}$ at $\phi=\pi/2$ [gray line in (d)] for the magnetic texture periods $\xi_m$ used in (d). 
    In all cases, $\mu=-3.75t$, $t_0=t$, and $L_y=200a$. 
    }\label{fig:7_phase} 
\end{figure}

We now focus on a wide junction in the trivial regime where the superconducting phase precipitates the topological phase transition. In \Cref{fig:7_phase}(a) we show the lowest energy levels as a function of $\phi$, and how they merge at zero energy around $\phi\lesssim\pi/2$. As expected, in the nontrivial phase we observe a zero-energy peak in the \gls{ldos} and the corresponding $1/\omega$ divergence of the polarized triplet, see \Cref{fig:7_phase}(b-c). 
For the trivial regime with $\phi\sim0$ the lowest energy \glspl{abs} has resonances at finite frequencies and the behavior of $p^\text{OTE}$ at zero frequency is linear (blue line). According to the map in \Cref{fig:6_topodiagram}(b), this regime is dominated by ESE pairing. The topological phase transition occurs around $\phi\lesssim\pi/2$ with a suppression of $F^\text{ESE}$ and an enhancement of $p^\text{OTE}$. 
The maximum induced topological gap is reached around $\phi=3\pi/4$, green line in \Cref{fig:7_phase}(a), with $p^\text{OTE}$ displaying a sharp resonance at zero energy. 
As the phase difference approaches $\phi=\pi$, the topological gap quickly reduces until the zero-energy states split at $\phi=\pi$. Our analysis of the polarized OTE pairing around zero frequency reveals that the $1/\omega$ behavior is broken at $\phi=\pi$ by a linear term. The hybridized Majorana modes at $\phi=\pi$ thus display $1/(\omega\pm\varepsilon)$ resonances. 

We now consider the opposite situation: a narrow junction in the nontrivial regime [\cref{fig:7_phase}(d-f)]. At $\phi=0$ the Majorana edge states are hybridized due to the overlap between their wavefunctions. The \gls{ldos} features two peaks at the hybridization energies, and $p^\text{OTE}_{11}$ is also resonant at those energies but linear around zero frequency. A finite $\phi\neq 0$ modifies the wavefunctions of the edge states, enhancing their overlap across the interface. Consequently, the hybridization energy increases, see \cref{fig:7_phase}(e). The polarized triplet displays a linear behavior at low energy, with the slope decreasing as the hybridization is increased [\cref{fig:7_phase}(f)]. 

\begin{figure*}
	\centering
	\subfloat{\includegraphics[width=\linewidth]{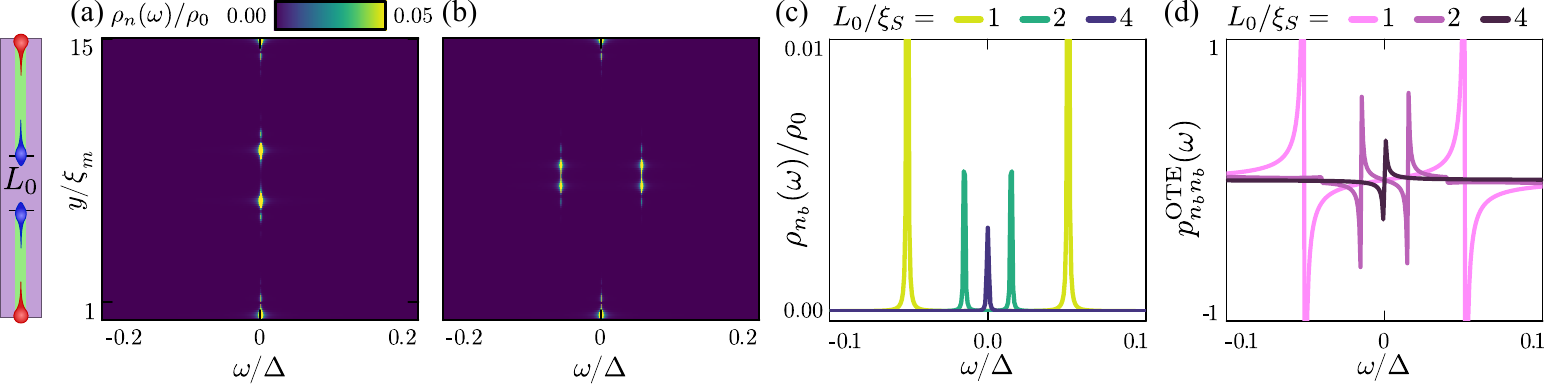}}
	\caption{Two topological segments separated by a nonmagnetic region. 
    (a,b) \gls{ldos} as a function of the energy for $L_0=4\xi_S$ (a) and $L_0=\xi_S$ (b). %Colorbar saturates at $0.05$.
    (c,d) \gls{ldos} (c) and polarized triplet (d) at the inner edge site $n_b=y_b/a$ for different values of $L_0$. 
    Parameters are $\mu=-3.75t$, $t_0=t$, $\Delta=0.2t$, $\xi_m=10a$, and $L_y=150$. }\label{fig:8_barrier-dos} 
\end{figure*}

%%%%%%%%%%%%%%%%%%%%%%%%%%%%%%%
%  BARRIER
%%%%%%%%%%%%%%%%%%%%%%%%%%%%%%%

\section{Magnetic texture interrupted by a nonmagnetic barrier\label{sec:barrier}}

We now consider the situation where the magnetic texture coupling the superconductors is interrupted by a nonmagnetic region, see \Cref{fig:2_sketch}(c,d). This region represents, for example, the experimental case where the superconductors come into direct contact bypassing the magnetic texture. In the topological regime, the 1D topological superconductor emerging at the interface between superconductors is divided into two nontrivial segments, as long as the length of the two surviving magnetic regions is several times the superconducting coherence length $\xi_S$, which is itself comparable to the magnetic length $\xi_m$. 

\subsection{Majorana bound states at the inner edges}

We model the nonmagnetic region as an interface segment of $N_0$ sites without magnetic hopping, ranging from $y_b$ to $y_t$, with $a<y_b<y_t<L_y$, see \Cref{eq:barrierModel} and \Cref{fig:2_sketch}(c,d). This region determines two new topological boundaries or \textit{inner edges}, located at the interface for $y=y_b$ and $y_t$, each hosting one \gls{mbs}. 
In the nontrivial regime, the junction interface is formed by two 1D topological superconductors, each hosting a pair of \glspl{mbs} located at one of the external edges and one of the inner ones, respectively, red and blue regions of \Cref{fig:2_sketch}(c,d). That is, one of the topological superconductor hosts Majorana modes at $y=a$ and $y_b$, while the other one has them at $y_t$ and $L_y$. 
The outermost \glspl{mbs} are localized at the edges of the system and their wavefunction decays toward the center of the system. By contrast, the wavefunctions of the inner \glspl{mbs} can extend in both directions: inside the magnetic and nonmagnetic barriers. 
Consequently, depending on the width $L_0=N_0 a $ of the nonmagnetic region the \glspl{mbs} at the inner edges emerge at zero energy (if $L_0 \gg \xi_S$, yielding two decoupled topological superconductors) or hybridize and acquire a finite energy $\varepsilon>0$ (if $L_0\gtrsim\xi_S$). 
\Cref{fig:8_barrier-dos}(a,b) shows the \gls{ldos} as a function of the energy along the interface at $\phi=0$. Around zero energy, we can clearly see that the outer \glspl{mbs} emerge exactly at zero energy in both cases, but the inner ones can split at finite energies $\pm \varepsilon$. In the first case, \Cref{fig:8_barrier-dos}(a), the inner \glspl{mbs} are separated by a long trivial region of length $L_0\approx4\xi_S$, and are hence decoupled. By contrast, the second map, \Cref{fig:8_barrier-dos}(b), shows the case where they become hybridized due to a narrow trivial region, $L_0 \approx \xi_S$. 

We evaluate the \gls{ldos} around $\omega=0$ at the inner edge $y=y_b$ in \Cref{fig:8_barrier-dos}(c) for three different values of the barrier length. As the Majorana state hybridizes, the local odd-frequency polarized triplets $p_{y_b y_b}^{\rm{OTE}}$ [\Cref{fig:8_barrier-dos}(d)] transition from the $1/\omega$ resonance for the decoupled case ($L_0\gg\xi_S$) into a linear zero-frequency behavior with resonances around $\omega=\varepsilon$ for the hybridized ones. 

We can conclude that the outer Majorana modes preserve their self-conjugation property, as their associated odd-frequency equal-spin triplet pairing features a $1/\omega$ divergence. By contrast, when the nonmagnetic barrier is narrow enough, the inner \glspl{mbs} hybridize and their associated $p^\text{OTE}$ develops a linear behavior around $\omega\sim0$, see \Cref{fig:8_barrier-dos}(d). 

\begin{figure}[b]
\subfloat{\includegraphics[width=\linewidth]{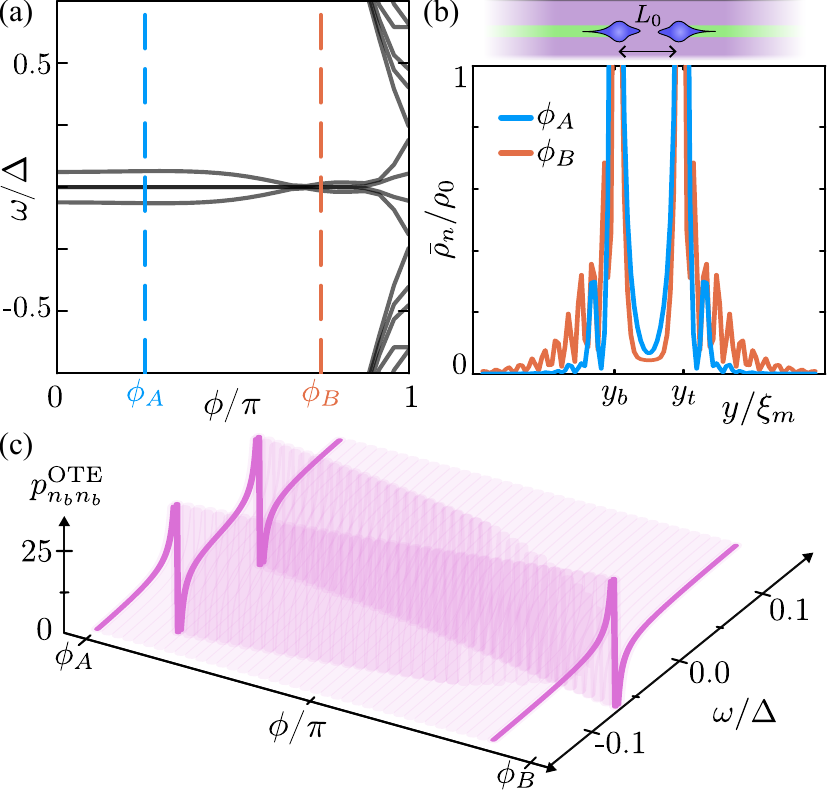}}
\caption{Phase stabilization of inner Majorana modes. 
    (a) Phase dependence of the lowest energy levels for a nontrivial junction with nonmagnetic barrier. 
    (b) Plot of $\bar{\rho}_n$ along the interface, in a segment around the nonmagnetic barrier, for $\phi_A=\pi/4$ [blue dashed line in panel (a)] and $\phi_B=3\pi/4$ (red lines). 
    (c) Energy and phase dependence of $p^\text{OTE}_{n_bn_b}$. 
    In all cases, $\mu=-3.75t$, $t_0=t$, $t_m=0.8t$, and $L_0=3\xi_S$. 
}\label{fig:9_inner}
\end{figure}

\subsection{Phase effect on the hybridization of inner states}

In the previous section we showed how the superconducting phase difference can be detrimental for the purity of the pair of \glspl{mbs} emerging in a single topological superconductor. The reason is that the phase affects the localization of the Majorana states and increases the overlap of their wavefunctions in narrow junctions. We now explore in \Cref{fig:9_inner} the effect of the phase on the \glspl{mbs} emerging at the edges of the nonmagnetic barrier. We only consider junctions in the nontrivial regime with barriers widths comparable to the localization length of the edge states. Consequently, at $\phi=0$ the inner edge states are hybridized. However, the superconducting phase now reduces the hybridization energy, see \Cref{fig:9_inner}(a). That is, the phase is shortening the tails of the \glspl{mbs} inside the nonmagnetic regions, reducing the overlap between inner states. At the same time, however, increasing the phase delocalizes the same Majorana edge states inside the magnetic regions. This effect is clearly shown in \Cref{fig:9_inner}(b) where we compare the wavefunctions of the lowest energy states at different phases $\phi=\pi/4$ (blue) and $\phi=3\pi/4$ (orange). 
To do so, we define an energy-averaged \gls{ldos}, 
$
    \bar{\rho}_n = \int_{-2\varepsilon}^{2\varepsilon} \md \omega \rho_n(\omega)
$, with $\varepsilon>0$ being the hybridization energy. 
At higher phases, the orange line in \Cref{fig:9_inner}(b) shows a longer localization length outside the nonmagnetic region ($y<y_b$ and $y>y_t$), but smaller values inside the magnetic barrier ($y_b<y<y_t$). 
The hybridization of the inner \glspl{mbs} in the presence of a nonmagnetic barrier is thus \textit{reduced} by the phase difference. 
In other words, while the phase difference has a detrimental effect for topological \gls{jj} hosting one pair of Majorana modes, it actually helps on setups with two topological superconductors by decoupling their \glspl{mbs}. 
While the phase helps localize the inner Majorana modes for wider barriers ($L_0\approx4\xi_S$), the hybridization of such states oscillates with the phase for narrower ones. In all cases, the nontrivial regime breaks down at $\phi=\pi$.

We now explore how the phase difference can improve the self-conjugation property of the inner \glspl{mbs} on \glspl{jj} with magnetic textures interrupted by nonmagnetic barriers. 
In \Cref{fig:9_inner}(c) we plot $p^\text{OTE}$ at the inner edge $y_b$ as a function of the energy and phase. The linear behavior of hybridized state at low phases evolves, by increasing the slope at low energies, into an almost perfect zero-energy resonance at higher phases.  
This is an example of how the phase difference can induce a topological phase transition, and also help stabilize the inner Majorana states (i.e., recover the self-conjugation property) at junctions featuring two topological superconductors.

%%%%%%%%%%%%%%%%%%%%%%%%%%%%%%%
%  CONCLUSIONS
%%%%%%%%%%%%%%%%%%%%%%%%%%%%%%%

\section{Conclusions\label{sec:conc}}

Topological superconductivity can emerge at Josephson junctions mediated by a magnetic texture. 
%In the nontrivial phase, a pair of Majorana states appears at the edges of the junction interface. 
We followed a microscopic tight-binding Green function formalism and computed the density of states and the induced pairing by analyzing the anomalous Green function. 
In the trivial regime, the superconductors forming the junction induce a dominant even-frequency singlet pairing at the interface. 
By contrast, we demonstrated that in the nontrivial regime a pair of Majorana states appears at the edges of the junction interface and their associated pairing becomes a spin-polarized odd-frequency triplet state. 
%In the nontrivial regime the odd-frequency triplet state becomes dominant over the even-frequency singlet pairing induced by the superconductors forming the junction. 
When the junction is wide enough, the Majorana bound states are decoupled and their associated odd-frequency triplet pairing displays a characteristic sharp $1/\omega$ resonance at zero frequency. Such a resonant behavior is a consequence of the self-conjugated property that defines Majorana fermions. The odd-frequency pairing in this case is also associated to the junction topological invariant $\mathcal{W}$. 
However, the Majorana modes in a narrow junction in the nontrivial regime hybridize acquiring a finite energy $\varepsilon>0$. Such states are no longer purely self-conjugated and their corresponding odd-frequency triplet pairing features resonances %of smaller amplitude 
at $1/(\omega\pm\varepsilon)$ and a linear behavior around zero frequency. 

%We have also studied a setup where a nonmagnetic barrier interrupts the magnetic texture and splits the topological superconductor into two topological segments, each featuring a pair of Majorana bound states. 
When a nonmagnetic barrier interrupts the magnetic texture, two new Majorana modes appear at the junction. 
Depending on how the barrier width compares to the localization length of the inner Majorana states, these \glspl{mbs}, which belong to different topological segments, can hybridize. When they do hybridize, their associated odd-frequency triplet pairings develop a linear behavior at low energy, like in the narrow junction, although their partner Majorana states at the outer edges remain fixed at zero frequency, featuring $1/\omega$ odd-frequency triplet state (i.e., the outermost \glspl{mbs} are self-conjugated). 
%When the barrier width is comparable to the localization length of the Majorana states, the Majorana modes at the inner edges of each segment hybridize and their associated odd-frequency triplet pairings develop a linear behavior at zero frequency. These inner states are thus no longer purely self-conjugated, although their partner Majorana states at the outer edges remain fixed at zero frequency in a $1/\omega$ odd-frequency triplet state (i.e., they are self-conjugated). 

We also demonstrated that the superconducting phase difference across the Josephson junction can be used as an experimental tuning knob to control the emergence of the nontrivial phase. In a setup with a single topological superconductor along the interface (that is, without nonmagnetic barrier) the phase difference usually enhances the hybridization of Majorana states in narrow junctions. By contrast, in junctions with nonmagnetic barriers of widths comparable to the Majorana localization length, where the inner Majorana modes hybridize, the phase difference contributes to decoupling the states and helps them recover their self-conjugation property. We have checked this by exploring how the odd-frequency triplet for those states reduces the linear component at zero frequency and recovers the $1/\omega$ resonant behavior. 

%Our work explores the potential of Josephson junctions mediated by magnetic textures as Majorana platforms. 

%These results highlight the central role of odd-frequency triplet correlations as a probe of topological superconductivity in magnetically engineered Josephson junctions. We analyzed the conditions for Majorana modes to display their defining property of self-conjugation and how Majorana states from different topological segments along the junction interact among each other. 

\acknowledgments

%\subsection{Author Contributions}
%
%\subsection{Financial Disclosure}
I.~S., R.~S.~S. and P.~B. acknowledge support by the Spanish CM ``Talento Program'' project No.~2019-T1/IND-14088, No.~2022-T1/IND-24070 and No.~2023-5A/IND-28927, the Agencia Estatal de Investigaci\'on project No.~PID2020-117992GA-I00, No.~PID2022-140552NA-I00, No.~PID2024-157821NB-I00 and No.~CNS2022-135950 and through the ``María de Maeztu'' Programme for Units of Excellence in R\&D (CEX2023-001316-M) and the Severo Ochoa Programme No. CEX2024-001445-S. J.~C. acknowledges financial support from the Swedish Research Council (Vetenskapsr\aa det Grant No.~2021-04121). 
%
%\subsection{Conflicts of Interest}
The authors declare no conflicts of interest.

\bibliography{spiral-correlators}

\begin{thebibliography}{149}%
\makeatletter
\providecommand \@ifxundefined [1]{%
 \@ifx{#1\undefined}
}%
\providecommand \@ifnum [1]{%
 \ifnum #1\expandafter \@firstoftwo
 \else \expandafter \@secondoftwo
 \fi
}%
\providecommand \@ifx [1]{%
 \ifx #1\expandafter \@firstoftwo
 \else \expandafter \@secondoftwo
 \fi
}%
\providecommand \natexlab [1]{#1}%
\providecommand \enquote  [1]{``#1''}%
\providecommand \bibnamefont  [1]{#1}%
\providecommand \bibfnamefont [1]{#1}%
\providecommand \citenamefont [1]{#1}%
\providecommand \href@noop [0]{\@secondoftwo}%
\providecommand \href [0]{\begingroup \@sanitize@url \@href}%
\providecommand \@href[1]{\@@startlink{#1}\@@href}%
\providecommand \@@href[1]{\endgroup#1\@@endlink}%
\providecommand \@sanitize@url [0]{\catcode `\\12\catcode `\$12\catcode `\&12\catcode `\#12\catcode `\^12\catcode `\_12\catcode `\%12\relax}%
\providecommand \@@startlink[1]{}%
\providecommand \@@endlink[0]{}%
\providecommand \url  [0]{\begingroup\@sanitize@url \@url }%
\providecommand \@url [1]{\endgroup\@href {#1}{\urlprefix }}%
\providecommand \urlprefix  [0]{URL }%
\providecommand \Eprint [0]{\href }%
\providecommand \doibase [0]{https://doi.org/}%
\providecommand \selectlanguage [0]{\@gobble}%
\providecommand \bibinfo  [0]{\@secondoftwo}%
\providecommand \bibfield  [0]{\@secondoftwo}%
\providecommand \translation [1]{[#1]}%
\providecommand \BibitemOpen [0]{}%
\providecommand \bibitemStop [0]{}%
\providecommand \bibitemNoStop [0]{.\EOS\space}%
\providecommand \EOS [0]{\spacefactor3000\relax}%
\providecommand \BibitemShut  [1]{\csname bibitem#1\endcsname}%
\let\auto@bib@innerbib\@empty
%</preamble>
\bibitem [{\citenamefont {Tanaka}\ \emph {et~al.}(2012)\citenamefont {Tanaka}, \citenamefont {Sato},\ and\ \citenamefont {Nagaosa}}]{Tanaka_reviewJPSJ}%
  \BibitemOpen
  \bibfield  {author} {\bibinfo {author} {\bibfnamefont {Y.}~\bibnamefont {Tanaka}}, \bibinfo {author} {\bibfnamefont {M.}~\bibnamefont {Sato}},\ and\ \bibinfo {author} {\bibfnamefont {N.}~\bibnamefont {Nagaosa}},\ }\bibfield  {title} {\bibinfo {title} {Symmetry and topology in superconductors –odd-frequency pairing and edge states–},\ }\href {https://doi.org/10.1143/JPSJ.81.011013} {\bibfield  {journal} {\bibinfo  {journal} {Journal of the Physical Society of Japan}\ }\textbf {\bibinfo {volume} {81}},\ \bibinfo {pages} {011013} (\bibinfo {year} {2012})}\BibitemShut {NoStop}%
\bibitem [{\citenamefont {Sato}\ and\ \citenamefont {Fujimoto}(2016)}]{Sato2016}%
  \BibitemOpen
  \bibfield  {author} {\bibinfo {author} {\bibfnamefont {M.}~\bibnamefont {Sato}}\ and\ \bibinfo {author} {\bibfnamefont {S.}~\bibnamefont {Fujimoto}},\ }\bibfield  {title} {\bibinfo {title} {Majorana fermions and topology in superconductors},\ }\href {https://doi.org/10.7566/JPSJ.85.072001} {\bibfield  {journal} {\bibinfo  {journal} {J. Phys. Soc. Jpn.}\ }\textbf {\bibinfo {volume} {85}},\ \bibinfo {pages} {072001} (\bibinfo {year} {2016})}\BibitemShut {NoStop}%
\bibitem [{\citenamefont {Sato}\ and\ \citenamefont {Ando}(2017)}]{Sato_2017}%
  \BibitemOpen
  \bibfield  {author} {\bibinfo {author} {\bibfnamefont {M.}~\bibnamefont {Sato}}\ and\ \bibinfo {author} {\bibfnamefont {Y.}~\bibnamefont {Ando}},\ }\bibfield  {title} {\bibinfo {title} {Topological superconductors: a review},\ }\href {https://dx.doi.org/10.1088/1361-6633/aa6ac7} {\bibfield  {journal} {\bibinfo  {journal} {Rep. Prog. Phys.}\ }\textbf {\bibinfo {volume} {80}},\ \bibinfo {pages} {076501} (\bibinfo {year} {2017})}\BibitemShut {NoStop}%
\bibitem [{\citenamefont {Aguado}(2017)}]{Aguado_review2017}%
  \BibitemOpen
  \bibfield  {author} {\bibinfo {author} {\bibfnamefont {R.}~\bibnamefont {Aguado}},\ }\bibfield  {title} {\bibinfo {title} {Majorana quasiparticles in condensed matter},\ }\href {https://doi.org/10.1393/ncr/i2017-10141-9} {\bibfield  {journal} {\bibinfo  {journal} {La Rivista del Nuovo Cimento}\ }\textbf {\bibinfo {volume} {40}},\ \bibinfo {pages} {523} (\bibinfo {year} {2017})}\BibitemShut {NoStop}%
\bibitem [{\citenamefont {Lutchyn}\ \emph {et~al.}(2018)\citenamefont {Lutchyn}, \citenamefont {Bakkers}, \citenamefont {Kouwenhoven}, \citenamefont {Krogstrup}, \citenamefont {Marcus},\ and\ \citenamefont {Oreg}}]{LutchynReview}%
  \BibitemOpen
  \bibfield  {author} {\bibinfo {author} {\bibfnamefont {R.~M.}\ \bibnamefont {Lutchyn}}, \bibinfo {author} {\bibfnamefont {E.~P. A.~M.}\ \bibnamefont {Bakkers}}, \bibinfo {author} {\bibfnamefont {L.~P.}\ \bibnamefont {Kouwenhoven}}, \bibinfo {author} {\bibfnamefont {P.}~\bibnamefont {Krogstrup}}, \bibinfo {author} {\bibfnamefont {C.~M.}\ \bibnamefont {Marcus}},\ and\ \bibinfo {author} {\bibfnamefont {Y.}~\bibnamefont {Oreg}},\ }\bibfield  {title} {\bibinfo {title} {{Majorana zero modes in superconductor{\textendash}semiconductor heterostructures}},\ }\href {https://doi.org/10.1038/s41578-018-0003-1} {\bibfield  {journal} {\bibinfo  {journal} {Nat. Rev. Mater.}\ }\textbf {\bibinfo {volume} {3}},\ \bibinfo {pages} {52} (\bibinfo {year} {2018})}\BibitemShut {NoStop}%
\bibitem [{\citenamefont {Frolov}\ \emph {et~al.}(2020)\citenamefont {Frolov}, \citenamefont {Manfra},\ and\ \citenamefont {Sau}}]{frolov2019quest}%
  \BibitemOpen
  \bibfield  {author} {\bibinfo {author} {\bibfnamefont {S.~M.}\ \bibnamefont {Frolov}}, \bibinfo {author} {\bibfnamefont {M.~J.}\ \bibnamefont {Manfra}},\ and\ \bibinfo {author} {\bibfnamefont {J.~D.}\ \bibnamefont {Sau}},\ }\bibfield  {title} {\bibinfo {title} {Topological superconductivity in hybrid devices},\ }\href {https://doi.org/10.1038/s41567-020-0925-6} {\bibfield  {journal} {\bibinfo  {journal} {Nat. Phys.}\ }\textbf {\bibinfo {volume} {16}},\ \bibinfo {pages} {718} (\bibinfo {year} {2020})}\BibitemShut {NoStop}%
\bibitem [{\citenamefont {Flensberg}\ \emph {et~al.}(2021)\citenamefont {Flensberg}, \citenamefont {von Oppen},\ and\ \citenamefont {Stern}}]{Flensberg_NRM2021}%
  \BibitemOpen
  \bibfield  {author} {\bibinfo {author} {\bibfnamefont {K.}~\bibnamefont {Flensberg}}, \bibinfo {author} {\bibfnamefont {F.}~\bibnamefont {von Oppen}},\ and\ \bibinfo {author} {\bibfnamefont {A.}~\bibnamefont {Stern}},\ }\bibfield  {title} {\bibinfo {title} {Engineered platforms for topological superconductivity and {Majorana} zero modes},\ }\href {https://doi.org/10.1038/s41578-021-00336-6} {\bibfield  {journal} {\bibinfo  {journal} {Nature Reviews Materials}\ }\textbf {\bibinfo {volume} {6}},\ \bibinfo {pages} {944} (\bibinfo {year} {2021})}\BibitemShut {NoStop}%
\bibitem [{\citenamefont {Tanaka}\ \emph {et~al.}(2024)\citenamefont {Tanaka}, \citenamefont {Tamura},\ and\ \citenamefont {Cayao}}]{Tanaka2024May}%
  \BibitemOpen
  \bibfield  {author} {\bibinfo {author} {\bibfnamefont {Y.}~\bibnamefont {Tanaka}}, \bibinfo {author} {\bibfnamefont {S.}~\bibnamefont {Tamura}},\ and\ \bibinfo {author} {\bibfnamefont {J.}~\bibnamefont {Cayao}},\ }\bibfield  {title} {\bibinfo {title} {Theory of {Majorana} zero modes in unconventional superconductors},\ }\href {https://doi.org/10.1093/ptep/ptae065} {\bibfield  {journal} {\bibinfo  {journal} {Progress of Theoretical and Experimental Physics}\ }\textbf {\bibinfo {volume} {2024}},\ \bibinfo {pages} {08C105} (\bibinfo {year} {2024})}\BibitemShut {NoStop}%
\bibitem [{\citenamefont {Fukaya}\ \emph {et~al.}(2025)\citenamefont {Fukaya}, \citenamefont {Lu}, \citenamefont {Yada}, \citenamefont {Tanaka},\ and\ \citenamefont {Cayao}}]{Fukaya_2025}%
  \BibitemOpen
  \bibfield  {author} {\bibinfo {author} {\bibfnamefont {Y.}~\bibnamefont {Fukaya}}, \bibinfo {author} {\bibfnamefont {B.}~\bibnamefont {Lu}}, \bibinfo {author} {\bibfnamefont {K.}~\bibnamefont {Yada}}, \bibinfo {author} {\bibfnamefont {Y.}~\bibnamefont {Tanaka}},\ and\ \bibinfo {author} {\bibfnamefont {J.}~\bibnamefont {Cayao}},\ }\bibfield  {title} {\bibinfo {title} {Superconducting phenomena in systems with unconventional magnets},\ }\href {http://dx.doi.org/10.1088/1361-648X/adf1cf} {\bibfield  {journal} {\bibinfo  {journal} {J. Phys.: Condens. Matter}\ }\textbf {\bibinfo {volume} {37}},\ \bibinfo {pages} {313003} (\bibinfo {year} {2025})}\BibitemShut {NoStop}%
\bibitem [{\citenamefont {Seoane~Souto}\ and\ \citenamefont {Aguado}(2024)}]{Souto_chapter}%
  \BibitemOpen
  \bibfield  {author} {\bibinfo {author} {\bibfnamefont {R.}~\bibnamefont {Seoane~Souto}}\ and\ \bibinfo {author} {\bibfnamefont {R.}~\bibnamefont {Aguado}},\ }\bibinfo {title} {Subgap states in semiconductor-superconductor devices for quantum technologies: {A}ndreev qubits and minimal {M}ajorana chains},\ in\ \href {https://doi.org/10.1007/978-3-031-55657-9_3} {\emph {\bibinfo {booktitle} {New Trends and Platforms for Quantum Technologies}}},\ \bibinfo {editor} {edited by\ \bibinfo {editor} {\bibfnamefont {R.}~\bibnamefont {Aguado}}, \bibinfo {editor} {\bibfnamefont {R.}~\bibnamefont {Citro}}, \bibinfo {editor} {\bibfnamefont {M.}~\bibnamefont {Lewenstein}},\ and\ \bibinfo {editor} {\bibfnamefont {M.}~\bibnamefont {Stern}}}\ (\bibinfo  {publisher} {Springer Nature Switzerland},\ \bibinfo {address} {Cham},\ \bibinfo {year} {2024})\ pp.\ \bibinfo {pages} {133--223}\BibitemShut {NoStop}%
\bibitem [{\citenamefont {Sarma}\ \emph {et~al.}(2015)\citenamefont {Sarma}, \citenamefont {Freedman},\ and\ \citenamefont {Nayak}}]{Sarma_NPJ2015}%
  \BibitemOpen
  \bibfield  {author} {\bibinfo {author} {\bibfnamefont {S.~D.}\ \bibnamefont {Sarma}}, \bibinfo {author} {\bibfnamefont {M.}~\bibnamefont {Freedman}},\ and\ \bibinfo {author} {\bibfnamefont {C.}~\bibnamefont {Nayak}},\ }\bibfield  {title} {\bibinfo {title} {Majorana zero modes and topological quantum computation},\ }\href {https://doi.org/10.1038/npjqi.2015.1} {\bibfield  {journal} {\bibinfo  {journal} {npj Quantum Information}\ }\textbf {\bibinfo {volume} {1}},\ \bibinfo {pages} {15001} (\bibinfo {year} {2015})}\BibitemShut {NoStop}%
\bibitem [{\citenamefont {Lahtinen}\ and\ \citenamefont {Pachos}(2017)}]{Lahtinen_2017}%
  \BibitemOpen
  \bibfield  {author} {\bibinfo {author} {\bibfnamefont {V.}~\bibnamefont {Lahtinen}}\ and\ \bibinfo {author} {\bibfnamefont {J.~K.}\ \bibnamefont {Pachos}},\ }\bibfield  {title} {\bibinfo {title} {A short introduction to topological quantum computation},\ }\href {https://doi.org/10.21468/SciPostPhys.3.3.021} {\bibfield  {journal} {\bibinfo  {journal} {SciPost Phys.}\ }\textbf {\bibinfo {volume} {3}},\ \bibinfo {pages} {021} (\bibinfo {year} {2017})}\BibitemShut {NoStop}%
\bibitem [{\citenamefont {Beenakker}(2020{\natexlab{a}})}]{beenakker2019search}%
  \BibitemOpen
  \bibfield  {author} {\bibinfo {author} {\bibfnamefont {C.~W.~J.}\ \bibnamefont {Beenakker}},\ }\bibfield  {title} {\bibinfo {title} {{Search for non-{A}belian {M}ajorana braiding statistics in superconductors}},\ }\href {https://doi.org/10.21468/SciPostPhysLectNotes.15} {\bibfield  {journal} {\bibinfo  {journal} {SciPost Phys. Lect. Notes}\ ,\ \bibinfo {pages} {15}} (\bibinfo {year} {2020}{\natexlab{a}})}\BibitemShut {NoStop}%
\bibitem [{\citenamefont {Aguado}\ and\ \citenamefont {Kouwenhoven}(2020)}]{aguado2020majorana}%
  \BibitemOpen
  \bibfield  {author} {\bibinfo {author} {\bibfnamefont {R.}~\bibnamefont {Aguado}}\ and\ \bibinfo {author} {\bibfnamefont {L.~P.}\ \bibnamefont {Kouwenhoven}},\ }\bibfield  {title} {\bibinfo {title} {Majorana qubits for topological quantum computing},\ }\href {https://doi.org/10.1063/PT.3.4499} {\bibfield  {journal} {\bibinfo  {journal} {Physics Today}\ }\textbf {\bibinfo {volume} {73}},\ \bibinfo {pages} {44} (\bibinfo {year} {2020})}\BibitemShut {NoStop}%
\bibitem [{\citenamefont {Marra}(2022)}]{Marra_2022}%
  \BibitemOpen
  \bibfield  {author} {\bibinfo {author} {\bibfnamefont {P.}~\bibnamefont {Marra}},\ }\bibfield  {title} {\bibinfo {title} {Majorana nanowires for topological quantum computation},\ }\href {https://doi.org/10.1063/5.0102999} {\bibfield  {journal} {\bibinfo  {journal} {J. Appl. Phys.}\ }\textbf {\bibinfo {volume} {132}},\ \bibinfo {pages} {231101} (\bibinfo {year} {2022})}\BibitemShut {NoStop}%
\bibitem [{\citenamefont {Aasen}\ \emph {et~al.}(2025)\citenamefont {Aasen}, \citenamefont {Aghaee}, \citenamefont {Alam}, \citenamefont {Andrzejczuk}, \citenamefont {Antipov}, \citenamefont {Astafev}, \citenamefont {Avilovas}, \citenamefont {Barzegar}, \citenamefont {Bauer}, \citenamefont {Becker}, \citenamefont {Bello-Rivas}, \citenamefont {Bhaskar}, \citenamefont {Bocharov}, \citenamefont {Boddapati}, \citenamefont {Bohn}, \citenamefont {Bommer}, \citenamefont {Bonderson}, \citenamefont {Borovsky}, \citenamefont {Bourdet}, \citenamefont {Boutin}, \citenamefont {Brown}, \citenamefont {Campbell}, \citenamefont {Casparis}, \citenamefont {Chakravarthi}, \citenamefont {Chao}, \citenamefont {Chapman}, \citenamefont {Chatoor}, \citenamefont {Christensen}, \citenamefont {Codd}, \citenamefont {Cole}, \citenamefont {Cooper}, \citenamefont {Corsetti}, \citenamefont {Cui}, \citenamefont {van Dam}, \citenamefont {Dandachi}, \citenamefont {Daraeizadeh}, \citenamefont {Dumitrascu}, \citenamefont {Ekefjärd},
  \citenamefont {Fallahi}, \citenamefont {Galletti}, \citenamefont {Gardner}, \citenamefont {Gatta}, \citenamefont {Gavranovic}, \citenamefont {Goulding}, \citenamefont {Govender}, \citenamefont {Griggio}, \citenamefont {Grigoryan}, \citenamefont {Grijalva}, \citenamefont {Gronin}, \citenamefont {Gukelberger}, \citenamefont {Haah}, \citenamefont {Hamdast}, \citenamefont {Hansen}, \citenamefont {Hastings}, \citenamefont {Heedt}, \citenamefont {Ho}, \citenamefont {Hogaboam}, \citenamefont {Holgaard}, \citenamefont {Hoogdalem}, \citenamefont {Indrapiromkul}, \citenamefont {Ingerslev}, \citenamefont {Ivancevic}, \citenamefont {Jablonski}, \citenamefont {Jensen}, \citenamefont {Jhoja}, \citenamefont {Jones}, \citenamefont {Kalashnikov}, \citenamefont {Kallaher}, \citenamefont {Kalra}, \citenamefont {Karimi}, \citenamefont {Karzig}, \citenamefont {Kimes}, \citenamefont {Kliuchnikov}, \citenamefont {Kloster}, \citenamefont {Knapp}, \citenamefont {Knee}, \citenamefont {Koski}, \citenamefont {Kostamo}, \citenamefont
  {Kuesel}, \citenamefont {Lackey}, \citenamefont {Laeven}, \citenamefont {Lai}, \citenamefont {de~Lange}, \citenamefont {Larsen}, \citenamefont {Lee}, \citenamefont {Lee}, \citenamefont {Leum}, \citenamefont {Li}, \citenamefont {Lindemann}, \citenamefont {Lucas}, \citenamefont {Lutchyn}, \citenamefont {Madsen}, \citenamefont {Madulid}, \citenamefont {Manfra}, \citenamefont {Markussen}, \citenamefont {Martinez}, \citenamefont {Mattila}, \citenamefont {Mattinson}, \citenamefont {McNeil}, \citenamefont {Mei}, \citenamefont {Mishmash}, \citenamefont {Mohandas}, \citenamefont {Mollgaard}, \citenamefont {de~Moor}, \citenamefont {Morgan}, \citenamefont {Moussa}, \citenamefont {Narla}, \citenamefont {Nayak}, \citenamefont {Nielsen}, \citenamefont {Nielsen}, \citenamefont {Nolet}, \citenamefont {Nystrom}, \citenamefont {O'Farrell}, \citenamefont {Otani}, \citenamefont {Paetznick}, \citenamefont {Papon}, \citenamefont {Paz}, \citenamefont {Petersson}, \citenamefont {Petit}, \citenamefont {Pikulin}, \citenamefont
  {Pons}, \citenamefont {Quinn}, \citenamefont {Rajpalke}, \citenamefont {Ramirez}, \citenamefont {Rasmussen}, \citenamefont {Razmadze}, \citenamefont {Reichardt}, \citenamefont {Ren}, \citenamefont {Reneris}, \citenamefont {Riccomini}, \citenamefont {Sadovskyy}, \citenamefont {Sainiemi}, \citenamefont {Saldaña}, \citenamefont {Sanlorenzo}, \citenamefont {Schaal}, \citenamefont {Schmidgall}, \citenamefont {Sfiligoj}, \citenamefont {da~Silva}, \citenamefont {Singh}, \citenamefont {Sinha}, \citenamefont {Soeken}, \citenamefont {Sohr}, \citenamefont {Stankevic}, \citenamefont {Stek}, \citenamefont {Strøm-Hansen}, \citenamefont {Stuppard}, \citenamefont {Sundaram}, \citenamefont {Suominen}, \citenamefont {Suter}, \citenamefont {Suzuki}, \citenamefont {Svore}, \citenamefont {Teicher}, \citenamefont {Thiyagarajah}, \citenamefont {Tholapi}, \citenamefont {Thomas}, \citenamefont {Tom}, \citenamefont {Toomey}, \citenamefont {Tracy}, \citenamefont {Troyer}, \citenamefont {Turley}, \citenamefont {Turner},
  \citenamefont {Upadhyay}, \citenamefont {Urban}, \citenamefont {Vaschillo}, \citenamefont {Viazmitinov}, \citenamefont {Vogel}, \citenamefont {Wang}, \citenamefont {Watson}, \citenamefont {Webster}, \citenamefont {Weston}, \citenamefont {Williamson}, \citenamefont {Winkler}, \citenamefont {van Woerkom}, \citenamefont {Wütz}, \citenamefont {Yang}, \citenamefont {Yu}, \citenamefont {Yucelen}, \citenamefont {Zamorano}, \citenamefont {Zeisel}, \citenamefont {Zheng}, \citenamefont {Zilke},\ and\ \citenamefont {Zimmerman}}]{aasen2025roadmap}%
  \BibitemOpen
  \bibfield  {author} {\bibinfo {author} {\bibfnamefont {D.}~\bibnamefont {Aasen}}, \bibinfo {author} {\bibfnamefont {M.}~\bibnamefont {Aghaee}}, \bibinfo {author} {\bibfnamefont {Z.}~\bibnamefont {Alam}}, \bibinfo {author} {\bibfnamefont {M.}~\bibnamefont {Andrzejczuk}}, \bibinfo {author} {\bibfnamefont {A.}~\bibnamefont {Antipov}}, \bibinfo {author} {\bibfnamefont {M.}~\bibnamefont {Astafev}}, \bibinfo {author} {\bibfnamefont {L.}~\bibnamefont {Avilovas}}, \bibinfo {author} {\bibfnamefont {A.}~\bibnamefont {Barzegar}}, \bibinfo {author} {\bibfnamefont {B.}~\bibnamefont {Bauer}}, \bibinfo {author} {\bibfnamefont {J.}~\bibnamefont {Becker}}, \bibinfo {author} {\bibfnamefont {J.~M.}\ \bibnamefont {Bello-Rivas}}, \bibinfo {author} {\bibfnamefont {U.}~\bibnamefont {Bhaskar}}, \bibinfo {author} {\bibfnamefont {A.}~\bibnamefont {Bocharov}}, \bibinfo {author} {\bibfnamefont {S.}~\bibnamefont {Boddapati}}, \bibinfo {author} {\bibfnamefont {D.}~\bibnamefont {Bohn}}, \bibinfo {author} {\bibfnamefont {J.}~\bibnamefont
  {Bommer}}, \bibinfo {author} {\bibfnamefont {P.}~\bibnamefont {Bonderson}}, \bibinfo {author} {\bibfnamefont {J.}~\bibnamefont {Borovsky}}, \bibinfo {author} {\bibfnamefont {L.}~\bibnamefont {Bourdet}}, \bibinfo {author} {\bibfnamefont {S.}~\bibnamefont {Boutin}}, \bibinfo {author} {\bibfnamefont {T.}~\bibnamefont {Brown}}, \bibinfo {author} {\bibfnamefont {G.}~\bibnamefont {Campbell}}, \bibinfo {author} {\bibfnamefont {L.}~\bibnamefont {Casparis}}, \bibinfo {author} {\bibfnamefont {S.}~\bibnamefont {Chakravarthi}}, \bibinfo {author} {\bibfnamefont {R.}~\bibnamefont {Chao}}, \bibinfo {author} {\bibfnamefont {B.~J.}\ \bibnamefont {Chapman}}, \bibinfo {author} {\bibfnamefont {S.}~\bibnamefont {Chatoor}}, \bibinfo {author} {\bibfnamefont {A.~W.}\ \bibnamefont {Christensen}}, \bibinfo {author} {\bibfnamefont {P.}~\bibnamefont {Codd}}, \bibinfo {author} {\bibfnamefont {W.}~\bibnamefont {Cole}}, \bibinfo {author} {\bibfnamefont {P.}~\bibnamefont {Cooper}}, \bibinfo {author} {\bibfnamefont {F.}~\bibnamefont
  {Corsetti}}, \bibinfo {author} {\bibfnamefont {A.}~\bibnamefont {Cui}}, \bibinfo {author} {\bibfnamefont {W.}~\bibnamefont {van Dam}}, \bibinfo {author} {\bibfnamefont {T.~E.}\ \bibnamefont {Dandachi}}, \bibinfo {author} {\bibfnamefont {S.}~\bibnamefont {Daraeizadeh}}, \bibinfo {author} {\bibfnamefont {A.}~\bibnamefont {Dumitrascu}}, \bibinfo {author} {\bibfnamefont {A.}~\bibnamefont {Ekefjärd}}, \bibinfo {author} {\bibfnamefont {S.}~\bibnamefont {Fallahi}}, \bibinfo {author} {\bibfnamefont {L.}~\bibnamefont {Galletti}}, \bibinfo {author} {\bibfnamefont {G.}~\bibnamefont {Gardner}}, \bibinfo {author} {\bibfnamefont {R.}~\bibnamefont {Gatta}}, \bibinfo {author} {\bibfnamefont {H.}~\bibnamefont {Gavranovic}}, \bibinfo {author} {\bibfnamefont {M.}~\bibnamefont {Goulding}}, \bibinfo {author} {\bibfnamefont {D.}~\bibnamefont {Govender}}, \bibinfo {author} {\bibfnamefont {F.}~\bibnamefont {Griggio}}, \bibinfo {author} {\bibfnamefont {R.}~\bibnamefont {Grigoryan}}, \bibinfo {author} {\bibfnamefont
  {S.}~\bibnamefont {Grijalva}}, \bibinfo {author} {\bibfnamefont {S.}~\bibnamefont {Gronin}}, \bibinfo {author} {\bibfnamefont {J.}~\bibnamefont {Gukelberger}}, \bibinfo {author} {\bibfnamefont {J.}~\bibnamefont {Haah}}, \bibinfo {author} {\bibfnamefont {M.}~\bibnamefont {Hamdast}}, \bibinfo {author} {\bibfnamefont {E.~B.}\ \bibnamefont {Hansen}}, \bibinfo {author} {\bibfnamefont {M.}~\bibnamefont {Hastings}}, \bibinfo {author} {\bibfnamefont {S.}~\bibnamefont {Heedt}}, \bibinfo {author} {\bibfnamefont {S.}~\bibnamefont {Ho}}, \bibinfo {author} {\bibfnamefont {J.}~\bibnamefont {Hogaboam}}, \bibinfo {author} {\bibfnamefont {L.}~\bibnamefont {Holgaard}}, \bibinfo {author} {\bibfnamefont {K.~V.}\ \bibnamefont {Hoogdalem}}, \bibinfo {author} {\bibfnamefont {J.}~\bibnamefont {Indrapiromkul}}, \bibinfo {author} {\bibfnamefont {H.}~\bibnamefont {Ingerslev}}, \bibinfo {author} {\bibfnamefont {L.}~\bibnamefont {Ivancevic}}, \bibinfo {author} {\bibfnamefont {S.}~\bibnamefont {Jablonski}}, \bibinfo {author}
  {\bibfnamefont {T.}~\bibnamefont {Jensen}}, \bibinfo {author} {\bibfnamefont {J.}~\bibnamefont {Jhoja}}, \bibinfo {author} {\bibfnamefont {J.}~\bibnamefont {Jones}}, \bibinfo {author} {\bibfnamefont {K.}~\bibnamefont {Kalashnikov}}, \bibinfo {author} {\bibfnamefont {R.}~\bibnamefont {Kallaher}}, \bibinfo {author} {\bibfnamefont {R.}~\bibnamefont {Kalra}}, \bibinfo {author} {\bibfnamefont {F.}~\bibnamefont {Karimi}}, \bibinfo {author} {\bibfnamefont {T.}~\bibnamefont {Karzig}}, \bibinfo {author} {\bibfnamefont {S.}~\bibnamefont {Kimes}}, \bibinfo {author} {\bibfnamefont {V.}~\bibnamefont {Kliuchnikov}}, \bibinfo {author} {\bibfnamefont {M.~E.}\ \bibnamefont {Kloster}}, \bibinfo {author} {\bibfnamefont {C.}~\bibnamefont {Knapp}}, \bibinfo {author} {\bibfnamefont {D.}~\bibnamefont {Knee}}, \bibinfo {author} {\bibfnamefont {J.}~\bibnamefont {Koski}}, \bibinfo {author} {\bibfnamefont {P.}~\bibnamefont {Kostamo}}, \bibinfo {author} {\bibfnamefont {J.}~\bibnamefont {Kuesel}}, \bibinfo {author} {\bibfnamefont
  {B.}~\bibnamefont {Lackey}}, \bibinfo {author} {\bibfnamefont {T.}~\bibnamefont {Laeven}}, \bibinfo {author} {\bibfnamefont {J.}~\bibnamefont {Lai}}, \bibinfo {author} {\bibfnamefont {G.}~\bibnamefont {de~Lange}}, \bibinfo {author} {\bibfnamefont {T.}~\bibnamefont {Larsen}}, \bibinfo {author} {\bibfnamefont {J.}~\bibnamefont {Lee}}, \bibinfo {author} {\bibfnamefont {K.}~\bibnamefont {Lee}}, \bibinfo {author} {\bibfnamefont {G.}~\bibnamefont {Leum}}, \bibinfo {author} {\bibfnamefont {K.}~\bibnamefont {Li}}, \bibinfo {author} {\bibfnamefont {T.}~\bibnamefont {Lindemann}}, \bibinfo {author} {\bibfnamefont {M.}~\bibnamefont {Lucas}}, \bibinfo {author} {\bibfnamefont {R.}~\bibnamefont {Lutchyn}}, \bibinfo {author} {\bibfnamefont {M.~H.}\ \bibnamefont {Madsen}}, \bibinfo {author} {\bibfnamefont {N.}~\bibnamefont {Madulid}}, \bibinfo {author} {\bibfnamefont {M.}~\bibnamefont {Manfra}}, \bibinfo {author} {\bibfnamefont {S.~B.}\ \bibnamefont {Markussen}}, \bibinfo {author} {\bibfnamefont {E.}~\bibnamefont
  {Martinez}}, \bibinfo {author} {\bibfnamefont {M.}~\bibnamefont {Mattila}}, \bibinfo {author} {\bibfnamefont {J.}~\bibnamefont {Mattinson}}, \bibinfo {author} {\bibfnamefont {R.}~\bibnamefont {McNeil}}, \bibinfo {author} {\bibfnamefont {A.~R.}\ \bibnamefont {Mei}}, \bibinfo {author} {\bibfnamefont {R.~V.}\ \bibnamefont {Mishmash}}, \bibinfo {author} {\bibfnamefont {G.}~\bibnamefont {Mohandas}}, \bibinfo {author} {\bibfnamefont {C.}~\bibnamefont {Mollgaard}}, \bibinfo {author} {\bibfnamefont {M.}~\bibnamefont {de~Moor}}, \bibinfo {author} {\bibfnamefont {T.}~\bibnamefont {Morgan}}, \bibinfo {author} {\bibfnamefont {G.}~\bibnamefont {Moussa}}, \bibinfo {author} {\bibfnamefont {A.}~\bibnamefont {Narla}}, \bibinfo {author} {\bibfnamefont {C.}~\bibnamefont {Nayak}}, \bibinfo {author} {\bibfnamefont {J.~H.}\ \bibnamefont {Nielsen}}, \bibinfo {author} {\bibfnamefont {W.~H.~P.}\ \bibnamefont {Nielsen}}, \bibinfo {author} {\bibfnamefont {F.}~\bibnamefont {Nolet}}, \bibinfo {author} {\bibfnamefont {M.}~\bibnamefont
  {Nystrom}}, \bibinfo {author} {\bibfnamefont {E.}~\bibnamefont {O'Farrell}}, \bibinfo {author} {\bibfnamefont {K.}~\bibnamefont {Otani}}, \bibinfo {author} {\bibfnamefont {A.}~\bibnamefont {Paetznick}}, \bibinfo {author} {\bibfnamefont {C.}~\bibnamefont {Papon}}, \bibinfo {author} {\bibfnamefont {A.}~\bibnamefont {Paz}}, \bibinfo {author} {\bibfnamefont {K.}~\bibnamefont {Petersson}}, \bibinfo {author} {\bibfnamefont {L.}~\bibnamefont {Petit}}, \bibinfo {author} {\bibfnamefont {D.}~\bibnamefont {Pikulin}}, \bibinfo {author} {\bibfnamefont {D.~O.~F.}\ \bibnamefont {Pons}}, \bibinfo {author} {\bibfnamefont {S.}~\bibnamefont {Quinn}}, \bibinfo {author} {\bibfnamefont {M.}~\bibnamefont {Rajpalke}}, \bibinfo {author} {\bibfnamefont {A.~A.}\ \bibnamefont {Ramirez}}, \bibinfo {author} {\bibfnamefont {K.}~\bibnamefont {Rasmussen}}, \bibinfo {author} {\bibfnamefont {D.}~\bibnamefont {Razmadze}}, \bibinfo {author} {\bibfnamefont {B.}~\bibnamefont {Reichardt}}, \bibinfo {author} {\bibfnamefont {Y.}~\bibnamefont
  {Ren}}, \bibinfo {author} {\bibfnamefont {K.}~\bibnamefont {Reneris}}, \bibinfo {author} {\bibfnamefont {R.}~\bibnamefont {Riccomini}}, \bibinfo {author} {\bibfnamefont {I.}~\bibnamefont {Sadovskyy}}, \bibinfo {author} {\bibfnamefont {L.}~\bibnamefont {Sainiemi}}, \bibinfo {author} {\bibfnamefont {J.~C.~E.}\ \bibnamefont {Saldaña}}, \bibinfo {author} {\bibfnamefont {I.}~\bibnamefont {Sanlorenzo}}, \bibinfo {author} {\bibfnamefont {S.}~\bibnamefont {Schaal}}, \bibinfo {author} {\bibfnamefont {E.}~\bibnamefont {Schmidgall}}, \bibinfo {author} {\bibfnamefont {C.}~\bibnamefont {Sfiligoj}}, \bibinfo {author} {\bibfnamefont {M.~P.}\ \bibnamefont {da~Silva}}, \bibinfo {author} {\bibfnamefont {S.}~\bibnamefont {Singh}}, \bibinfo {author} {\bibfnamefont {S.}~\bibnamefont {Sinha}}, \bibinfo {author} {\bibfnamefont {M.}~\bibnamefont {Soeken}}, \bibinfo {author} {\bibfnamefont {P.}~\bibnamefont {Sohr}}, \bibinfo {author} {\bibfnamefont {T.}~\bibnamefont {Stankevic}}, \bibinfo {author} {\bibfnamefont {L.}~\bibnamefont
  {Stek}}, \bibinfo {author} {\bibfnamefont {P.}~\bibnamefont {Strøm-Hansen}}, \bibinfo {author} {\bibfnamefont {E.}~\bibnamefont {Stuppard}}, \bibinfo {author} {\bibfnamefont {A.}~\bibnamefont {Sundaram}}, \bibinfo {author} {\bibfnamefont {H.}~\bibnamefont {Suominen}}, \bibinfo {author} {\bibfnamefont {J.}~\bibnamefont {Suter}}, \bibinfo {author} {\bibfnamefont {S.}~\bibnamefont {Suzuki}}, \bibinfo {author} {\bibfnamefont {K.}~\bibnamefont {Svore}}, \bibinfo {author} {\bibfnamefont {S.}~\bibnamefont {Teicher}}, \bibinfo {author} {\bibfnamefont {N.}~\bibnamefont {Thiyagarajah}}, \bibinfo {author} {\bibfnamefont {R.}~\bibnamefont {Tholapi}}, \bibinfo {author} {\bibfnamefont {M.}~\bibnamefont {Thomas}}, \bibinfo {author} {\bibfnamefont {D.}~\bibnamefont {Tom}}, \bibinfo {author} {\bibfnamefont {E.}~\bibnamefont {Toomey}}, \bibinfo {author} {\bibfnamefont {J.}~\bibnamefont {Tracy}}, \bibinfo {author} {\bibfnamefont {M.}~\bibnamefont {Troyer}}, \bibinfo {author} {\bibfnamefont {M.}~\bibnamefont {Turley}},
  \bibinfo {author} {\bibfnamefont {M.~D.}\ \bibnamefont {Turner}}, \bibinfo {author} {\bibfnamefont {S.}~\bibnamefont {Upadhyay}}, \bibinfo {author} {\bibfnamefont {I.}~\bibnamefont {Urban}}, \bibinfo {author} {\bibfnamefont {A.}~\bibnamefont {Vaschillo}}, \bibinfo {author} {\bibfnamefont {D.}~\bibnamefont {Viazmitinov}}, \bibinfo {author} {\bibfnamefont {D.}~\bibnamefont {Vogel}}, \bibinfo {author} {\bibfnamefont {Z.}~\bibnamefont {Wang}}, \bibinfo {author} {\bibfnamefont {J.}~\bibnamefont {Watson}}, \bibinfo {author} {\bibfnamefont {A.}~\bibnamefont {Webster}}, \bibinfo {author} {\bibfnamefont {J.}~\bibnamefont {Weston}}, \bibinfo {author} {\bibfnamefont {T.}~\bibnamefont {Williamson}}, \bibinfo {author} {\bibfnamefont {G.~W.}\ \bibnamefont {Winkler}}, \bibinfo {author} {\bibfnamefont {D.~J.}\ \bibnamefont {van Woerkom}}, \bibinfo {author} {\bibfnamefont {B.~P.}\ \bibnamefont {Wütz}}, \bibinfo {author} {\bibfnamefont {C.~K.}\ \bibnamefont {Yang}}, \bibinfo {author} {\bibfnamefont {R.}~\bibnamefont {Yu}},
  \bibinfo {author} {\bibfnamefont {E.}~\bibnamefont {Yucelen}}, \bibinfo {author} {\bibfnamefont {J.~H.}\ \bibnamefont {Zamorano}}, \bibinfo {author} {\bibfnamefont {R.}~\bibnamefont {Zeisel}}, \bibinfo {author} {\bibfnamefont {G.}~\bibnamefont {Zheng}}, \bibinfo {author} {\bibfnamefont {J.}~\bibnamefont {Zilke}},\ and\ \bibinfo {author} {\bibfnamefont {A.}~\bibnamefont {Zimmerman}},\ }\bibfield  {title} {\bibinfo {title} {Roadmap to fault tolerant quantum computation using topological qubit arrays},\ }\href {https://arxiv.org/abs/2502.12252} {\bibfield  {journal} {\bibinfo  {journal} {arXiv:2502.12252}\ } (\bibinfo {year} {2025})}\BibitemShut {NoStop}%
\bibitem [{\citenamefont {Prada}\ \emph {et~al.}(2020)\citenamefont {Prada}, \citenamefont {San-Jose}, \citenamefont {de~Moor}, \citenamefont {Geresdi}, \citenamefont {Lee}, \citenamefont {Klinovaja}, \citenamefont {Loss}, \citenamefont {Nyg{\aa}rd}, \citenamefont {Aguado},\ and\ \citenamefont {Kouwenhoven}}]{prada2020}%
  \BibitemOpen
  \bibfield  {author} {\bibinfo {author} {\bibfnamefont {E.}~\bibnamefont {Prada}}, \bibinfo {author} {\bibfnamefont {P.}~\bibnamefont {San-Jose}}, \bibinfo {author} {\bibfnamefont {M.~W.}\ \bibnamefont {de~Moor}}, \bibinfo {author} {\bibfnamefont {A.}~\bibnamefont {Geresdi}}, \bibinfo {author} {\bibfnamefont {E.~J.}\ \bibnamefont {Lee}}, \bibinfo {author} {\bibfnamefont {J.}~\bibnamefont {Klinovaja}}, \bibinfo {author} {\bibfnamefont {D.}~\bibnamefont {Loss}}, \bibinfo {author} {\bibfnamefont {J.}~\bibnamefont {Nyg{\aa}rd}}, \bibinfo {author} {\bibfnamefont {R.}~\bibnamefont {Aguado}},\ and\ \bibinfo {author} {\bibfnamefont {L.~P.}\ \bibnamefont {Kouwenhoven}},\ }\bibfield  {title} {\bibinfo {title} {From {Andreev} to {Majorana} bound states in hybrid superconductor--semiconductor nanowires},\ }\href {https://doi.org/10.1038/s42254-020-0228-y} {\bibfield  {journal} {\bibinfo  {journal} {Nature Reviews Physics}\ }\textbf {\bibinfo {volume} {2}},\ \bibinfo {pages} {575} (\bibinfo {year} {2020})}\BibitemShut
  {NoStop}%
\bibitem [{\citenamefont {Beenakker}(2020{\natexlab{b}})}]{beenakker2020search}%
  \BibitemOpen
  \bibfield  {author} {\bibinfo {author} {\bibfnamefont {C.~W.~J.}\ \bibnamefont {Beenakker}},\ }\bibfield  {title} {\bibinfo {title} {Search for non-{A}belian {M}ajorana braiding statistics in superconductors},\ }\href {https://doi.org/10.21468/SciPostPhysLectNotes.15} {\bibfield  {journal} {\bibinfo  {journal} {SciPost Phys. Lect. Notes}\ ,\ \bibinfo {pages} {15}} (\bibinfo {year} {2020}{\natexlab{b}})}\BibitemShut {NoStop}%
\bibitem [{\citenamefont {Cayao}\ \emph {et~al.}(2017)\citenamefont {Cayao}, \citenamefont {San-Jose}, \citenamefont {Black-Schaffer}, \citenamefont {Aguado},\ and\ \citenamefont {Prada}}]{Cayao2017Nov}%
  \BibitemOpen
  \bibfield  {author} {\bibinfo {author} {\bibfnamefont {J.}~\bibnamefont {Cayao}}, \bibinfo {author} {\bibfnamefont {P.}~\bibnamefont {San-Jose}}, \bibinfo {author} {\bibfnamefont {A.~M.}\ \bibnamefont {Black-Schaffer}}, \bibinfo {author} {\bibfnamefont {R.}~\bibnamefont {Aguado}},\ and\ \bibinfo {author} {\bibfnamefont {E.}~\bibnamefont {Prada}},\ }\bibfield  {title} {\bibinfo {title} {Majorana splitting from critical currents in {J}osephson junctions},\ }\href {https://doi.org/10.1103/PhysRevB.96.205425} {\bibfield  {journal} {\bibinfo  {journal} {Phys. Rev. B}\ }\textbf {\bibinfo {volume} {96}},\ \bibinfo {pages} {205425} (\bibinfo {year} {2017})}\BibitemShut {NoStop}%
\bibitem [{\citenamefont {Clarke}(2017)}]{Clarke2017Nov}%
  \BibitemOpen
  \bibfield  {author} {\bibinfo {author} {\bibfnamefont {D.~J.}\ \bibnamefont {Clarke}},\ }\bibfield  {title} {\bibinfo {title} {Experimentally accessible topological quality factor for wires with zero energy modes},\ }\href {https://doi.org/10.1103/PhysRevB.96.201109} {\bibfield  {journal} {\bibinfo  {journal} {Phys. Rev. B}\ }\textbf {\bibinfo {volume} {96}},\ \bibinfo {pages} {201109} (\bibinfo {year} {2017})}\BibitemShut {NoStop}%
\bibitem [{\citenamefont {Prada}\ \emph {et~al.}(2017)\citenamefont {Prada}, \citenamefont {Aguado},\ and\ \citenamefont {San-Jose}}]{Prada2017Aug}%
  \BibitemOpen
  \bibfield  {author} {\bibinfo {author} {\bibfnamefont {E.}~\bibnamefont {Prada}}, \bibinfo {author} {\bibfnamefont {R.}~\bibnamefont {Aguado}},\ and\ \bibinfo {author} {\bibfnamefont {P.}~\bibnamefont {San-Jose}},\ }\bibfield  {title} {\bibinfo {title} {Measuring {Majorana} nonlocality and spin structure with a quantum dot},\ }\href {https://doi.org/10.1103/PhysRevB.96.085418} {\bibfield  {journal} {\bibinfo  {journal} {Phys. Rev. B}\ }\textbf {\bibinfo {volume} {96}},\ \bibinfo {pages} {085418} (\bibinfo {year} {2017})}\BibitemShut {NoStop}%
\bibitem [{\citenamefont {Deng}\ \emph {et~al.}(2018)\citenamefont {Deng}, \citenamefont {Vaitiek\ifmmode~\dot{e}\else \.{e}\fi{}nas}, \citenamefont {Prada}, \citenamefont {San-Jose}, \citenamefont {Nyg\aa{}rd}, \citenamefont {Krogstrup}, \citenamefont {Aguado},\ and\ \citenamefont {Marcus}}]{Deng2018Aug}%
  \BibitemOpen
  \bibfield  {author} {\bibinfo {author} {\bibfnamefont {M.-T.}\ \bibnamefont {Deng}}, \bibinfo {author} {\bibfnamefont {S.}~\bibnamefont {Vaitiek\ifmmode~\dot{e}\else \.{e}\fi{}nas}}, \bibinfo {author} {\bibfnamefont {E.}~\bibnamefont {Prada}}, \bibinfo {author} {\bibfnamefont {P.}~\bibnamefont {San-Jose}}, \bibinfo {author} {\bibfnamefont {J.}~\bibnamefont {Nyg\aa{}rd}}, \bibinfo {author} {\bibfnamefont {P.}~\bibnamefont {Krogstrup}}, \bibinfo {author} {\bibfnamefont {R.}~\bibnamefont {Aguado}},\ and\ \bibinfo {author} {\bibfnamefont {C.~M.}\ \bibnamefont {Marcus}},\ }\bibfield  {title} {\bibinfo {title} {Nonlocality of {Majorana} modes in hybrid nanowires},\ }\href {https://doi.org/10.1103/PhysRevB.98.085125} {\bibfield  {journal} {\bibinfo  {journal} {Phys. Rev. B}\ }\textbf {\bibinfo {volume} {98}},\ \bibinfo {pages} {085125} (\bibinfo {year} {2018})}\BibitemShut {NoStop}%
\bibitem [{\citenamefont {Smirnov}(2022)}]{Smirnov2022May}%
  \BibitemOpen
  \bibfield  {author} {\bibinfo {author} {\bibfnamefont {S.}~\bibnamefont {Smirnov}},\ }\bibfield  {title} {\bibinfo {title} {Revealing universal {Majorana} fractionalization using differential shot noise and conductance in nonequilibrium states controlled by tunneling phases},\ }\href {https://doi.org/10.1103/PhysRevB.105.205430} {\bibfield  {journal} {\bibinfo  {journal} {Phys. Rev. B}\ }\textbf {\bibinfo {volume} {105}},\ \bibinfo {pages} {205430} (\bibinfo {year} {2022})}\BibitemShut {NoStop}%
\bibitem [{\citenamefont {Smirnov}(2024)}]{Smirnov2024May}%
  \BibitemOpen
  \bibfield  {author} {\bibinfo {author} {\bibfnamefont {S.}~\bibnamefont {Smirnov}},\ }\bibfield  {title} {\bibinfo {title} {Nonequilibrium finite frequency resonances in differential quantum noise driven by {Majorana} interference},\ }\href {https://doi.org/10.1103/PhysRevB.109.195410} {\bibfield  {journal} {\bibinfo  {journal} {Phys. Rev. B}\ }\textbf {\bibinfo {volume} {109}},\ \bibinfo {pages} {195410} (\bibinfo {year} {2024})}\BibitemShut {NoStop}%
\bibitem [{\citenamefont {Vimal}\ and\ \citenamefont {Cayao}(2024)}]{Vimal2024Dec}%
  \BibitemOpen
  \bibfield  {author} {\bibinfo {author} {\bibfnamefont {V.~K.}\ \bibnamefont {Vimal}}\ and\ \bibinfo {author} {\bibfnamefont {J.}~\bibnamefont {Cayao}},\ }\bibfield  {title} {\bibinfo {title} {Entanglement measures of {Majorana} bound states},\ }\href {https://doi.org/10.1103/PhysRevB.110.224510} {\bibfield  {journal} {\bibinfo  {journal} {Phys. Rev. B}\ }\textbf {\bibinfo {volume} {110}},\ \bibinfo {pages} {224510} (\bibinfo {year} {2024})}\BibitemShut {NoStop}%
\bibitem [{\citenamefont {Dutta}\ \emph {et~al.}(2024)\citenamefont {Dutta}, \citenamefont {Cayao}, \citenamefont {Black-Schaffer},\ and\ \citenamefont {Burset}}]{Dutta2024Mar}%
  \BibitemOpen
  \bibfield  {author} {\bibinfo {author} {\bibfnamefont {P.}~\bibnamefont {Dutta}}, \bibinfo {author} {\bibfnamefont {J.}~\bibnamefont {Cayao}}, \bibinfo {author} {\bibfnamefont {A.~M.}\ \bibnamefont {Black-Schaffer}},\ and\ \bibinfo {author} {\bibfnamefont {P.}~\bibnamefont {Burset}},\ }\bibfield  {title} {\bibinfo {title} {Nonlocality of {Majorana} bound states revealed by electron waiting times in a topological {Andreev} interferometer},\ }\href {https://doi.org/10.1103/PhysRevResearch.6.L012062} {\bibfield  {journal} {\bibinfo  {journal} {Phys. Rev. Res.}\ }\textbf {\bibinfo {volume} {6}},\ \bibinfo {pages} {L012062} (\bibinfo {year} {2024})}\BibitemShut {NoStop}%
\bibitem [{\citenamefont {Cayao}\ \emph {et~al.}(2024)\citenamefont {Cayao}, \citenamefont {Nagaosa},\ and\ \citenamefont {Tanaka}}]{Cayao2024Feb}%
  \BibitemOpen
  \bibfield  {author} {\bibinfo {author} {\bibfnamefont {J.}~\bibnamefont {Cayao}}, \bibinfo {author} {\bibfnamefont {N.}~\bibnamefont {Nagaosa}},\ and\ \bibinfo {author} {\bibfnamefont {Y.}~\bibnamefont {Tanaka}},\ }\bibfield  {title} {\bibinfo {title} {Enhancing the {Josephson} diode effect with {Majorana} bound states},\ }\href {https://doi.org/10.1103/PhysRevB.109.L081405} {\bibfield  {journal} {\bibinfo  {journal} {Phys. Rev. B}\ }\textbf {\bibinfo {volume} {109}},\ \bibinfo {pages} {L081405} (\bibinfo {year} {2024})}\BibitemShut {NoStop}%
\bibitem [{\citenamefont {Mondal}\ \emph {et~al.}(2025)\citenamefont {Mondal}, \citenamefont {Fu},\ and\ \citenamefont {Cayao}}]{mondal2025JDE}%
  \BibitemOpen
  \bibfield  {author} {\bibinfo {author} {\bibfnamefont {S.}~\bibnamefont {Mondal}}, \bibinfo {author} {\bibfnamefont {P.-H.}\ \bibnamefont {Fu}},\ and\ \bibinfo {author} {\bibfnamefont {J.}~\bibnamefont {Cayao}},\ }\bibfield  {title} {\bibinfo {title} {Josephson diode effect with {Andreev} and {Majorana} bound states},\ }\href {https://arxiv.org/abs/2503.08318} {\bibfield  {journal} {\bibinfo  {journal} {arXiv:2503.08318}\ } (\bibinfo {year} {2025})}\BibitemShut {NoStop}%
\bibitem [{\citenamefont {Kells}\ \emph {et~al.}(2012)\citenamefont {Kells}, \citenamefont {Meidan},\ and\ \citenamefont {Brouwer}}]{Kells2012Sep}%
  \BibitemOpen
  \bibfield  {author} {\bibinfo {author} {\bibfnamefont {G.}~\bibnamefont {Kells}}, \bibinfo {author} {\bibfnamefont {D.}~\bibnamefont {Meidan}},\ and\ \bibinfo {author} {\bibfnamefont {P.~W.}\ \bibnamefont {Brouwer}},\ }\bibfield  {title} {\bibinfo {title} {Near-zero-energy end states in topologically trivial spin-orbit coupled superconducting nanowires with a smooth confinement},\ }\href {https://doi.org/10.1103/PhysRevB.86.100503} {\bibfield  {journal} {\bibinfo  {journal} {Phys. Rev. B}\ }\textbf {\bibinfo {volume} {86}},\ \bibinfo {pages} {100503} (\bibinfo {year} {2012})}\BibitemShut {NoStop}%
\bibitem [{\citenamefont {Prada}\ \emph {et~al.}(2012)\citenamefont {Prada}, \citenamefont {San-Jose},\ and\ \citenamefont {Aguado}}]{Prada2012Nov}%
  \BibitemOpen
  \bibfield  {author} {\bibinfo {author} {\bibfnamefont {E.}~\bibnamefont {Prada}}, \bibinfo {author} {\bibfnamefont {P.}~\bibnamefont {San-Jose}},\ and\ \bibinfo {author} {\bibfnamefont {R.}~\bibnamefont {Aguado}},\ }\bibfield  {title} {\bibinfo {title} {Transport spectroscopy of $ns$ nanowire junctions with {Majorana} fermions},\ }\href {https://doi.org/10.1103/PhysRevB.86.180503} {\bibfield  {journal} {\bibinfo  {journal} {Phys. Rev. B}\ }\textbf {\bibinfo {volume} {86}},\ \bibinfo {pages} {180503} (\bibinfo {year} {2012})}\BibitemShut {NoStop}%
\bibitem [{\citenamefont {San-Jos\'{e}}\ \emph {et~al.}(2016)\citenamefont {San-Jos\'{e}}, \citenamefont {Cayao}, \citenamefont {Prada},\ and\ \citenamefont {Aguado}}]{SanJoseSciRep}%
  \BibitemOpen
  \bibfield  {author} {\bibinfo {author} {\bibfnamefont {P.}~\bibnamefont {San-Jos\'{e}}}, \bibinfo {author} {\bibfnamefont {J.}~\bibnamefont {Cayao}}, \bibinfo {author} {\bibfnamefont {E.}~\bibnamefont {Prada}},\ and\ \bibinfo {author} {\bibfnamefont {R.}~\bibnamefont {Aguado}},\ }\bibfield  {title} {\bibinfo {title} {Majorana bound states from exceptional points in non-topological superconductors},\ }\href {http://dx.doi.org/10.1038/srep21427} {\bibfield  {journal} {\bibinfo  {journal} {Sci. Rep.}\ }\textbf {\bibinfo {volume} {6}},\ \bibinfo {pages} {21427} (\bibinfo {year} {2016})}\BibitemShut {NoStop}%
\bibitem [{\citenamefont {Cayao}\ \emph {et~al.}(2015)\citenamefont {Cayao}, \citenamefont {Prada}, \citenamefont {San-Jose},\ and\ \citenamefont {Aguado}}]{Cayao2015Jan}%
  \BibitemOpen
  \bibfield  {author} {\bibinfo {author} {\bibfnamefont {J.}~\bibnamefont {Cayao}}, \bibinfo {author} {\bibfnamefont {E.}~\bibnamefont {Prada}}, \bibinfo {author} {\bibfnamefont {P.}~\bibnamefont {San-Jose}},\ and\ \bibinfo {author} {\bibfnamefont {R.}~\bibnamefont {Aguado}},\ }\bibfield  {title} {\bibinfo {title} {{SNS} junctions in nanowires with spin-orbit coupling: Role of confinement and helicity on the subgap spectrum},\ }\href {https://doi.org/10.1103/PhysRevB.91.024514} {\bibfield  {journal} {\bibinfo  {journal} {Phys. Rev. B}\ }\textbf {\bibinfo {volume} {91}},\ \bibinfo {pages} {024514} (\bibinfo {year} {2015})}\BibitemShut {NoStop}%
\bibitem [{\citenamefont {Liu}\ \emph {et~al.}(2017)\citenamefont {Liu}, \citenamefont {Sau}, \citenamefont {Stanescu},\ and\ \citenamefont {Das~Sarma}}]{Liu_PRB_2017}%
  \BibitemOpen
  \bibfield  {author} {\bibinfo {author} {\bibfnamefont {C.-X.}\ \bibnamefont {Liu}}, \bibinfo {author} {\bibfnamefont {J.~D.}\ \bibnamefont {Sau}}, \bibinfo {author} {\bibfnamefont {T.~D.}\ \bibnamefont {Stanescu}},\ and\ \bibinfo {author} {\bibfnamefont {S.}~\bibnamefont {Das~Sarma}},\ }\bibfield  {title} {\bibinfo {title} {Andreev bound states versus majorana bound states in quantum dot-nanowire-superconductor hybrid structures: Trivial versus topological zero-bias conductance peaks},\ }\href {https://doi.org/10.1103/PhysRevB.96.075161} {\bibfield  {journal} {\bibinfo  {journal} {Phys. Rev. B}\ }\textbf {\bibinfo {volume} {96}},\ \bibinfo {pages} {075161} (\bibinfo {year} {2017})}\BibitemShut {NoStop}%
\bibitem [{\citenamefont {Liu}\ \emph {et~al.}(2018)\citenamefont {Liu}, \citenamefont {Sau},\ and\ \citenamefont {Das~Sarma}}]{Liu_PRB_2018}%
  \BibitemOpen
  \bibfield  {author} {\bibinfo {author} {\bibfnamefont {C.-X.}\ \bibnamefont {Liu}}, \bibinfo {author} {\bibfnamefont {J.~D.}\ \bibnamefont {Sau}},\ and\ \bibinfo {author} {\bibfnamefont {S.}~\bibnamefont {Das~Sarma}},\ }\bibfield  {title} {\bibinfo {title} {Distinguishing topological majorana bound states from trivial andreev bound states: Proposed tests through differential tunneling conductance spectroscopy},\ }\href {https://doi.org/10.1103/PhysRevB.97.214502} {\bibfield  {journal} {\bibinfo  {journal} {Phys. Rev. B}\ }\textbf {\bibinfo {volume} {97}},\ \bibinfo {pages} {214502} (\bibinfo {year} {2018})}\BibitemShut {NoStop}%
\bibitem [{\citenamefont {Awoga}\ \emph {et~al.}(2019)\citenamefont {Awoga}, \citenamefont {Cayao},\ and\ \citenamefont {Black-Schaffer}}]{Awoga2019Sep}%
  \BibitemOpen
  \bibfield  {author} {\bibinfo {author} {\bibfnamefont {O.~A.}\ \bibnamefont {Awoga}}, \bibinfo {author} {\bibfnamefont {J.}~\bibnamefont {Cayao}},\ and\ \bibinfo {author} {\bibfnamefont {A.~M.}\ \bibnamefont {Black-Schaffer}},\ }\bibfield  {title} {\bibinfo {title} {Supercurrent detection of topologically trivial zero-energy states in nanowire junctions},\ }\href {https://doi.org/10.1103/PhysRevLett.123.117001} {\bibfield  {journal} {\bibinfo  {journal} {Phys. Rev. Lett.}\ }\textbf {\bibinfo {volume} {123}},\ \bibinfo {pages} {117001} (\bibinfo {year} {2019})}\BibitemShut {NoStop}%
\bibitem [{\citenamefont {Vuik}\ \emph {et~al.}(2019)\citenamefont {Vuik}, \citenamefont {Nijholt}, \citenamefont {Akhmerov},\ and\ \citenamefont {Wimmer}}]{Vuik_SPP_2019}%
  \BibitemOpen
  \bibfield  {author} {\bibinfo {author} {\bibfnamefont {A.}~\bibnamefont {Vuik}}, \bibinfo {author} {\bibfnamefont {B.}~\bibnamefont {Nijholt}}, \bibinfo {author} {\bibfnamefont {A.~R.}\ \bibnamefont {Akhmerov}},\ and\ \bibinfo {author} {\bibfnamefont {M.}~\bibnamefont {Wimmer}},\ }\bibfield  {title} {\bibinfo {title} {{Reproducing topological properties with quasi-Majorana states}},\ }\href {https://doi.org/10.21468/SciPostPhys.7.5.061} {\bibfield  {journal} {\bibinfo  {journal} {SciPost Phys.}\ }\textbf {\bibinfo {volume} {7}},\ \bibinfo {pages} {061} (\bibinfo {year} {2019})}\BibitemShut {NoStop}%
\bibitem [{\citenamefont {Cayao}\ and\ \citenamefont {Black-Schaffer}(2021)}]{Cayao2021Jul}%
  \BibitemOpen
  \bibfield  {author} {\bibinfo {author} {\bibfnamefont {J.}~\bibnamefont {Cayao}}\ and\ \bibinfo {author} {\bibfnamefont {A.~M.}\ \bibnamefont {Black-Schaffer}},\ }\bibfield  {title} {\bibinfo {title} {Distinguishing trivial and topological zero-energy states in long nanowire junctions},\ }\href {https://doi.org/10.1103/PhysRevB.104.L020501} {\bibfield  {journal} {\bibinfo  {journal} {Phys. Rev. B}\ }\textbf {\bibinfo {volume} {104}},\ \bibinfo {pages} {L020501} (\bibinfo {year} {2021})}\BibitemShut {NoStop}%
\bibitem [{\citenamefont {Cayao}\ and\ \citenamefont {Burset}(2021)}]{Burset_2021}%
  \BibitemOpen
  \bibfield  {author} {\bibinfo {author} {\bibfnamefont {J.}~\bibnamefont {Cayao}}\ and\ \bibinfo {author} {\bibfnamefont {P.}~\bibnamefont {Burset}},\ }\bibfield  {title} {\bibinfo {title} {Confinement-induced zero-bias peaks in conventional superconductor hybrids},\ }\href {https://doi.org/10.1103/PhysRevB.104.134507} {\bibfield  {journal} {\bibinfo  {journal} {Phys. Rev. B}\ }\textbf {\bibinfo {volume} {104}},\ \bibinfo {pages} {134507} (\bibinfo {year} {2021})}\BibitemShut {NoStop}%
\bibitem [{\citenamefont {de~Mendon\ifmmode~\mbox{\c{c}}\else \c{c}\fi{}a}\ \emph {et~al.}(2023)\citenamefont {de~Mendon\ifmmode~\mbox{\c{c}}\else \c{c}\fi{}a}, \citenamefont {Manesco}, \citenamefont {Sandler},\ and\ \citenamefont {Dias~da Silva}}]{deMendonca2023May}%
  \BibitemOpen
  \bibfield  {author} {\bibinfo {author} {\bibfnamefont {B.~S.}\ \bibnamefont {de~Mendon\ifmmode~\mbox{\c{c}}\else \c{c}\fi{}a}}, \bibinfo {author} {\bibfnamefont {A.~L.~R.}\ \bibnamefont {Manesco}}, \bibinfo {author} {\bibfnamefont {N.}~\bibnamefont {Sandler}},\ and\ \bibinfo {author} {\bibfnamefont {L.~G. G.~V.}\ \bibnamefont {Dias~da Silva}},\ }\bibfield  {title} {\bibinfo {title} {Near zero energy {Caroli}--de {Gennes}--{Matricon} vortex states in the presence of impurities},\ }\href {https://doi.org/10.1103/PhysRevB.107.184509} {\bibfield  {journal} {\bibinfo  {journal} {Phys. Rev. B}\ }\textbf {\bibinfo {volume} {107}},\ \bibinfo {pages} {184509} (\bibinfo {year} {2023})}\BibitemShut {NoStop}%
\bibitem [{\citenamefont {Baldo}\ \emph {et~al.}(2023)\citenamefont {Baldo}, \citenamefont {Da~Silva}, \citenamefont {Black-Schaffer},\ and\ \citenamefont {Cayao}}]{baldo2023zero}%
  \BibitemOpen
  \bibfield  {author} {\bibinfo {author} {\bibfnamefont {L.}~\bibnamefont {Baldo}}, \bibinfo {author} {\bibfnamefont {L.~G.~D.}\ \bibnamefont {Da~Silva}}, \bibinfo {author} {\bibfnamefont {A.~M.}\ \bibnamefont {Black-Schaffer}},\ and\ \bibinfo {author} {\bibfnamefont {J.}~\bibnamefont {Cayao}},\ }\bibfield  {title} {\bibinfo {title} {Zero-frequency supercurrent susceptibility signatures of trivial and topological zero-energy states in nanowire junctions},\ }\href {https://doi.org/10.1088/1361-6668/acb670} {\bibfield  {journal} {\bibinfo  {journal} {Supercond. Sci. Technol.}\ }\textbf {\bibinfo {volume} {36}},\ \bibinfo {pages} {034003} (\bibinfo {year} {2023})}\BibitemShut {NoStop}%
\bibitem [{\citenamefont {Awoga}\ \emph {et~al.}(2023)\citenamefont {Awoga}, \citenamefont {Leijnse}, \citenamefont {Black-Schaffer},\ and\ \citenamefont {Cayao}}]{Awoga2023May}%
  \BibitemOpen
  \bibfield  {author} {\bibinfo {author} {\bibfnamefont {O.~A.}\ \bibnamefont {Awoga}}, \bibinfo {author} {\bibfnamefont {M.}~\bibnamefont {Leijnse}}, \bibinfo {author} {\bibfnamefont {A.~M.}\ \bibnamefont {Black-Schaffer}},\ and\ \bibinfo {author} {\bibfnamefont {J.}~\bibnamefont {Cayao}},\ }\bibfield  {title} {\bibinfo {title} {Mitigating disorder-induced zero-energy states in weakly coupled superconductor-semiconductor hybrid systems},\ }\href {https://doi.org/10.1103/PhysRevB.107.184519} {\bibfield  {journal} {\bibinfo  {journal} {Phys. Rev. B}\ }\textbf {\bibinfo {volume} {107}},\ \bibinfo {pages} {184519} (\bibinfo {year} {2023})}\BibitemShut {NoStop}%
\bibitem [{\citenamefont {Hess}\ \emph {et~al.}(2023)\citenamefont {Hess}, \citenamefont {Legg}, \citenamefont {Loss},\ and\ \citenamefont {Klinovaja}}]{hes23}%
  \BibitemOpen
  \bibfield  {author} {\bibinfo {author} {\bibfnamefont {R.}~\bibnamefont {Hess}}, \bibinfo {author} {\bibfnamefont {H.~F.}\ \bibnamefont {Legg}}, \bibinfo {author} {\bibfnamefont {D.}~\bibnamefont {Loss}},\ and\ \bibinfo {author} {\bibfnamefont {J.}~\bibnamefont {Klinovaja}},\ }\bibfield  {title} {\bibinfo {title} {Trivial {Andreev} band mimicking topological bulk gap reopening in the nonlocal conductance of long {Rashba} nanowires},\ }\href {https://doi.org/10.1103/PhysRevLett.130.207001} {\bibfield  {journal} {\bibinfo  {journal} {Phys. Rev. Lett.}\ }\textbf {\bibinfo {volume} {130}},\ \bibinfo {pages} {207001} (\bibinfo {year} {2023})}\BibitemShut {NoStop}%
\bibitem [{\citenamefont {Cayao}(2024{\natexlab{a}})}]{Cayao2024Aug}%
  \BibitemOpen
  \bibfield  {author} {\bibinfo {author} {\bibfnamefont {J.}~\bibnamefont {Cayao}},\ }\bibfield  {title} {\bibinfo {title} {Non-{Hermitian} zero-energy pinning of {Andreev} and {Majorana} bound states in superconductor-semiconductor systems},\ }\href {https://doi.org/10.1103/PhysRevB.110.085414} {\bibfield  {journal} {\bibinfo  {journal} {Phys. Rev. B}\ }\textbf {\bibinfo {volume} {110}},\ \bibinfo {pages} {085414} (\bibinfo {year} {2024}{\natexlab{a}})}\BibitemShut {NoStop}%
\bibitem [{\citenamefont {Ahmed}\ \emph {et~al.}(2025{\natexlab{a}})\citenamefont {Ahmed}, \citenamefont {Tanaka},\ and\ \citenamefont {Cayao}}]{ahmed2025}%
  \BibitemOpen
  \bibfield  {author} {\bibinfo {author} {\bibfnamefont {E.}~\bibnamefont {Ahmed}}, \bibinfo {author} {\bibfnamefont {Y.}~\bibnamefont {Tanaka}},\ and\ \bibinfo {author} {\bibfnamefont {J.}~\bibnamefont {Cayao}},\ }\bibfield  {title} {\bibinfo {title} {Anomalous proximity effect under {Andreev} and {Majorana} bound states},\ }\href {https://doi.org/10.1007/s10948-025-07057-9} {\bibfield  {journal} {\bibinfo  {journal} {J. Supercond. Nov. Magn.}\ }\textbf {\bibinfo {volume} {38}},\ \bibinfo {pages} {220} (\bibinfo {year} {2025}{\natexlab{a}})}\BibitemShut {NoStop}%
\bibitem [{\citenamefont {Braunecker}\ \emph {et~al.}(2010)\citenamefont {Braunecker}, \citenamefont {Japaridze}, \citenamefont {Klinovaja},\ and\ \citenamefont {Loss}}]{Braunecker2010}%
  \BibitemOpen
  \bibfield  {author} {\bibinfo {author} {\bibfnamefont {B.}~\bibnamefont {Braunecker}}, \bibinfo {author} {\bibfnamefont {G.~I.}\ \bibnamefont {Japaridze}}, \bibinfo {author} {\bibfnamefont {J.}~\bibnamefont {Klinovaja}},\ and\ \bibinfo {author} {\bibfnamefont {D.}~\bibnamefont {Loss}},\ }\bibfield  {title} {\bibinfo {title} {Spin-selective {Peierls} transition in interacting one-dimensional conductors with spin-orbit interaction},\ }\href {https://doi.org/10.1103/PhysRevB.82.045127} {\bibfield  {journal} {\bibinfo  {journal} {Phys. Rev. B}\ }\textbf {\bibinfo {volume} {82}},\ \bibinfo {pages} {045127} (\bibinfo {year} {2010})}\BibitemShut {NoStop}%
\bibitem [{\citenamefont {Choy}\ \emph {et~al.}(2011)\citenamefont {Choy}, \citenamefont {Edge}, \citenamefont {Akhmerov},\ and\ \citenamefont {Beenakker}}]{Beenakker_2011}%
  \BibitemOpen
  \bibfield  {author} {\bibinfo {author} {\bibfnamefont {T.-P.}\ \bibnamefont {Choy}}, \bibinfo {author} {\bibfnamefont {J.~M.}\ \bibnamefont {Edge}}, \bibinfo {author} {\bibfnamefont {A.~R.}\ \bibnamefont {Akhmerov}},\ and\ \bibinfo {author} {\bibfnamefont {C.~W.~J.}\ \bibnamefont {Beenakker}},\ }\bibfield  {title} {\bibinfo {title} {Majorana fermions emerging from magnetic nanoparticles on a superconductor without spin-orbit coupling},\ }\href {https://doi.org/10.1103/PhysRevB.84.195442} {\bibfield  {journal} {\bibinfo  {journal} {Phys. Rev. B}\ }\textbf {\bibinfo {volume} {84}},\ \bibinfo {pages} {195442} (\bibinfo {year} {2011})}\BibitemShut {NoStop}%
\bibitem [{\citenamefont {Egger}\ and\ \citenamefont {Flensberg}(2012)}]{Egger_2012_PRB}%
  \BibitemOpen
  \bibfield  {author} {\bibinfo {author} {\bibfnamefont {R.}~\bibnamefont {Egger}}\ and\ \bibinfo {author} {\bibfnamefont {K.}~\bibnamefont {Flensberg}},\ }\bibfield  {title} {\bibinfo {title} {{Emerging {Dirac} and {Majorana} fermions for carbon nanotubes with proximity-induced pairing and spiral magnetic field}},\ }\href {https://doi.org/10.1103/PhysRevB.85.235462} {\bibfield  {journal} {\bibinfo  {journal} {Phys. Rev. B}\ }\textbf {\bibinfo {volume} {85}},\ \bibinfo {pages} {235462} (\bibinfo {year} {2012})}\BibitemShut {NoStop}%
\bibitem [{\citenamefont {Braunecker}\ and\ \citenamefont {Simon}(2013)}]{Braunecker_2013}%
  \BibitemOpen
  \bibfield  {author} {\bibinfo {author} {\bibfnamefont {B.}~\bibnamefont {Braunecker}}\ and\ \bibinfo {author} {\bibfnamefont {P.}~\bibnamefont {Simon}},\ }\bibfield  {title} {\bibinfo {title} {Interplay between classical magnetic moments and superconductivity in quantum one-dimensional conductors: Toward a self-sustained topological {Majorana} phase},\ }\href {https://doi.org/10.1103/PhysRevLett.111.147202} {\bibfield  {journal} {\bibinfo  {journal} {Phys. Rev. Lett.}\ }\textbf {\bibinfo {volume} {111}},\ \bibinfo {pages} {147202} (\bibinfo {year} {2013})}\BibitemShut {NoStop}%
\bibitem [{\citenamefont {Klinovaja}\ \emph {et~al.}(2013)\citenamefont {Klinovaja}, \citenamefont {Stano}, \citenamefont {Yazdani},\ and\ \citenamefont {Loss}}]{Klinovaja_2013_PRL}%
  \BibitemOpen
  \bibfield  {author} {\bibinfo {author} {\bibfnamefont {J.}~\bibnamefont {Klinovaja}}, \bibinfo {author} {\bibfnamefont {P.}~\bibnamefont {Stano}}, \bibinfo {author} {\bibfnamefont {A.}~\bibnamefont {Yazdani}},\ and\ \bibinfo {author} {\bibfnamefont {D.}~\bibnamefont {Loss}},\ }\bibfield  {title} {\bibinfo {title} {{Topological superconductivity and Majorana fermions in RKKY systems}},\ }\href {https://doi.org/10.1103/PhysRevLett.111.186805} {\bibfield  {journal} {\bibinfo  {journal} {Phys. Rev. Lett.}\ }\textbf {\bibinfo {volume} {111}},\ \bibinfo {pages} {186805} (\bibinfo {year} {2013})}\BibitemShut {NoStop}%
\bibitem [{\citenamefont {Vazifeh}\ and\ \citenamefont {Franz}(2013)}]{Franz_2013}%
  \BibitemOpen
  \bibfield  {author} {\bibinfo {author} {\bibfnamefont {M.~M.}\ \bibnamefont {Vazifeh}}\ and\ \bibinfo {author} {\bibfnamefont {M.}~\bibnamefont {Franz}},\ }\bibfield  {title} {\bibinfo {title} {Self-organized topological state with {Majorana} fermions},\ }\href {https://doi.org/10.1103/PhysRevLett.111.206802} {\bibfield  {journal} {\bibinfo  {journal} {Phys. Rev. Lett.}\ }\textbf {\bibinfo {volume} {111}},\ \bibinfo {pages} {206802} (\bibinfo {year} {2013})}\BibitemShut {NoStop}%
\bibitem [{\citenamefont {Nadj-Perge}\ \emph {et~al.}(2013)\citenamefont {Nadj-Perge}, \citenamefont {Drozdov}, \citenamefont {Bernevig},\ and\ \citenamefont {Yazdani}}]{Perge_PRB2013}%
  \BibitemOpen
  \bibfield  {author} {\bibinfo {author} {\bibfnamefont {S.}~\bibnamefont {Nadj-Perge}}, \bibinfo {author} {\bibfnamefont {I.~K.}\ \bibnamefont {Drozdov}}, \bibinfo {author} {\bibfnamefont {B.~A.}\ \bibnamefont {Bernevig}},\ and\ \bibinfo {author} {\bibfnamefont {A.}~\bibnamefont {Yazdani}},\ }\bibfield  {title} {\bibinfo {title} {Proposal for realizing {Majorana} fermions in chains of magnetic atoms on a superconductor},\ }\href {https://doi.org/10.1103/PhysRevB.88.020407} {\bibfield  {journal} {\bibinfo  {journal} {Phys. Rev. B}\ }\textbf {\bibinfo {volume} {88}},\ \bibinfo {pages} {020407} (\bibinfo {year} {2013})}\BibitemShut {NoStop}%
\bibitem [{\citenamefont {Pientka}\ \emph {et~al.}(2013)\citenamefont {Pientka}, \citenamefont {Glazman},\ and\ \citenamefont {von Oppen}}]{Pientka_PRB2013}%
  \BibitemOpen
  \bibfield  {author} {\bibinfo {author} {\bibfnamefont {F.}~\bibnamefont {Pientka}}, \bibinfo {author} {\bibfnamefont {L.~I.}\ \bibnamefont {Glazman}},\ and\ \bibinfo {author} {\bibfnamefont {F.}~\bibnamefont {von Oppen}},\ }\bibfield  {title} {\bibinfo {title} {Topological superconducting phase in helical {Shiba} chains},\ }\href {https://doi.org/10.1103/PhysRevB.88.155420} {\bibfield  {journal} {\bibinfo  {journal} {Phys. Rev. B}\ }\textbf {\bibinfo {volume} {88}},\ \bibinfo {pages} {155420} (\bibinfo {year} {2013})}\BibitemShut {NoStop}%
\bibitem [{\citenamefont {P\"oyh\"onen}\ \emph {et~al.}(2014)\citenamefont {P\"oyh\"onen}, \citenamefont {Weststr\"om}, \citenamefont {R\"ontynen},\ and\ \citenamefont {Ojanen}}]{Ojanen_2014}%
  \BibitemOpen
  \bibfield  {author} {\bibinfo {author} {\bibfnamefont {K.}~\bibnamefont {P\"oyh\"onen}}, \bibinfo {author} {\bibfnamefont {A.}~\bibnamefont {Weststr\"om}}, \bibinfo {author} {\bibfnamefont {J.}~\bibnamefont {R\"ontynen}},\ and\ \bibinfo {author} {\bibfnamefont {T.}~\bibnamefont {Ojanen}},\ }\bibfield  {title} {\bibinfo {title} {Majorana states in helical {Shiba} chains and ladders},\ }\href {https://doi.org/10.1103/PhysRevB.89.115109} {\bibfield  {journal} {\bibinfo  {journal} {Phys. Rev. B}\ }\textbf {\bibinfo {volume} {89}},\ \bibinfo {pages} {115109} (\bibinfo {year} {2014})}\BibitemShut {NoStop}%
\bibitem [{\citenamefont {Heimes}\ \emph {et~al.}(2014)\citenamefont {Heimes}, \citenamefont {Kotetes},\ and\ \citenamefont {Sch\"on}}]{Heimes_2014}%
  \BibitemOpen
  \bibfield  {author} {\bibinfo {author} {\bibfnamefont {A.}~\bibnamefont {Heimes}}, \bibinfo {author} {\bibfnamefont {P.}~\bibnamefont {Kotetes}},\ and\ \bibinfo {author} {\bibfnamefont {G.}~\bibnamefont {Sch\"on}},\ }\bibfield  {title} {\bibinfo {title} {Majorana fermions from {Shiba} states in an antiferromagnetic chain on top of a superconductor},\ }\href {https://doi.org/10.1103/PhysRevB.90.060507} {\bibfield  {journal} {\bibinfo  {journal} {Phys. Rev. B}\ }\textbf {\bibinfo {volume} {90}},\ \bibinfo {pages} {060507} (\bibinfo {year} {2014})}\BibitemShut {NoStop}%
\bibitem [{\citenamefont {Xiao}\ and\ \citenamefont {An}(2015)}]{Xiao2015}%
  \BibitemOpen
  \bibfield  {author} {\bibinfo {author} {\bibfnamefont {J.}~\bibnamefont {Xiao}}\ and\ \bibinfo {author} {\bibfnamefont {J.}~\bibnamefont {An}},\ }\bibfield  {title} {\bibinfo {title} {{Chiral symmetries and Majorana fermions in coupled magnetic atomic chains on a superconductor}},\ }\href {https://doi.org/10.1088/1367-2630/17/11/113034} {\bibfield  {journal} {\bibinfo  {journal} {New J. Phys.}\ }\textbf {\bibinfo {volume} {17}},\ \bibinfo {pages} {113034} (\bibinfo {year} {2015})}\BibitemShut {NoStop}%
\bibitem [{\citenamefont {Schecter}\ \emph {et~al.}(2016)\citenamefont {Schecter}, \citenamefont {Flensberg}, \citenamefont {Christensen}, \citenamefont {Andersen},\ and\ \citenamefont {Paaske}}]{Paaske_2016}%
  \BibitemOpen
  \bibfield  {author} {\bibinfo {author} {\bibfnamefont {M.}~\bibnamefont {Schecter}}, \bibinfo {author} {\bibfnamefont {K.}~\bibnamefont {Flensberg}}, \bibinfo {author} {\bibfnamefont {M.~H.}\ \bibnamefont {Christensen}}, \bibinfo {author} {\bibfnamefont {B.~M.}\ \bibnamefont {Andersen}},\ and\ \bibinfo {author} {\bibfnamefont {J.}~\bibnamefont {Paaske}},\ }\bibfield  {title} {\bibinfo {title} {Self-organized topological superconductivity in a {Yu}-{Shiba}-{Rusinov} chain},\ }\href {https://doi.org/10.1103/PhysRevB.93.140503} {\bibfield  {journal} {\bibinfo  {journal} {Phys. Rev. B}\ }\textbf {\bibinfo {volume} {93}},\ \bibinfo {pages} {140503} (\bibinfo {year} {2016})}\BibitemShut {NoStop}%
\bibitem [{\citenamefont {Christensen}\ \emph {et~al.}(2016)\citenamefont {Christensen}, \citenamefont {Schecter}, \citenamefont {Flensberg}, \citenamefont {Andersen},\ and\ \citenamefont {Paaske}}]{Paaske_2016b}%
  \BibitemOpen
  \bibfield  {author} {\bibinfo {author} {\bibfnamefont {M.~H.}\ \bibnamefont {Christensen}}, \bibinfo {author} {\bibfnamefont {M.}~\bibnamefont {Schecter}}, \bibinfo {author} {\bibfnamefont {K.}~\bibnamefont {Flensberg}}, \bibinfo {author} {\bibfnamefont {B.~M.}\ \bibnamefont {Andersen}},\ and\ \bibinfo {author} {\bibfnamefont {J.}~\bibnamefont {Paaske}},\ }\bibfield  {title} {\bibinfo {title} {Spiral magnetic order and topological superconductivity in a chain of magnetic adatoms on a two-dimensional superconductor},\ }\href {https://doi.org/10.1103/PhysRevB.94.144509} {\bibfield  {journal} {\bibinfo  {journal} {Phys. Rev. B}\ }\textbf {\bibinfo {volume} {94}},\ \bibinfo {pages} {144509} (\bibinfo {year} {2016})}\BibitemShut {NoStop}%
\bibitem [{\citenamefont {Marra}\ and\ \citenamefont {Cuoco}(2017)}]{Cuoco_2017}%
  \BibitemOpen
  \bibfield  {author} {\bibinfo {author} {\bibfnamefont {P.}~\bibnamefont {Marra}}\ and\ \bibinfo {author} {\bibfnamefont {M.}~\bibnamefont {Cuoco}},\ }\bibfield  {title} {\bibinfo {title} {Controlling {Majorana} states in topologically inhomogeneous superconductors},\ }\href {https://doi.org/10.1103/PhysRevB.95.140504} {\bibfield  {journal} {\bibinfo  {journal} {Phys. Rev. B}\ }\textbf {\bibinfo {volume} {95}},\ \bibinfo {pages} {140504} (\bibinfo {year} {2017})}\BibitemShut {NoStop}%
\bibitem [{\citenamefont {Kim}\ \emph {et~al.}(2018)\citenamefont {Kim}, \citenamefont {Palacio-Morales}, \citenamefont {Posske}, \citenamefont {R{\ifmmode\acute{o}\else\'{o}\fi}zsa}, \citenamefont {Palot{\ifmmode\acute{a}\else\'{a}\fi}s}, \citenamefont {Szunyogh}, \citenamefont {Thorwart},\ and\ \citenamefont {Wiesendanger}}]{Kim2018}%
  \BibitemOpen
  \bibfield  {author} {\bibinfo {author} {\bibfnamefont {H.}~\bibnamefont {Kim}}, \bibinfo {author} {\bibfnamefont {A.}~\bibnamefont {Palacio-Morales}}, \bibinfo {author} {\bibfnamefont {T.}~\bibnamefont {Posske}}, \bibinfo {author} {\bibfnamefont {L.}~\bibnamefont {R{\ifmmode\acute{o}\else\'{o}\fi}zsa}}, \bibinfo {author} {\bibfnamefont {K.}~\bibnamefont {Palot{\ifmmode\acute{a}\else\'{a}\fi}s}}, \bibinfo {author} {\bibfnamefont {L.}~\bibnamefont {Szunyogh}}, \bibinfo {author} {\bibfnamefont {M.}~\bibnamefont {Thorwart}},\ and\ \bibinfo {author} {\bibfnamefont {R.}~\bibnamefont {Wiesendanger}},\ }\bibfield  {title} {\bibinfo {title} {{Toward tailoring Majorana bound states in artificially constructed magnetic atom chains on elemental superconductors}},\ }\bibfield  {journal} {\bibinfo  {journal} {Sci. Adv.}\ }\textbf {\bibinfo {volume} {4}},\ \href {https://doi.org/10.1126/sciadv.aar5251} {10.1126/sciadv.aar5251} (\bibinfo {year} {2018})\BibitemShut {NoStop}%
\bibitem [{\citenamefont {Cayao}\ \emph {et~al.}(2018)\citenamefont {Cayao}, \citenamefont {Black-Schaffer}, \citenamefont {Prada},\ and\ \citenamefont {Aguado}}]{cayao2018andreev}%
  \BibitemOpen
  \bibfield  {author} {\bibinfo {author} {\bibfnamefont {J.}~\bibnamefont {Cayao}}, \bibinfo {author} {\bibfnamefont {A.~M.}\ \bibnamefont {Black-Schaffer}}, \bibinfo {author} {\bibfnamefont {E.}~\bibnamefont {Prada}},\ and\ \bibinfo {author} {\bibfnamefont {R.}~\bibnamefont {Aguado}},\ }\bibfield  {title} {\bibinfo {title} {Andreev spectrum and supercurrents in nanowire-based {SNS} junctions containing {M}ajorana bound states},\ }\href {https://doi.org/10.3762/bjnano.9.127} {\bibfield  {journal} {\bibinfo  {journal} {Beilstein J. Nanotechnol.}\ }\textbf {\bibinfo {volume} {9}},\ \bibinfo {pages} {1339} (\bibinfo {year} {2018})}\BibitemShut {NoStop}%
\bibitem [{\citenamefont {Kornich}\ \emph {et~al.}(2020)\citenamefont {Kornich}, \citenamefont {Vavilov}, \citenamefont {Friesen}, \citenamefont {Eriksson},\ and\ \citenamefont {Coppersmith}}]{Kornich_2020}%
  \BibitemOpen
  \bibfield  {author} {\bibinfo {author} {\bibfnamefont {V.}~\bibnamefont {Kornich}}, \bibinfo {author} {\bibfnamefont {M.~G.}\ \bibnamefont {Vavilov}}, \bibinfo {author} {\bibfnamefont {M.}~\bibnamefont {Friesen}}, \bibinfo {author} {\bibfnamefont {M.~A.}\ \bibnamefont {Eriksson}},\ and\ \bibinfo {author} {\bibfnamefont {S.~N.}\ \bibnamefont {Coppersmith}},\ }\bibfield  {title} {\bibinfo {title} {Majorana bound states in nanowire-superconductor hybrid systems in periodic magnetic fields},\ }\href {https://doi.org/10.1103/PhysRevB.101.125414} {\bibfield  {journal} {\bibinfo  {journal} {Phys. Rev. B}\ }\textbf {\bibinfo {volume} {101}},\ \bibinfo {pages} {125414} (\bibinfo {year} {2020})}\BibitemShut {NoStop}%
\bibitem [{\citenamefont {Escribano}\ \emph {et~al.}(2021)\citenamefont {Escribano}, \citenamefont {Prada}, \citenamefont {Oreg},\ and\ \citenamefont {Yeyati}}]{escribano2021}%
  \BibitemOpen
  \bibfield  {author} {\bibinfo {author} {\bibfnamefont {S.~D.}\ \bibnamefont {Escribano}}, \bibinfo {author} {\bibfnamefont {E.}~\bibnamefont {Prada}}, \bibinfo {author} {\bibfnamefont {Y.}~\bibnamefont {Oreg}},\ and\ \bibinfo {author} {\bibfnamefont {A.~L.}\ \bibnamefont {Yeyati}},\ }\bibfield  {title} {\bibinfo {title} {Tunable proximity effects and topological superconductivity in ferromagnetic hybrid nanowires},\ }\href {https://doi.org/10.1103/PhysRevB.104.L041404} {\bibfield  {journal} {\bibinfo  {journal} {Phys. Rev. B}\ }\textbf {\bibinfo {volume} {104}},\ \bibinfo {pages} {L041404} (\bibinfo {year} {2021})}\BibitemShut {NoStop}%
\bibitem [{\citenamefont {Liu}\ \emph {et~al.}(2021)\citenamefont {Liu}, \citenamefont {Schuwalow}, \citenamefont {Liu}, \citenamefont {Vilkelis}, \citenamefont {Manesco}, \citenamefont {Krogstrup},\ and\ \citenamefont {Wimmer}}]{Liu_PRB2021}%
  \BibitemOpen
  \bibfield  {author} {\bibinfo {author} {\bibfnamefont {C.-X.}\ \bibnamefont {Liu}}, \bibinfo {author} {\bibfnamefont {S.}~\bibnamefont {Schuwalow}}, \bibinfo {author} {\bibfnamefont {Y.}~\bibnamefont {Liu}}, \bibinfo {author} {\bibfnamefont {K.}~\bibnamefont {Vilkelis}}, \bibinfo {author} {\bibfnamefont {A.~L.~R.}\ \bibnamefont {Manesco}}, \bibinfo {author} {\bibfnamefont {P.}~\bibnamefont {Krogstrup}},\ and\ \bibinfo {author} {\bibfnamefont {M.}~\bibnamefont {Wimmer}},\ }\bibfield  {title} {\bibinfo {title} {Electronic properties of {InAs}/{EuS}/{Al} hybrid nanowires},\ }\href {https://doi.org/10.1103/PhysRevB.104.014516} {\bibfield  {journal} {\bibinfo  {journal} {Phys. Rev. B}\ }\textbf {\bibinfo {volume} {104}},\ \bibinfo {pages} {014516} (\bibinfo {year} {2021})}\BibitemShut {NoStop}%
\bibitem [{\citenamefont {Langbehn}\ \emph {et~al.}(2021)\citenamefont {Langbehn}, \citenamefont {Acero~Gonz\'alez}, \citenamefont {Brouwer},\ and\ \citenamefont {von Oppen}}]{Langbehn_PRB2021}%
  \BibitemOpen
  \bibfield  {author} {\bibinfo {author} {\bibfnamefont {J.}~\bibnamefont {Langbehn}}, \bibinfo {author} {\bibfnamefont {S.}~\bibnamefont {Acero~Gonz\'alez}}, \bibinfo {author} {\bibfnamefont {P.~W.}\ \bibnamefont {Brouwer}},\ and\ \bibinfo {author} {\bibfnamefont {F.}~\bibnamefont {von Oppen}},\ }\bibfield  {title} {\bibinfo {title} {Topological superconductivity in tripartite superconductor-ferromagnet-semiconductor nanowires},\ }\href {https://doi.org/10.1103/PhysRevB.103.165301} {\bibfield  {journal} {\bibinfo  {journal} {Phys. Rev. B}\ }\textbf {\bibinfo {volume} {103}},\ \bibinfo {pages} {165301} (\bibinfo {year} {2021})}\BibitemShut {NoStop}%
\bibitem [{\citenamefont {Awoga}\ \emph {et~al.}(2022)\citenamefont {Awoga}, \citenamefont {Cayao},\ and\ \citenamefont {Black-Schaffer}}]{Awoga2022Apr}%
  \BibitemOpen
  \bibfield  {author} {\bibinfo {author} {\bibfnamefont {O.~A.}\ \bibnamefont {Awoga}}, \bibinfo {author} {\bibfnamefont {J.}~\bibnamefont {Cayao}},\ and\ \bibinfo {author} {\bibfnamefont {A.~M.}\ \bibnamefont {Black-Schaffer}},\ }\bibfield  {title} {\bibinfo {title} {Robust topological superconductivity in weakly coupled nanowire-superconductor hybrid structures},\ }\href {https://doi.org/10.1103/PhysRevB.105.144509} {\bibfield  {journal} {\bibinfo  {journal} {Phys. Rev. B}\ }\textbf {\bibinfo {volume} {105}},\ \bibinfo {pages} {144509} (\bibinfo {year} {2022})}\BibitemShut {NoStop}%
\bibitem [{\citenamefont {Escribano}\ \emph {et~al.}(2022)\citenamefont {Escribano}, \citenamefont {Maiani}, \citenamefont {Leijnse}, \citenamefont {Flensberg}, \citenamefont {Oreg}, \citenamefont {Levy~Yeyati}, \citenamefont {Prada},\ and\ \citenamefont {Seoane~Souto}}]{Escribano_NPJ2022}%
  \BibitemOpen
  \bibfield  {author} {\bibinfo {author} {\bibfnamefont {S.~D.}\ \bibnamefont {Escribano}}, \bibinfo {author} {\bibfnamefont {A.}~\bibnamefont {Maiani}}, \bibinfo {author} {\bibfnamefont {M.}~\bibnamefont {Leijnse}}, \bibinfo {author} {\bibfnamefont {K.}~\bibnamefont {Flensberg}}, \bibinfo {author} {\bibfnamefont {Y.}~\bibnamefont {Oreg}}, \bibinfo {author} {\bibfnamefont {A.}~\bibnamefont {Levy~Yeyati}}, \bibinfo {author} {\bibfnamefont {E.}~\bibnamefont {Prada}},\ and\ \bibinfo {author} {\bibfnamefont {R.}~\bibnamefont {Seoane~Souto}},\ }\bibfield  {title} {\bibinfo {title} {Semiconductor-ferromagnet-superconductor planar heterostructures for {1D} topological superconductivity},\ }\href {https://doi.org/10.1038/s41535-022-00489-9} {\bibfield  {journal} {\bibinfo  {journal} {npj Quantum Materials}\ }\textbf {\bibinfo {volume} {7}},\ \bibinfo {pages} {81} (\bibinfo {year} {2022})}\BibitemShut {NoStop}%
\bibitem [{\citenamefont {Jardine}\ \emph {et~al.}(2021)\citenamefont {Jardine}, \citenamefont {Stenger}, \citenamefont {Jiang}, \citenamefont {de~Jong}, \citenamefont {Wang}, \citenamefont {Jayich},\ and\ \citenamefont {Frolov}}]{Frolov_2021}%
  \BibitemOpen
  \bibfield  {author} {\bibinfo {author} {\bibfnamefont {M.~J.~A.}\ \bibnamefont {Jardine}}, \bibinfo {author} {\bibfnamefont {J.~P.~T.}\ \bibnamefont {Stenger}}, \bibinfo {author} {\bibfnamefont {Y.}~\bibnamefont {Jiang}}, \bibinfo {author} {\bibfnamefont {E.~J.}\ \bibnamefont {de~Jong}}, \bibinfo {author} {\bibfnamefont {W.}~\bibnamefont {Wang}}, \bibinfo {author} {\bibfnamefont {A.~C.~B.}\ \bibnamefont {Jayich}},\ and\ \bibinfo {author} {\bibfnamefont {S.~M.}\ \bibnamefont {Frolov}},\ }\bibfield  {title} {\bibinfo {title} {{Integrating micromagnets and hybrid nanowires for topological quantum computing}},\ }\href {https://doi.org/10.21468/SciPostPhys.11.5.090} {\bibfield  {journal} {\bibinfo  {journal} {SciPost Phys.}\ }\textbf {\bibinfo {volume} {11}},\ \bibinfo {pages} {090} (\bibinfo {year} {2021})}\BibitemShut {NoStop}%
\bibitem [{\citenamefont {Mondal}\ \emph {et~al.}(2023)\citenamefont {Mondal}, \citenamefont {Ghosh}, \citenamefont {Nag},\ and\ \citenamefont {Saha}}]{Mondal_PRB2023}%
  \BibitemOpen
  \bibfield  {author} {\bibinfo {author} {\bibfnamefont {D.}~\bibnamefont {Mondal}}, \bibinfo {author} {\bibfnamefont {A.~K.}\ \bibnamefont {Ghosh}}, \bibinfo {author} {\bibfnamefont {T.}~\bibnamefont {Nag}},\ and\ \bibinfo {author} {\bibfnamefont {A.}~\bibnamefont {Saha}},\ }\bibfield  {title} {\bibinfo {title} {{Engineering anomalous Floquet Majorana modes and their time evolution in a helical Shiba chain}},\ }\href {https://doi.org/10.1103/PhysRevB.108.L081403} {\bibfield  {journal} {\bibinfo  {journal} {Phys. Rev. B}\ }\textbf {\bibinfo {volume} {108}},\ \bibinfo {pages} {L081403} (\bibinfo {year} {2023})}\BibitemShut {NoStop}%
\bibitem [{\citenamefont {Mizushima}\ \emph {et~al.}(2025)\citenamefont {Mizushima}, \citenamefont {Tanaka},\ and\ \citenamefont {Cayao}}]{mizushima2025}%
  \BibitemOpen
  \bibfield  {author} {\bibinfo {author} {\bibfnamefont {T.}~\bibnamefont {Mizushima}}, \bibinfo {author} {\bibfnamefont {Y.}~\bibnamefont {Tanaka}},\ and\ \bibinfo {author} {\bibfnamefont {J.}~\bibnamefont {Cayao}},\ }\bibfield  {title} {\bibinfo {title} {Detecting topological phase transition in superconductor-semiconductor hybrids by electronic {Raman} spectroscopy},\ }\href {https://arxiv.org/abs/2502.19841} {\bibfield  {journal} {\bibinfo  {journal} {arXiv: 2502.19841}\ } (\bibinfo {year} {2025})}\BibitemShut {NoStop}%
\bibitem [{\citenamefont {Mourik}\ \emph {et~al.}(2012)\citenamefont {Mourik}, \citenamefont {Zuo}, \citenamefont {Frolov}, \citenamefont {Plissard}, \citenamefont {Bakkers},\ and\ \citenamefont {Kouwenhoven}}]{Mourik_Science2012}%
  \BibitemOpen
  \bibfield  {author} {\bibinfo {author} {\bibfnamefont {V.}~\bibnamefont {Mourik}}, \bibinfo {author} {\bibfnamefont {K.}~\bibnamefont {Zuo}}, \bibinfo {author} {\bibfnamefont {S.~M.}\ \bibnamefont {Frolov}}, \bibinfo {author} {\bibfnamefont {S.~R.}\ \bibnamefont {Plissard}}, \bibinfo {author} {\bibfnamefont {E.~P. A.~M.}\ \bibnamefont {Bakkers}},\ and\ \bibinfo {author} {\bibfnamefont {L.~P.}\ \bibnamefont {Kouwenhoven}},\ }\bibfield  {title} {\bibinfo {title} {{Signatures of Majorana fermions in hybrid superconductor-semiconductor nanowire devices}},\ }\href {https://doi.org/10.1126/science.1222360} {\bibfield  {journal} {\bibinfo  {journal} {Science}\ }\textbf {\bibinfo {volume} {336}},\ \bibinfo {pages} {1003} (\bibinfo {year} {2012})}\BibitemShut {NoStop}%
\bibitem [{\citenamefont {Nadj-Perge}\ \emph {et~al.}(2014)\citenamefont {Nadj-Perge}, \citenamefont {Drozdov}, \citenamefont {Li}, \citenamefont {Chen}, \citenamefont {Jeon}, \citenamefont {Seo}, \citenamefont {MacDonald}, \citenamefont {Bernevig},\ and\ \citenamefont {Yazdani}}]{Nadj-Perge2014}%
  \BibitemOpen
  \bibfield  {author} {\bibinfo {author} {\bibfnamefont {S.}~\bibnamefont {Nadj-Perge}}, \bibinfo {author} {\bibfnamefont {I.~K.}\ \bibnamefont {Drozdov}}, \bibinfo {author} {\bibfnamefont {J.}~\bibnamefont {Li}}, \bibinfo {author} {\bibfnamefont {H.}~\bibnamefont {Chen}}, \bibinfo {author} {\bibfnamefont {S.}~\bibnamefont {Jeon}}, \bibinfo {author} {\bibfnamefont {J.}~\bibnamefont {Seo}}, \bibinfo {author} {\bibfnamefont {A.~H.}\ \bibnamefont {MacDonald}}, \bibinfo {author} {\bibfnamefont {B.~A.}\ \bibnamefont {Bernevig}},\ and\ \bibinfo {author} {\bibfnamefont {A.}~\bibnamefont {Yazdani}},\ }\bibfield  {title} {\bibinfo {title} {{Observation of Majorana fermions in ferromagnetic atomic chains on a superconductor}},\ }\href {https://www.science.org/doi/abs/10.1126/science.1259327} {\bibfield  {journal} {\bibinfo  {journal} {Science}\ }\textbf {\bibinfo {volume} {346}},\ \bibinfo {pages} {602} (\bibinfo {year} {2014})}\BibitemShut {NoStop}%
\bibitem [{\citenamefont {Ruby}\ \emph {et~al.}(2015)\citenamefont {Ruby}, \citenamefont {Pientka}, \citenamefont {Peng}, \citenamefont {von Oppen}, \citenamefont {Heinrich},\ and\ \citenamefont {Franke}}]{Ruby_PRL_2015}%
  \BibitemOpen
  \bibfield  {author} {\bibinfo {author} {\bibfnamefont {M.}~\bibnamefont {Ruby}}, \bibinfo {author} {\bibfnamefont {F.}~\bibnamefont {Pientka}}, \bibinfo {author} {\bibfnamefont {Y.}~\bibnamefont {Peng}}, \bibinfo {author} {\bibfnamefont {F.}~\bibnamefont {von Oppen}}, \bibinfo {author} {\bibfnamefont {B.~W.}\ \bibnamefont {Heinrich}},\ and\ \bibinfo {author} {\bibfnamefont {K.~J.}\ \bibnamefont {Franke}},\ }\bibfield  {title} {\bibinfo {title} {End states and subgap structure in proximity-coupled chains of magnetic adatoms},\ }\href {https://doi.org/10.1103/PhysRevLett.115.197204} {\bibfield  {journal} {\bibinfo  {journal} {Phys. Rev. Lett.}\ }\textbf {\bibinfo {volume} {115}},\ \bibinfo {pages} {197204} (\bibinfo {year} {2015})}\BibitemShut {NoStop}%
\bibitem [{\citenamefont {Jeon}\ \emph {et~al.}(2017)\citenamefont {Jeon}, \citenamefont {Xie}, \citenamefont {Li}, \citenamefont {Wang}, \citenamefont {Bernevig},\ and\ \citenamefont {Yazdani}}]{Jeon_Science_2017}%
  \BibitemOpen
  \bibfield  {author} {\bibinfo {author} {\bibfnamefont {S.}~\bibnamefont {Jeon}}, \bibinfo {author} {\bibfnamefont {Y.}~\bibnamefont {Xie}}, \bibinfo {author} {\bibfnamefont {J.}~\bibnamefont {Li}}, \bibinfo {author} {\bibfnamefont {Z.}~\bibnamefont {Wang}}, \bibinfo {author} {\bibfnamefont {B.~A.}\ \bibnamefont {Bernevig}},\ and\ \bibinfo {author} {\bibfnamefont {A.}~\bibnamefont {Yazdani}},\ }\bibfield  {title} {\bibinfo {title} {{Distinguishing a Majorana zero mode using spin-resolved measurements}},\ }\href {https://doi.org/10.1126/science.aan3670} {\bibfield  {journal} {\bibinfo  {journal} {Science}\ }\textbf {\bibinfo {volume} {358}},\ \bibinfo {pages} {772} (\bibinfo {year} {2017})}\BibitemShut {NoStop}%
\bibitem [{\citenamefont {Manna}\ \emph {et~al.}(2020)\citenamefont {Manna}, \citenamefont {Wei}, \citenamefont {Xie}, \citenamefont {Law}, \citenamefont {Lee},\ and\ \citenamefont {Moodera}}]{manna2020}%
  \BibitemOpen
  \bibfield  {author} {\bibinfo {author} {\bibfnamefont {S.}~\bibnamefont {Manna}}, \bibinfo {author} {\bibfnamefont {P.}~\bibnamefont {Wei}}, \bibinfo {author} {\bibfnamefont {Y.}~\bibnamefont {Xie}}, \bibinfo {author} {\bibfnamefont {K.~T.}\ \bibnamefont {Law}}, \bibinfo {author} {\bibfnamefont {P.~A.}\ \bibnamefont {Lee}},\ and\ \bibinfo {author} {\bibfnamefont {J.~S.}\ \bibnamefont {Moodera}},\ }\bibfield  {title} {\bibinfo {title} {Signature of a pair of {Majorana} zero modes in superconducting gold surface states},\ }\href {https://doi.org/10.1073/pnas.1919753117} {\bibfield  {journal} {\bibinfo  {journal} {Proceedings of the National Academy of Sciences}\ }\textbf {\bibinfo {volume} {117}},\ \bibinfo {pages} {8775} (\bibinfo {year} {2020})}\BibitemShut {NoStop}%
\bibitem [{\citenamefont {Maiani}\ \emph {et~al.}(2021)\citenamefont {Maiani}, \citenamefont {Seoane~Souto}, \citenamefont {Leijnse},\ and\ \citenamefont {Flensberg}}]{maiani2021}%
  \BibitemOpen
  \bibfield  {author} {\bibinfo {author} {\bibfnamefont {A.}~\bibnamefont {Maiani}}, \bibinfo {author} {\bibfnamefont {R.}~\bibnamefont {Seoane~Souto}}, \bibinfo {author} {\bibfnamefont {M.}~\bibnamefont {Leijnse}},\ and\ \bibinfo {author} {\bibfnamefont {K.}~\bibnamefont {Flensberg}},\ }\bibfield  {title} {\bibinfo {title} {Topological superconductivity in semiconductor--superconductor--magnetic-insulator heterostructures},\ }\href {https://doi.org/10.1103/PhysRevB.103.104508} {\bibfield  {journal} {\bibinfo  {journal} {Phys. Rev. B}\ }\textbf {\bibinfo {volume} {103}},\ \bibinfo {pages} {104508} (\bibinfo {year} {2021})}\BibitemShut {NoStop}%
\bibitem [{\citenamefont {Woods}\ and\ \citenamefont {Stanescu}(2021)}]{Woods_PRB2021}%
  \BibitemOpen
  \bibfield  {author} {\bibinfo {author} {\bibfnamefont {B.~D.}\ \bibnamefont {Woods}}\ and\ \bibinfo {author} {\bibfnamefont {T.~D.}\ \bibnamefont {Stanescu}},\ }\bibfield  {title} {\bibinfo {title} {Electrostatic effects and topological superconductivity in semiconductor--superconductor--magnetic-insulator hybrid wires},\ }\href {https://doi.org/10.1103/PhysRevB.104.195433} {\bibfield  {journal} {\bibinfo  {journal} {Phys. Rev. B}\ }\textbf {\bibinfo {volume} {104}},\ \bibinfo {pages} {195433} (\bibinfo {year} {2021})}\BibitemShut {NoStop}%
\bibitem [{\citenamefont {Khindanov}\ \emph {et~al.}(2021)\citenamefont {Khindanov}, \citenamefont {Alicea}, \citenamefont {Lee}, \citenamefont {Cole},\ and\ \citenamefont {Antipov}}]{Khindanov_PRB2021}%
  \BibitemOpen
  \bibfield  {author} {\bibinfo {author} {\bibfnamefont {A.}~\bibnamefont {Khindanov}}, \bibinfo {author} {\bibfnamefont {J.}~\bibnamefont {Alicea}}, \bibinfo {author} {\bibfnamefont {P.}~\bibnamefont {Lee}}, \bibinfo {author} {\bibfnamefont {W.~S.}\ \bibnamefont {Cole}},\ and\ \bibinfo {author} {\bibfnamefont {A.~E.}\ \bibnamefont {Antipov}},\ }\bibfield  {title} {\bibinfo {title} {Topological superconductivity in nanowires proximate to a diffusive superconductor--magnetic-insulator bilayer},\ }\href {https://doi.org/10.1103/PhysRevB.103.134506} {\bibfield  {journal} {\bibinfo  {journal} {Phys. Rev. B}\ }\textbf {\bibinfo {volume} {103}},\ \bibinfo {pages} {134506} (\bibinfo {year} {2021})}\BibitemShut {NoStop}%
\bibitem [{\citenamefont {Vaitiek{\.e}nas}\ \emph {et~al.}(2021)\citenamefont {Vaitiek{\.e}nas}, \citenamefont {Liu}, \citenamefont {Krogstrup},\ and\ \citenamefont {Marcus}}]{vaitiekenas2021}%
  \BibitemOpen
  \bibfield  {author} {\bibinfo {author} {\bibfnamefont {S.}~\bibnamefont {Vaitiek{\.e}nas}}, \bibinfo {author} {\bibfnamefont {Y.}~\bibnamefont {Liu}}, \bibinfo {author} {\bibfnamefont {P.}~\bibnamefont {Krogstrup}},\ and\ \bibinfo {author} {\bibfnamefont {C.}~\bibnamefont {Marcus}},\ }\bibfield  {title} {\bibinfo {title} {Zero-bias peaks at zero magnetic field in ferromagnetic hybrid nanowires},\ }\href {https://doi.org/10.1038/s41567-020-1017-3} {\bibfield  {journal} {\bibinfo  {journal} {Nature Physics}\ }\textbf {\bibinfo {volume} {17}},\ \bibinfo {pages} {43} (\bibinfo {year} {2021})}\BibitemShut {NoStop}%
\bibitem [{\citenamefont {Vaitiek\ifmmode~\dot{e}\else \.{e}\fi{}nas}\ \emph {et~al.}(2022)\citenamefont {Vaitiek\ifmmode~\dot{e}\else \.{e}\fi{}nas}, \citenamefont {Souto}, \citenamefont {Liu}, \citenamefont {Krogstrup}, \citenamefont {Flensberg}, \citenamefont {Leijnse},\ and\ \citenamefont {Marcus}}]{vaitiekenas2022}%
  \BibitemOpen
  \bibfield  {author} {\bibinfo {author} {\bibfnamefont {S.}~\bibnamefont {Vaitiek\ifmmode~\dot{e}\else \.{e}\fi{}nas}}, \bibinfo {author} {\bibfnamefont {R.~S.}\ \bibnamefont {Souto}}, \bibinfo {author} {\bibfnamefont {Y.}~\bibnamefont {Liu}}, \bibinfo {author} {\bibfnamefont {P.}~\bibnamefont {Krogstrup}}, \bibinfo {author} {\bibfnamefont {K.}~\bibnamefont {Flensberg}}, \bibinfo {author} {\bibfnamefont {M.}~\bibnamefont {Leijnse}},\ and\ \bibinfo {author} {\bibfnamefont {C.~M.}\ \bibnamefont {Marcus}},\ }\bibfield  {title} {\bibinfo {title} {Evidence for spin-polarized bound states in semiconductor--superconductor--ferromagnetic-insulator islands},\ }\href {https://doi.org/10.1103/PhysRevB.105.L041304} {\bibfield  {journal} {\bibinfo  {journal} {Phys. Rev. B}\ }\textbf {\bibinfo {volume} {105}},\ \bibinfo {pages} {L041304} (\bibinfo {year} {2022})}\BibitemShut {NoStop}%
\bibitem [{\citenamefont {Razmadze}\ \emph {et~al.}(2023)\citenamefont {Razmadze}, \citenamefont {Souto}, \citenamefont {Galletti}, \citenamefont {Maiani}, \citenamefont {Liu}, \citenamefont {Krogstrup}, \citenamefont {Schrade}, \citenamefont {Gyenis}, \citenamefont {Marcus},\ and\ \citenamefont {Vaitiek\ifmmode~\dot{e}\else \.{e}\fi{}nas}}]{Razmadze_PRB2023}%
  \BibitemOpen
  \bibfield  {author} {\bibinfo {author} {\bibfnamefont {D.}~\bibnamefont {Razmadze}}, \bibinfo {author} {\bibfnamefont {R.~S.}\ \bibnamefont {Souto}}, \bibinfo {author} {\bibfnamefont {L.}~\bibnamefont {Galletti}}, \bibinfo {author} {\bibfnamefont {A.}~\bibnamefont {Maiani}}, \bibinfo {author} {\bibfnamefont {Y.}~\bibnamefont {Liu}}, \bibinfo {author} {\bibfnamefont {P.}~\bibnamefont {Krogstrup}}, \bibinfo {author} {\bibfnamefont {C.}~\bibnamefont {Schrade}}, \bibinfo {author} {\bibfnamefont {A.}~\bibnamefont {Gyenis}}, \bibinfo {author} {\bibfnamefont {C.~M.}\ \bibnamefont {Marcus}},\ and\ \bibinfo {author} {\bibfnamefont {S.}~\bibnamefont {Vaitiek\ifmmode~\dot{e}\else \.{e}\fi{}nas}},\ }\bibfield  {title} {\bibinfo {title} {Supercurrent reversal in ferromagnetic hybrid nanowire {Josephson} junctions},\ }\href {https://doi.org/10.1103/PhysRevB.107.L081301} {\bibfield  {journal} {\bibinfo  {journal} {Phys. Rev. B}\ }\textbf {\bibinfo {volume} {107}},\ \bibinfo {pages} {L081301} (\bibinfo {year}
  {2023})}\BibitemShut {NoStop}%
\bibitem [{\citenamefont {Desjardins}\ \emph {et~al.}(2019)\citenamefont {Desjardins}, \citenamefont {Contamin}, \citenamefont {Delbecq}, \citenamefont {Dartiailh}, \citenamefont {Bruhat}, \citenamefont {Cubaynes}, \citenamefont {Viennot}, \citenamefont {Mallet}, \citenamefont {Rohart}, \citenamefont {Thiaville}, \citenamefont {Cottet},\ and\ \citenamefont {Kontos}}]{Desjardins_NatMater_2019}%
  \BibitemOpen
  \bibfield  {author} {\bibinfo {author} {\bibfnamefont {M.~M.}\ \bibnamefont {Desjardins}}, \bibinfo {author} {\bibfnamefont {L.~C.}\ \bibnamefont {Contamin}}, \bibinfo {author} {\bibfnamefont {M.~R.}\ \bibnamefont {Delbecq}}, \bibinfo {author} {\bibfnamefont {M.~C.}\ \bibnamefont {Dartiailh}}, \bibinfo {author} {\bibfnamefont {L.~E.}\ \bibnamefont {Bruhat}}, \bibinfo {author} {\bibfnamefont {T.}~\bibnamefont {Cubaynes}}, \bibinfo {author} {\bibfnamefont {J.~J.}\ \bibnamefont {Viennot}}, \bibinfo {author} {\bibfnamefont {F.}~\bibnamefont {Mallet}}, \bibinfo {author} {\bibfnamefont {S.}~\bibnamefont {Rohart}}, \bibinfo {author} {\bibfnamefont {A.}~\bibnamefont {Thiaville}}, \bibinfo {author} {\bibfnamefont {A.}~\bibnamefont {Cottet}},\ and\ \bibinfo {author} {\bibfnamefont {T.}~\bibnamefont {Kontos}},\ }\bibfield  {title} {\bibinfo {title} {{Synthetic spin{\textendash}orbit interaction for Majorana devices}},\ }\href {https://doi.org/10.1038/s41563-019-0457-6} {\bibfield  {journal} {\bibinfo  {journal} {Nat.
  Mater.}\ }\textbf {\bibinfo {volume} {18}},\ \bibinfo {pages} {1060} (\bibinfo {year} {2019})}\BibitemShut {NoStop}%
\bibitem [{\citenamefont {Steffensen}\ \emph {et~al.}(2022)\citenamefont {Steffensen}, \citenamefont {Christensen}, \citenamefont {Andersen},\ and\ \citenamefont {Kotetes}}]{Steffensen_2022_PRR}%
  \BibitemOpen
  \bibfield  {author} {\bibinfo {author} {\bibfnamefont {D.}~\bibnamefont {Steffensen}}, \bibinfo {author} {\bibfnamefont {M.~H.}\ \bibnamefont {Christensen}}, \bibinfo {author} {\bibfnamefont {B.~M.}\ \bibnamefont {Andersen}},\ and\ \bibinfo {author} {\bibfnamefont {P.}~\bibnamefont {Kotetes}},\ }\bibfield  {title} {\bibinfo {title} {{Topological superconductivity induced by magnetic texture crystals}},\ }\href {https://doi.org/10.1103/PhysRevResearch.4.013225} {\bibfield  {journal} {\bibinfo  {journal} {Phys. Rev. Res.}\ }\textbf {\bibinfo {volume} {4}},\ \bibinfo {pages} {013225} (\bibinfo {year} {2022})}\BibitemShut {NoStop}%
\bibitem [{\citenamefont {Pientka}\ \emph {et~al.}(2017)\citenamefont {Pientka}, \citenamefont {Keselman}, \citenamefont {Berg}, \citenamefont {Yacoby}, \citenamefont {Stern},\ and\ \citenamefont {Halperin}}]{Pientka2017}%
  \BibitemOpen
  \bibfield  {author} {\bibinfo {author} {\bibfnamefont {F.}~\bibnamefont {Pientka}}, \bibinfo {author} {\bibfnamefont {A.}~\bibnamefont {Keselman}}, \bibinfo {author} {\bibfnamefont {E.}~\bibnamefont {Berg}}, \bibinfo {author} {\bibfnamefont {A.}~\bibnamefont {Yacoby}}, \bibinfo {author} {\bibfnamefont {A.}~\bibnamefont {Stern}},\ and\ \bibinfo {author} {\bibfnamefont {B.~I.}\ \bibnamefont {Halperin}},\ }\bibfield  {title} {\bibinfo {title} {Topological superconductivity in a planar {Josephson} junction},\ }\href {https://doi.org/10.1103/PhysRevX.7.021032} {\bibfield  {journal} {\bibinfo  {journal} {Phys. Rev. X}\ }\textbf {\bibinfo {volume} {7}},\ \bibinfo {pages} {021032} (\bibinfo {year} {2017})}\BibitemShut {NoStop}%
\bibitem [{\citenamefont {Hell}\ \emph {et~al.}(2017)\citenamefont {Hell}, \citenamefont {Leijnse},\ and\ \citenamefont {Flensberg}}]{Hell_PRL2017}%
  \BibitemOpen
  \bibfield  {author} {\bibinfo {author} {\bibfnamefont {M.}~\bibnamefont {Hell}}, \bibinfo {author} {\bibfnamefont {M.}~\bibnamefont {Leijnse}},\ and\ \bibinfo {author} {\bibfnamefont {K.}~\bibnamefont {Flensberg}},\ }\bibfield  {title} {\bibinfo {title} {Two-dimensional platform for networks of {Majorana} bound states},\ }\href {https://doi.org/10.1103/PhysRevLett.118.107701} {\bibfield  {journal} {\bibinfo  {journal} {Phys. Rev. Lett.}\ }\textbf {\bibinfo {volume} {118}},\ \bibinfo {pages} {107701} (\bibinfo {year} {2017})}\BibitemShut {NoStop}%
\bibitem [{\citenamefont {Laeven}\ \emph {et~al.}(2020)\citenamefont {Laeven}, \citenamefont {Nijholt}, \citenamefont {Wimmer},\ and\ \citenamefont {Akhmerov}}]{Laeven_PRL2020}%
  \BibitemOpen
  \bibfield  {author} {\bibinfo {author} {\bibfnamefont {T.}~\bibnamefont {Laeven}}, \bibinfo {author} {\bibfnamefont {B.}~\bibnamefont {Nijholt}}, \bibinfo {author} {\bibfnamefont {M.}~\bibnamefont {Wimmer}},\ and\ \bibinfo {author} {\bibfnamefont {A.~R.}\ \bibnamefont {Akhmerov}},\ }\bibfield  {title} {\bibinfo {title} {Enhanced proximity effect in zigzag-shaped {Majorana} {Josephson} junctions},\ }\href {https://doi.org/10.1103/PhysRevLett.125.086802} {\bibfield  {journal} {\bibinfo  {journal} {Phys. Rev. Lett.}\ }\textbf {\bibinfo {volume} {125}},\ \bibinfo {pages} {086802} (\bibinfo {year} {2020})}\BibitemShut {NoStop}%
\bibitem [{\citenamefont {Paudel}\ \emph {et~al.}(2021)\citenamefont {Paudel}, \citenamefont {Cole}, \citenamefont {Woods},\ and\ \citenamefont {Stanescu}}]{Paudel_PRB2021}%
  \BibitemOpen
  \bibfield  {author} {\bibinfo {author} {\bibfnamefont {P.~P.}\ \bibnamefont {Paudel}}, \bibinfo {author} {\bibfnamefont {T.}~\bibnamefont {Cole}}, \bibinfo {author} {\bibfnamefont {B.~D.}\ \bibnamefont {Woods}},\ and\ \bibinfo {author} {\bibfnamefont {T.~D.}\ \bibnamefont {Stanescu}},\ }\bibfield  {title} {\bibinfo {title} {{Enhanced topological superconductivity in spatially modulated planar Josephson junctions}},\ }\href {https://doi.org/10.1103/PhysRevB.104.155428} {\bibfield  {journal} {\bibinfo  {journal} {Phys. Rev. B}\ }\textbf {\bibinfo {volume} {104}},\ \bibinfo {pages} {155428} (\bibinfo {year} {2021})}\BibitemShut {NoStop}%
\bibitem [{\citenamefont {Lesser}\ and\ \citenamefont {Oreg}(2022{\natexlab{a}})}]{Lesser_JoPD2022}%
  \BibitemOpen
  \bibfield  {author} {\bibinfo {author} {\bibfnamefont {O.}~\bibnamefont {Lesser}}\ and\ \bibinfo {author} {\bibfnamefont {Y.}~\bibnamefont {Oreg}},\ }\bibfield  {title} {\bibinfo {title} {Majorana zero modes induced by superconducting phase bias},\ }\href {https://doi.org/10.1088/1361-6463/ac4a37} {\bibfield  {journal} {\bibinfo  {journal} {Journal of Physics D: Applied Physics}\ }\textbf {\bibinfo {volume} {55}},\ \bibinfo {pages} {164001} (\bibinfo {year} {2022}{\natexlab{a}})}\BibitemShut {NoStop}%
\bibitem [{\citenamefont {Kuiri}\ and\ \citenamefont {Nowak}(2023)}]{Kuiri_PRB2023}%
  \BibitemOpen
  \bibfield  {author} {\bibinfo {author} {\bibfnamefont {D.}~\bibnamefont {Kuiri}}\ and\ \bibinfo {author} {\bibfnamefont {M.~P.}\ \bibnamefont {Nowak}},\ }\bibfield  {title} {\bibinfo {title} {{Nonlocal transport signatures of topological superconductivity in a phase-biased planar Josephson junction}},\ }\href {https://doi.org/10.1103/PhysRevB.108.205405} {\bibfield  {journal} {\bibinfo  {journal} {Phys. Rev. B}\ }\textbf {\bibinfo {volume} {108}},\ \bibinfo {pages} {205405} (\bibinfo {year} {2023})}\BibitemShut {NoStop}%
\bibitem [{\citenamefont {Ai}\ \emph {et~al.}(2021)\citenamefont {Ai}, \citenamefont {Zhang}, \citenamefont {Yang}, \citenamefont {Xie}, \citenamefont {Yang}, \citenamefont {Jia}, \citenamefont {Zhang}, \citenamefont {Liu}, \citenamefont {Li}, \citenamefont {Leng}, \citenamefont {Cao}, \citenamefont {Sun}, \citenamefont {Zhang}, \citenamefont {Kou}, \citenamefont {Han}, \citenamefont {Xiu},\ and\ \citenamefont {Dong}}]{Ai2021}%
  \BibitemOpen
  \bibfield  {author} {\bibinfo {author} {\bibfnamefont {L.}~\bibnamefont {Ai}}, \bibinfo {author} {\bibfnamefont {E.}~\bibnamefont {Zhang}}, \bibinfo {author} {\bibfnamefont {J.}~\bibnamefont {Yang}}, \bibinfo {author} {\bibfnamefont {X.}~\bibnamefont {Xie}}, \bibinfo {author} {\bibfnamefont {Y.}~\bibnamefont {Yang}}, \bibinfo {author} {\bibfnamefont {Z.}~\bibnamefont {Jia}}, \bibinfo {author} {\bibfnamefont {Y.}~\bibnamefont {Zhang}}, \bibinfo {author} {\bibfnamefont {S.}~\bibnamefont {Liu}}, \bibinfo {author} {\bibfnamefont {Z.}~\bibnamefont {Li}}, \bibinfo {author} {\bibfnamefont {P.}~\bibnamefont {Leng}}, \bibinfo {author} {\bibfnamefont {X.}~\bibnamefont {Cao}}, \bibinfo {author} {\bibfnamefont {X.}~\bibnamefont {Sun}}, \bibinfo {author} {\bibfnamefont {T.}~\bibnamefont {Zhang}}, \bibinfo {author} {\bibfnamefont {X.}~\bibnamefont {Kou}}, \bibinfo {author} {\bibfnamefont {Z.}~\bibnamefont {Han}}, \bibinfo {author} {\bibfnamefont {F.}~\bibnamefont {Xiu}},\ and\ \bibinfo {author} {\bibfnamefont
  {S.}~\bibnamefont {Dong}},\ }\bibfield  {title} {\bibinfo {title} {{Van der Waals ferromagnetic Josephson junctions}},\ }\href {https://doi.org/10.1038/s41467-021-26946-w} {\bibfield  {journal} {\bibinfo  {journal} {Nat. Commun.}\ }\textbf {\bibinfo {volume} {12}},\ \bibinfo {pages} {1} (\bibinfo {year} {2021})}\BibitemShut {NoStop}%
\bibitem [{\citenamefont {Idzuchi}\ \emph {et~al.}(2021)\citenamefont {Idzuchi}, \citenamefont {Pientka}, \citenamefont {Huang}, \citenamefont {Harada}, \citenamefont {G{\ifmmode\ddot{u}\else\"{u}\fi}l}, \citenamefont {Shin}, \citenamefont {Nguyen}, \citenamefont {Jo}, \citenamefont {Shindo}, \citenamefont {Cava}, \citenamefont {Canfield},\ and\ \citenamefont {Kim}}]{Idzuchi2021}%
  \BibitemOpen
  \bibfield  {author} {\bibinfo {author} {\bibfnamefont {H.}~\bibnamefont {Idzuchi}}, \bibinfo {author} {\bibfnamefont {F.}~\bibnamefont {Pientka}}, \bibinfo {author} {\bibfnamefont {K.-F.}\ \bibnamefont {Huang}}, \bibinfo {author} {\bibfnamefont {K.}~\bibnamefont {Harada}}, \bibinfo {author} {\bibfnamefont {{\ifmmode\ddot{O}\else\"{O}\fi}.}~\bibnamefont {G{\ifmmode\ddot{u}\else\"{u}\fi}l}}, \bibinfo {author} {\bibfnamefont {Y.~J.}\ \bibnamefont {Shin}}, \bibinfo {author} {\bibfnamefont {L.~T.}\ \bibnamefont {Nguyen}}, \bibinfo {author} {\bibfnamefont {N.~H.}\ \bibnamefont {Jo}}, \bibinfo {author} {\bibfnamefont {D.}~\bibnamefont {Shindo}}, \bibinfo {author} {\bibfnamefont {R.~J.}\ \bibnamefont {Cava}}, \bibinfo {author} {\bibfnamefont {P.~C.}\ \bibnamefont {Canfield}},\ and\ \bibinfo {author} {\bibfnamefont {P.}~\bibnamefont {Kim}},\ }\bibfield  {title} {\bibinfo {title} {{Unconventional supercurrent phase in Ising superconductor Josephson junction with atomically thin magnetic insulator}},\ }\href
  {https://doi.org/10.1038/s41467-021-25608-1} {\bibfield  {journal} {\bibinfo  {journal} {Nat. Commun.}\ }\textbf {\bibinfo {volume} {12}},\ \bibinfo {pages} {1} (\bibinfo {year} {2021})}\BibitemShut {NoStop}%
\bibitem [{\citenamefont {Xi}\ \emph {et~al.}(2022)\citenamefont {Xi}, \citenamefont {Wang}, \citenamefont {Zhao}, \citenamefont {Park}, \citenamefont {Law}, \citenamefont {Berger}, \citenamefont {Forr\'{o}}, \citenamefont {Shan},\ and\ \citenamefont {Mak}}]{kang_van_2022}%
  \BibitemOpen
  \bibfield  {author} {\bibinfo {author} {\bibfnamefont {X.}~\bibnamefont {Xi}}, \bibinfo {author} {\bibfnamefont {Z.}~\bibnamefont {Wang}}, \bibinfo {author} {\bibfnamefont {W.}~\bibnamefont {Zhao}}, \bibinfo {author} {\bibfnamefont {J.-H.}\ \bibnamefont {Park}}, \bibinfo {author} {\bibfnamefont {K.~T.}\ \bibnamefont {Law}}, \bibinfo {author} {\bibfnamefont {H.}~\bibnamefont {Berger}}, \bibinfo {author} {\bibfnamefont {L.}~\bibnamefont {Forr\'{o}}}, \bibinfo {author} {\bibfnamefont {J.}~\bibnamefont {Shan}},\ and\ \bibinfo {author} {\bibfnamefont {K.~F.}\ \bibnamefont {Mak}},\ }\bibfield  {title} {\bibinfo {title} {Van der {Waals} {$\pi$} {Josephson} junctions},\ }\href {https://doi.org/10.1021/acs.nanolett.2c01640} {\bibfield  {journal} {\bibinfo  {journal} {Nano Letters}\ }\textbf {\bibinfo {volume} {22}},\ \bibinfo {pages} {5510} (\bibinfo {year} {2022})}\BibitemShut {NoStop}%
\bibitem [{\citenamefont {Wu}\ \emph {et~al.}(2022)\citenamefont {Wu}, \citenamefont {Wang}, \citenamefont {Xu}, \citenamefont {Sivakumar}, \citenamefont {Pasco}, \citenamefont {Filippozzi}, \citenamefont {Parkin}, \citenamefont {Zeng}, \citenamefont {McQueen},\ and\ \citenamefont {Ali}}]{wu_field-free_2022}%
  \BibitemOpen
  \bibfield  {author} {\bibinfo {author} {\bibfnamefont {H.}~\bibnamefont {Wu}}, \bibinfo {author} {\bibfnamefont {Y.}~\bibnamefont {Wang}}, \bibinfo {author} {\bibfnamefont {Y.}~\bibnamefont {Xu}}, \bibinfo {author} {\bibfnamefont {P.~K.}\ \bibnamefont {Sivakumar}}, \bibinfo {author} {\bibfnamefont {C.}~\bibnamefont {Pasco}}, \bibinfo {author} {\bibfnamefont {U.}~\bibnamefont {Filippozzi}}, \bibinfo {author} {\bibfnamefont {S.~S.~P.}\ \bibnamefont {Parkin}}, \bibinfo {author} {\bibfnamefont {Y.-J.}\ \bibnamefont {Zeng}}, \bibinfo {author} {\bibfnamefont {T.}~\bibnamefont {McQueen}},\ and\ \bibinfo {author} {\bibfnamefont {M.~N.}\ \bibnamefont {Ali}},\ }\bibfield  {title} {\bibinfo {title} {The field-free {Josephson} diode in a van der {Waals} heterostructure},\ }\href {https://doi.org/10.1038/s41586-022-04504-8} {\bibfield  {journal} {\bibinfo  {journal} {Nature}\ }\textbf {\bibinfo {volume} {604}},\ \bibinfo {pages} {653} (\bibinfo {year} {2022})}\BibitemShut {NoStop}%
\bibitem [{\citenamefont {Hu}\ \emph {et~al.}(2023)\citenamefont {Hu}, \citenamefont {Wang}, \citenamefont {Wang}, \citenamefont {Zhang}, \citenamefont {Feng}, \citenamefont {Wang}, \citenamefont {Niu}, \citenamefont {Zhang},\ and\ \citenamefont {Xiang}}]{hu_long-range-skin_2023}%
  \BibitemOpen
  \bibfield  {author} {\bibinfo {author} {\bibfnamefont {G.}~\bibnamefont {Hu}}, \bibinfo {author} {\bibfnamefont {C.}~\bibnamefont {Wang}}, \bibinfo {author} {\bibfnamefont {S.}~\bibnamefont {Wang}}, \bibinfo {author} {\bibfnamefont {Y.}~\bibnamefont {Zhang}}, \bibinfo {author} {\bibfnamefont {Y.}~\bibnamefont {Feng}}, \bibinfo {author} {\bibfnamefont {Z.}~\bibnamefont {Wang}}, \bibinfo {author} {\bibfnamefont {Q.}~\bibnamefont {Niu}}, \bibinfo {author} {\bibfnamefont {Z.}~\bibnamefont {Zhang}},\ and\ \bibinfo {author} {\bibfnamefont {B.}~\bibnamefont {Xiang}},\ }\bibfield  {title} {\bibinfo {title} {Long-range skin {Josephson} supercurrent across a van der {Waals} ferromagnet},\ }\href {https://doi.org/10.1038/s41467-023-37603-9} {\bibfield  {journal} {\bibinfo  {journal} {Nature Communications}\ }\textbf {\bibinfo {volume} {14}},\ \bibinfo {pages} {1779} (\bibinfo {year} {2023})}\BibitemShut {NoStop}%
\bibitem [{\citenamefont {Spuri}\ \emph {et~al.}(2024)\citenamefont {Spuri}, \citenamefont {Nikolic}, \citenamefont {Chakraborty}, \citenamefont {Klang}, \citenamefont {Alpern}, \citenamefont {Millo}, \citenamefont {Steinberg}, \citenamefont {Belzig}, \citenamefont {Scheer},\ and\ \citenamefont {Di~Bernardo}}]{Spuri2023}%
  \BibitemOpen
  \bibfield  {author} {\bibinfo {author} {\bibfnamefont {A.}~\bibnamefont {Spuri}}, \bibinfo {author} {\bibfnamefont {D.}~\bibnamefont {Nikolic}}, \bibinfo {author} {\bibfnamefont {S.}~\bibnamefont {Chakraborty}}, \bibinfo {author} {\bibfnamefont {M.}~\bibnamefont {Klang}}, \bibinfo {author} {\bibfnamefont {H.}~\bibnamefont {Alpern}}, \bibinfo {author} {\bibfnamefont {O.}~\bibnamefont {Millo}}, \bibinfo {author} {\bibfnamefont {H.}~\bibnamefont {Steinberg}}, \bibinfo {author} {\bibfnamefont {W.}~\bibnamefont {Belzig}}, \bibinfo {author} {\bibfnamefont {E.}~\bibnamefont {Scheer}},\ and\ \bibinfo {author} {\bibfnamefont {A.}~\bibnamefont {Di~Bernardo}},\ }\bibfield  {title} {\bibinfo {title} {Signature of long-ranged spin triplets across a two-dimensional superconductor/helimagnet van der waals interface},\ }\href {https://doi.org/10.1103/PhysRevResearch.6.L012046} {\bibfield  {journal} {\bibinfo  {journal} {Phys. Rev. Res.}\ }\textbf {\bibinfo {volume} {6}},\ \bibinfo {pages} {L012046} (\bibinfo {year}
  {2024})}\BibitemShut {NoStop}%
\bibitem [{\citenamefont {Sardinero}\ \emph {et~al.}(2024)\citenamefont {Sardinero}, \citenamefont {Seoane~Souto},\ and\ \citenamefont {Burset}}]{Sardinero2024}%
  \BibitemOpen
  \bibfield  {author} {\bibinfo {author} {\bibfnamefont {I.}~\bibnamefont {Sardinero}}, \bibinfo {author} {\bibfnamefont {R.}~\bibnamefont {Seoane~Souto}},\ and\ \bibinfo {author} {\bibfnamefont {P.}~\bibnamefont {Burset}},\ }\bibfield  {title} {\bibinfo {title} {Topological superconductivity in a magnetic-texture coupled {Josephson} junction},\ }\href {https://doi.org/10.1103/PhysRevB.110.L060505} {\bibfield  {journal} {\bibinfo  {journal} {Phys. Rev. B}\ }\textbf {\bibinfo {volume} {110}},\ \bibinfo {pages} {L060505} (\bibinfo {year} {2024})}\BibitemShut {NoStop}%
\bibitem [{\citenamefont {Gonz{\ifmmode\acute{a}\else\'{a}\fi}lez-S{\ifmmode\acute{a}\else\'{a}\fi}nchez}\ \emph {et~al.}(2025)\citenamefont {Gonz{\ifmmode\acute{a}\else\'{a}\fi}lez-S{\ifmmode\acute{a}\else\'{a}\fi}nchez}, \citenamefont {Sardinero}, \citenamefont {Cuadra}, \citenamefont {Spuri}, \citenamefont {Moreno}, \citenamefont {Suderow}, \citenamefont {Scheer}, \citenamefont {Burset}, \citenamefont {Di~Bernardo}, \citenamefont {Souto},\ and\ \citenamefont {Lee}}]{Gonzalez-Sanchez2025May}%
  \BibitemOpen
  \bibfield  {author} {\bibinfo {author} {\bibfnamefont {C.}~\bibnamefont {Gonz{\ifmmode\acute{a}\else\'{a}\fi}lez-S{\ifmmode\acute{a}\else\'{a}\fi}nchez}}, \bibinfo {author} {\bibfnamefont {I.}~\bibnamefont {Sardinero}}, \bibinfo {author} {\bibfnamefont {J.}~\bibnamefont {Cuadra}}, \bibinfo {author} {\bibfnamefont {A.}~\bibnamefont {Spuri}}, \bibinfo {author} {\bibfnamefont {J.~A.}\ \bibnamefont {Moreno}}, \bibinfo {author} {\bibfnamefont {H.}~\bibnamefont {Suderow}}, \bibinfo {author} {\bibfnamefont {E.}~\bibnamefont {Scheer}}, \bibinfo {author} {\bibfnamefont {P.}~\bibnamefont {Burset}}, \bibinfo {author} {\bibfnamefont {A.}~\bibnamefont {Di~Bernardo}}, \bibinfo {author} {\bibfnamefont {R.~S.}\ \bibnamefont {Souto}},\ and\ \bibinfo {author} {\bibfnamefont {E.~J.~H.}\ \bibnamefont {Lee}},\ }\bibfield  {title} {\bibinfo {title} {{Signatures of edge states in antiferromagnetic van der Waals Josephson junctions}},\ }\bibfield  {journal} {\bibinfo  {journal} {arXiv}\ }\href
  {https://doi.org/10.48550/arXiv.2505.18578} {10.48550/arXiv.2505.18578} (\bibinfo {year} {2025}),\ \Eprint {https://arxiv.org/abs/2505.18578} {2505.18578} \BibitemShut {NoStop}%
\bibitem [{\citenamefont {Cayao}\ \emph {et~al.}(2020)\citenamefont {Cayao}, \citenamefont {Triola},\ and\ \citenamefont {Black-Schaffer}}]{Cayao2020odd}%
  \BibitemOpen
  \bibfield  {author} {\bibinfo {author} {\bibfnamefont {J.}~\bibnamefont {Cayao}}, \bibinfo {author} {\bibfnamefont {C.}~\bibnamefont {Triola}},\ and\ \bibinfo {author} {\bibfnamefont {A.~M.}\ \bibnamefont {Black-Schaffer}},\ }\bibfield  {title} {\bibinfo {title} {Odd-frequency superconducting pairing in one-dimensional systems},\ }\href {https://doi.org/10.1140/epjst/e2019-900168-0} {\bibfield  {journal} {\bibinfo  {journal} {Eur. Phys. J.: Spec. Top.}\ }\textbf {\bibinfo {volume} {229}},\ \bibinfo {pages} {545} (\bibinfo {year} {2020})}\BibitemShut {NoStop}%
\bibitem [{\citenamefont {Linder}\ and\ \citenamefont {Balatsky}(2019)}]{Linder_2019}%
  \BibitemOpen
  \bibfield  {author} {\bibinfo {author} {\bibfnamefont {J.}~\bibnamefont {Linder}}\ and\ \bibinfo {author} {\bibfnamefont {A.~V.}\ \bibnamefont {Balatsky}},\ }\bibfield  {title} {\bibinfo {title} {Odd-frequency superconductivity},\ }\href {https://doi.org/10.1103/RevModPhys.91.045005} {\bibfield  {journal} {\bibinfo  {journal} {Rev. Mod. Phys.}\ }\textbf {\bibinfo {volume} {91}},\ \bibinfo {pages} {045005} (\bibinfo {year} {2019})}\BibitemShut {NoStop}%
\bibitem [{\citenamefont {Asano}\ and\ \citenamefont {Tanaka}(2013)}]{Asano2013Mar}%
  \BibitemOpen
  \bibfield  {author} {\bibinfo {author} {\bibfnamefont {Y.}~\bibnamefont {Asano}}\ and\ \bibinfo {author} {\bibfnamefont {Y.}~\bibnamefont {Tanaka}},\ }\bibfield  {title} {\bibinfo {title} {Majorana fermions and odd-frequency cooper pairs in a normal-metal nanowire proximity-coupled to a topological superconductor},\ }\href {https://doi.org/10.1103/PhysRevB.87.104513} {\bibfield  {journal} {\bibinfo  {journal} {Phys. Rev. B}\ }\textbf {\bibinfo {volume} {87}},\ \bibinfo {pages} {104513} (\bibinfo {year} {2013})}\BibitemShut {NoStop}%
\bibitem [{\citenamefont {Huang}\ \emph {et~al.}(2015)\citenamefont {Huang}, \citenamefont {W\"olfle},\ and\ \citenamefont {Balatsky}}]{Huang2015Sep}%
  \BibitemOpen
  \bibfield  {author} {\bibinfo {author} {\bibfnamefont {Z.}~\bibnamefont {Huang}}, \bibinfo {author} {\bibfnamefont {P.}~\bibnamefont {W\"olfle}},\ and\ \bibinfo {author} {\bibfnamefont {A.~V.}\ \bibnamefont {Balatsky}},\ }\bibfield  {title} {\bibinfo {title} {Odd-frequency pairing of interacting {Majorana} fermions},\ }\href {https://doi.org/10.1103/PhysRevB.92.121404} {\bibfield  {journal} {\bibinfo  {journal} {Phys. Rev. B}\ }\textbf {\bibinfo {volume} {92}},\ \bibinfo {pages} {121404} (\bibinfo {year} {2015})}\BibitemShut {NoStop}%
\bibitem [{\citenamefont {Cr\'epin}\ \emph {et~al.}(2015)\citenamefont {Cr\'epin}, \citenamefont {Burset},\ and\ \citenamefont {Trauzettel}}]{Crepin2015Sep}%
  \BibitemOpen
  \bibfield  {author} {\bibinfo {author} {\bibfnamefont {F.}~\bibnamefont {Cr\'epin}}, \bibinfo {author} {\bibfnamefont {P.}~\bibnamefont {Burset}},\ and\ \bibinfo {author} {\bibfnamefont {B.}~\bibnamefont {Trauzettel}},\ }\bibfield  {title} {\bibinfo {title} {Odd-frequency triplet superconductivity at the helical edge of a topological insulator},\ }\href {https://doi.org/10.1103/PhysRevB.92.100507} {\bibfield  {journal} {\bibinfo  {journal} {Phys. Rev. B}\ }\textbf {\bibinfo {volume} {92}},\ \bibinfo {pages} {100507} (\bibinfo {year} {2015})}\BibitemShut {NoStop}%
\bibitem [{\citenamefont {Kashuba}\ \emph {et~al.}(2017)\citenamefont {Kashuba}, \citenamefont {Sothmann}, \citenamefont {Burset},\ and\ \citenamefont {Trauzettel}}]{Kashuba2017May}%
  \BibitemOpen
  \bibfield  {author} {\bibinfo {author} {\bibfnamefont {O.}~\bibnamefont {Kashuba}}, \bibinfo {author} {\bibfnamefont {B.}~\bibnamefont {Sothmann}}, \bibinfo {author} {\bibfnamefont {P.}~\bibnamefont {Burset}},\ and\ \bibinfo {author} {\bibfnamefont {B.}~\bibnamefont {Trauzettel}},\ }\bibfield  {title} {\bibinfo {title} {Majorana {STM} as a perfect detector of odd-frequency superconductivity},\ }\href {https://doi.org/10.1103/PhysRevB.95.174516} {\bibfield  {journal} {\bibinfo  {journal} {Phys. Rev. B}\ }\textbf {\bibinfo {volume} {95}},\ \bibinfo {pages} {174516} (\bibinfo {year} {2017})}\BibitemShut {NoStop}%
\bibitem [{\citenamefont {Cayao}\ and\ \citenamefont {Black-Schaffer}(2017)}]{Cayao2017Oct}%
  \BibitemOpen
  \bibfield  {author} {\bibinfo {author} {\bibfnamefont {J.}~\bibnamefont {Cayao}}\ and\ \bibinfo {author} {\bibfnamefont {A.~M.}\ \bibnamefont {Black-Schaffer}},\ }\bibfield  {title} {\bibinfo {title} {Odd-frequency superconducting pairing and subgap density of states at the edge of a two-dimensional topological insulator without magnetism},\ }\href {https://doi.org/10.1103/PhysRevB.96.155426} {\bibfield  {journal} {\bibinfo  {journal} {Phys. Rev. B}\ }\textbf {\bibinfo {volume} {96}},\ \bibinfo {pages} {155426} (\bibinfo {year} {2017})}\BibitemShut {NoStop}%
\bibitem [{\citenamefont {Tamura}\ \emph {et~al.}(2020)\citenamefont {Tamura}, \citenamefont {Nakosai}, \citenamefont {Black-Schaffer}, \citenamefont {Tanaka},\ and\ \citenamefont {Cayao}}]{Tamura2020Jun}%
  \BibitemOpen
  \bibfield  {author} {\bibinfo {author} {\bibfnamefont {S.}~\bibnamefont {Tamura}}, \bibinfo {author} {\bibfnamefont {S.}~\bibnamefont {Nakosai}}, \bibinfo {author} {\bibfnamefont {A.~M.}\ \bibnamefont {Black-Schaffer}}, \bibinfo {author} {\bibfnamefont {Y.}~\bibnamefont {Tanaka}},\ and\ \bibinfo {author} {\bibfnamefont {J.}~\bibnamefont {Cayao}},\ }\bibfield  {title} {\bibinfo {title} {Bulk odd-frequency pairing in the superconducting {Su}-{Schrieffer}-{Heeger} model},\ }\href {https://doi.org/10.1103/PhysRevB.101.214507} {\bibfield  {journal} {\bibinfo  {journal} {Phys. Rev. B}\ }\textbf {\bibinfo {volume} {101}},\ \bibinfo {pages} {214507} (\bibinfo {year} {2020})}\BibitemShut {NoStop}%
\bibitem [{\citenamefont {Lee}\ \emph {et~al.}(2017)\citenamefont {Lee}, \citenamefont {Lutchyn},\ and\ \citenamefont {Maciejko}}]{Lee2017May}%
  \BibitemOpen
  \bibfield  {author} {\bibinfo {author} {\bibfnamefont {S.-P.}\ \bibnamefont {Lee}}, \bibinfo {author} {\bibfnamefont {R.~M.}\ \bibnamefont {Lutchyn}},\ and\ \bibinfo {author} {\bibfnamefont {J.}~\bibnamefont {Maciejko}},\ }\bibfield  {title} {\bibinfo {title} {Odd-frequency superconductivity in a nanowire coupled to {Majorana} zero modes},\ }\href {https://doi.org/10.1103/PhysRevB.95.184506} {\bibfield  {journal} {\bibinfo  {journal} {Phys. Rev. B}\ }\textbf {\bibinfo {volume} {95}},\ \bibinfo {pages} {184506} (\bibinfo {year} {2017})}\BibitemShut {NoStop}%
\bibitem [{\citenamefont {Keidel}\ \emph {et~al.}(2018)\citenamefont {Keidel}, \citenamefont {Burset},\ and\ \citenamefont {Trauzettel}}]{Keidel_PRB2018}%
  \BibitemOpen
  \bibfield  {author} {\bibinfo {author} {\bibfnamefont {F.}~\bibnamefont {Keidel}}, \bibinfo {author} {\bibfnamefont {P.}~\bibnamefont {Burset}},\ and\ \bibinfo {author} {\bibfnamefont {B.}~\bibnamefont {Trauzettel}},\ }\bibfield  {title} {\bibinfo {title} {Tunable hybridization of {Majorana} bound states at the quantum spin hall edge},\ }\href {https://doi.org/10.1103/PhysRevB.97.075408} {\bibfield  {journal} {\bibinfo  {journal} {Phys. Rev. B}\ }\textbf {\bibinfo {volume} {97}},\ \bibinfo {pages} {075408} (\bibinfo {year} {2018})}\BibitemShut {NoStop}%
\bibitem [{\citenamefont {Fleckenstein}\ \emph {et~al.}(2018)\citenamefont {Fleckenstein}, \citenamefont {Ziani},\ and\ \citenamefont {Trauzettel}}]{Fleckenstein2018Apr}%
  \BibitemOpen
  \bibfield  {author} {\bibinfo {author} {\bibfnamefont {C.}~\bibnamefont {Fleckenstein}}, \bibinfo {author} {\bibfnamefont {N.~T.}\ \bibnamefont {Ziani}},\ and\ \bibinfo {author} {\bibfnamefont {B.}~\bibnamefont {Trauzettel}},\ }\bibfield  {title} {\bibinfo {title} {Conductance signatures of odd-frequency superconductivity in quantum spin {Hall} systems using a quantum point contact},\ }\href {https://doi.org/10.1103/PhysRevB.97.134523} {\bibfield  {journal} {\bibinfo  {journal} {Phys. Rev. B}\ }\textbf {\bibinfo {volume} {97}},\ \bibinfo {pages} {134523} (\bibinfo {year} {2018})}\BibitemShut {NoStop}%
\bibitem [{\citenamefont {Tsintzis}\ \emph {et~al.}(2019)\citenamefont {Tsintzis}, \citenamefont {Black-Schaffer},\ and\ \citenamefont {Cayao}}]{Tsintzis2019Sep}%
  \BibitemOpen
  \bibfield  {author} {\bibinfo {author} {\bibfnamefont {A.}~\bibnamefont {Tsintzis}}, \bibinfo {author} {\bibfnamefont {A.~M.}\ \bibnamefont {Black-Schaffer}},\ and\ \bibinfo {author} {\bibfnamefont {J.}~\bibnamefont {Cayao}},\ }\bibfield  {title} {\bibinfo {title} {Odd-frequency superconducting pairing in {Kitaev}-based junctions},\ }\href {https://doi.org/10.1103/PhysRevB.100.115433} {\bibfield  {journal} {\bibinfo  {journal} {Phys. Rev. B}\ }\textbf {\bibinfo {volume} {100}},\ \bibinfo {pages} {115433} (\bibinfo {year} {2019})}\BibitemShut {NoStop}%
\bibitem [{\citenamefont {Ziani}\ \emph {et~al.}(2020)\citenamefont {Ziani}, \citenamefont {Fleckenstein}, \citenamefont {Vigliotti}, \citenamefont {Trauzettel},\ and\ \citenamefont {Sassetti}}]{Ziani2020May}%
  \BibitemOpen
  \bibfield  {author} {\bibinfo {author} {\bibfnamefont {N.~T.}\ \bibnamefont {Ziani}}, \bibinfo {author} {\bibfnamefont {C.}~\bibnamefont {Fleckenstein}}, \bibinfo {author} {\bibfnamefont {L.}~\bibnamefont {Vigliotti}}, \bibinfo {author} {\bibfnamefont {B.}~\bibnamefont {Trauzettel}},\ and\ \bibinfo {author} {\bibfnamefont {M.}~\bibnamefont {Sassetti}},\ }\bibfield  {title} {\bibinfo {title} {From fractional solitons to {Majorana} fermions in a paradigmatic model of topological superconductivity},\ }\href {https://doi.org/10.1103/PhysRevB.101.195303} {\bibfield  {journal} {\bibinfo  {journal} {Phys. Rev. B}\ }\textbf {\bibinfo {volume} {101}},\ \bibinfo {pages} {195303} (\bibinfo {year} {2020})}\BibitemShut {NoStop}%
\bibitem [{\citenamefont {Takagi}\ \emph {et~al.}(2020)\citenamefont {Takagi}, \citenamefont {Tamura},\ and\ \citenamefont {Tanaka}}]{Takagi2020Jan}%
  \BibitemOpen
  \bibfield  {author} {\bibinfo {author} {\bibfnamefont {D.}~\bibnamefont {Takagi}}, \bibinfo {author} {\bibfnamefont {S.}~\bibnamefont {Tamura}},\ and\ \bibinfo {author} {\bibfnamefont {Y.}~\bibnamefont {Tanaka}},\ }\bibfield  {title} {\bibinfo {title} {Odd-frequency pairing and proximity effect in {Kitaev} chain systems including a topological critical point},\ }\href {https://doi.org/10.1103/PhysRevB.101.024509} {\bibfield  {journal} {\bibinfo  {journal} {Phys. Rev. B}\ }\textbf {\bibinfo {volume} {101}},\ \bibinfo {pages} {024509} (\bibinfo {year} {2020})}\BibitemShut {NoStop}%
\bibitem [{\citenamefont {Kuzmanovski}\ \emph {et~al.}(2020)\citenamefont {Kuzmanovski}, \citenamefont {Black-Schaffer},\ and\ \citenamefont {Cayao}}]{Kuzmanovski2020Mar}%
  \BibitemOpen
  \bibfield  {author} {\bibinfo {author} {\bibfnamefont {D.}~\bibnamefont {Kuzmanovski}}, \bibinfo {author} {\bibfnamefont {A.~M.}\ \bibnamefont {Black-Schaffer}},\ and\ \bibinfo {author} {\bibfnamefont {J.}~\bibnamefont {Cayao}},\ }\bibfield  {title} {\bibinfo {title} {Suppression of odd-frequency pairing by phase disorder in a nanowire coupled to {Majorana} zero modes},\ }\href {https://doi.org/10.1103/PhysRevB.101.094506} {\bibfield  {journal} {\bibinfo  {journal} {Phys. Rev. B}\ }\textbf {\bibinfo {volume} {101}},\ \bibinfo {pages} {094506} (\bibinfo {year} {2020})}\BibitemShut {NoStop}%
\bibitem [{\citenamefont {Lu}\ \emph {et~al.}(2022)\citenamefont {Lu}, \citenamefont {Cheng}, \citenamefont {Burset},\ and\ \citenamefont {Tanaka}}]{Lu2022Dec}%
  \BibitemOpen
  \bibfield  {author} {\bibinfo {author} {\bibfnamefont {B.}~\bibnamefont {Lu}}, \bibinfo {author} {\bibfnamefont {G.}~\bibnamefont {Cheng}}, \bibinfo {author} {\bibfnamefont {P.}~\bibnamefont {Burset}},\ and\ \bibinfo {author} {\bibfnamefont {Y.}~\bibnamefont {Tanaka}},\ }\bibfield  {title} {\bibinfo {title} {Identifying majorana bound states at quantum spin hall edges using a metallic probe},\ }\href {https://doi.org/10.1103/PhysRevB.106.245427} {\bibfield  {journal} {\bibinfo  {journal} {Phys. Rev. B}\ }\textbf {\bibinfo {volume} {106}},\ \bibinfo {pages} {245427} (\bibinfo {year} {2022})}\BibitemShut {NoStop}%
\bibitem [{\citenamefont {Cayao}\ \emph {et~al.}(2022)\citenamefont {Cayao}, \citenamefont {Dutta}, \citenamefont {Burset},\ and\ \citenamefont {Black-Schaffer}}]{Cayao2022Sep}%
  \BibitemOpen
  \bibfield  {author} {\bibinfo {author} {\bibfnamefont {J.}~\bibnamefont {Cayao}}, \bibinfo {author} {\bibfnamefont {P.}~\bibnamefont {Dutta}}, \bibinfo {author} {\bibfnamefont {P.}~\bibnamefont {Burset}},\ and\ \bibinfo {author} {\bibfnamefont {A.~M.}\ \bibnamefont {Black-Schaffer}},\ }\bibfield  {title} {\bibinfo {title} {Phase-tunable electron transport assisted by odd-frequency {Cooper} pairs in topological josephson junctions},\ }\href {https://doi.org/10.1103/PhysRevB.106.L100502} {\bibfield  {journal} {\bibinfo  {journal} {Phys. Rev. B}\ }\textbf {\bibinfo {volume} {106}},\ \bibinfo {pages} {L100502} (\bibinfo {year} {2022})}\BibitemShut {NoStop}%
\bibitem [{\citenamefont {Yang}\ \emph {et~al.}(2023)\citenamefont {Yang}, \citenamefont {Burset},\ and\ \citenamefont {Lu}}]{Yang_2023}%
  \BibitemOpen
  \bibfield  {author} {\bibinfo {author} {\bibfnamefont {X.}~\bibnamefont {Yang}}, \bibinfo {author} {\bibfnamefont {P.}~\bibnamefont {Burset}},\ and\ \bibinfo {author} {\bibfnamefont {B.}~\bibnamefont {Lu}},\ }\bibfield  {title} {\bibinfo {title} {Phase-tunable multiple andreev reflections in a quantum spin hall strip},\ }\href {https://doi.org/10.1088/1361-6668/ace2f0} {\bibfield  {journal} {\bibinfo  {journal} {Superconductor Science and Technology}\ }\textbf {\bibinfo {volume} {36}},\ \bibinfo {pages} {085012} (\bibinfo {year} {2023})}\BibitemShut {NoStop}%
\bibitem [{\citenamefont {Ahmed}\ \emph {et~al.}(2025{\natexlab{b}})\citenamefont {Ahmed}, \citenamefont {Tamura}, \citenamefont {Tanaka},\ and\ \citenamefont {Cayao}}]{Ahmed2025Jan}%
  \BibitemOpen
  \bibfield  {author} {\bibinfo {author} {\bibfnamefont {E.}~\bibnamefont {Ahmed}}, \bibinfo {author} {\bibfnamefont {S.}~\bibnamefont {Tamura}}, \bibinfo {author} {\bibfnamefont {Y.}~\bibnamefont {Tanaka}},\ and\ \bibinfo {author} {\bibfnamefont {J.}~\bibnamefont {Cayao}},\ }\bibfield  {title} {\bibinfo {title} {Odd-frequency superconducting pairing due to multiple {Majorana} edge modes in driven topological superconductors},\ }\href {https://doi.org/10.1103/PhysRevB.111.024507} {\bibfield  {journal} {\bibinfo  {journal} {Phys. Rev. B}\ }\textbf {\bibinfo {volume} {111}},\ \bibinfo {pages} {024507} (\bibinfo {year} {2025}{\natexlab{b}})}\BibitemShut {NoStop}%
\bibitem [{\citenamefont {Cayao}(2024{\natexlab{b}})}]{Cayao2024Sep}%
  \BibitemOpen
  \bibfield  {author} {\bibinfo {author} {\bibfnamefont {J.}~\bibnamefont {Cayao}},\ }\bibfield  {title} {\bibinfo {title} {Emergent pair symmetries in systems with poor man's {Majorana} modes},\ }\href {https://doi.org/10.1103/PhysRevB.110.125408} {\bibfield  {journal} {\bibinfo  {journal} {Phys. Rev. B}\ }\textbf {\bibinfo {volume} {110}},\ \bibinfo {pages} {125408} (\bibinfo {year} {2024}{\natexlab{b}})}\BibitemShut {NoStop}%
\bibitem [{\citenamefont {Ahmed}\ \emph {et~al.}(2025{\natexlab{c}})\citenamefont {Ahmed}, \citenamefont {Tamura}, \citenamefont {Tanaka},\ and\ \citenamefont {Cayao}}]{Ahmed2025Jun}%
  \BibitemOpen
  \bibfield  {author} {\bibinfo {author} {\bibfnamefont {E.}~\bibnamefont {Ahmed}}, \bibinfo {author} {\bibfnamefont {S.}~\bibnamefont {Tamura}}, \bibinfo {author} {\bibfnamefont {Y.}~\bibnamefont {Tanaka}},\ and\ \bibinfo {author} {\bibfnamefont {J.}~\bibnamefont {Cayao}},\ }\bibfield  {title} {\bibinfo {title} {Odd-frequency pairing due to {Majorana} and trivial {Andreev} bound states},\ }\href {https://doi.org/10.1103/fksg-x8pr} {\bibfield  {journal} {\bibinfo  {journal} {Phys. Rev. B}\ }\textbf {\bibinfo {volume} {111}},\ \bibinfo {pages} {224508} (\bibinfo {year} {2025}{\natexlab{c}})}\BibitemShut {NoStop}%
\bibitem [{\citenamefont {Tamura}\ \emph {et~al.}(2019)\citenamefont {Tamura}, \citenamefont {Hoshino},\ and\ \citenamefont {Tanaka}}]{Tamura2019May}%
  \BibitemOpen
  \bibfield  {author} {\bibinfo {author} {\bibfnamefont {S.}~\bibnamefont {Tamura}}, \bibinfo {author} {\bibfnamefont {S.}~\bibnamefont {Hoshino}},\ and\ \bibinfo {author} {\bibfnamefont {Y.}~\bibnamefont {Tanaka}},\ }\bibfield  {title} {\bibinfo {title} {Odd-frequency pairs in chiral symmetric systems: Spectral bulk-boundary correspondence and topological criticality},\ }\href {https://doi.org/10.1103/PhysRevB.99.184512} {\bibfield  {journal} {\bibinfo  {journal} {Phys. Rev. B}\ }\textbf {\bibinfo {volume} {99}},\ \bibinfo {pages} {184512} (\bibinfo {year} {2019})}\BibitemShut {NoStop}%
\bibitem [{\citenamefont {Tamura}\ \emph {et~al.}(2021)\citenamefont {Tamura}, \citenamefont {Hoshino},\ and\ \citenamefont {Tanaka}}]{Tamura2021Oct}%
  \BibitemOpen
  \bibfield  {author} {\bibinfo {author} {\bibfnamefont {S.}~\bibnamefont {Tamura}}, \bibinfo {author} {\bibfnamefont {S.}~\bibnamefont {Hoshino}},\ and\ \bibinfo {author} {\bibfnamefont {Y.}~\bibnamefont {Tanaka}},\ }\bibfield  {title} {\bibinfo {title} {Generalization of spectral bulk-boundary correspondence},\ }\href {https://doi.org/10.1103/PhysRevB.104.165125} {\bibfield  {journal} {\bibinfo  {journal} {Phys. Rev. B}\ }\textbf {\bibinfo {volume} {104}},\ \bibinfo {pages} {165125} (\bibinfo {year} {2021})}\BibitemShut {NoStop}%
\bibitem [{\citenamefont {Bergeret}\ \emph {et~al.}(2005)\citenamefont {Bergeret}, \citenamefont {Volkov},\ and\ \citenamefont {Efetov}}]{BergeretRMP2005}%
  \BibitemOpen
  \bibfield  {author} {\bibinfo {author} {\bibfnamefont {F.~S.}\ \bibnamefont {Bergeret}}, \bibinfo {author} {\bibfnamefont {A.~F.}\ \bibnamefont {Volkov}},\ and\ \bibinfo {author} {\bibfnamefont {K.~B.}\ \bibnamefont {Efetov}},\ }\bibfield  {title} {\bibinfo {title} {Odd triplet superconductivity and related phenomena in superconductor-ferromagnet structures},\ }\href {https://doi.org/10.1103/RevModPhys.77.1321} {\bibfield  {journal} {\bibinfo  {journal} {Rev. Mod. Phys.}\ }\textbf {\bibinfo {volume} {77}},\ \bibinfo {pages} {1321} (\bibinfo {year} {2005})}\BibitemShut {NoStop}%
\bibitem [{\citenamefont {Eschrig}\ \emph {et~al.}(2007)\citenamefont {Eschrig}, \citenamefont {L{\"o}fwander}, \citenamefont {Cuevas}, \citenamefont {Kopu},\ and\ \citenamefont {Sch{\"o}n}}]{eschrig2007symmetries}%
  \BibitemOpen
  \bibfield  {author} {\bibinfo {author} {\bibfnamefont {M.}~\bibnamefont {Eschrig}}, \bibinfo {author} {\bibfnamefont {T.}~\bibnamefont {L{\"o}fwander}}, \bibinfo {author} {\bibfnamefont {J.}~\bibnamefont {Cuevas}}, \bibinfo {author} {\bibfnamefont {J.}~\bibnamefont {Kopu}},\ and\ \bibinfo {author} {\bibfnamefont {G.}~\bibnamefont {Sch{\"o}n}},\ }\bibfield  {title} {\bibinfo {title} {Symmetries of pairing correlations in superconductor--ferromagnet nanostructures},\ }\href {https://doi.org/10.1007/s10909-007-9329-6} {\bibfield  {journal} {\bibinfo  {journal} {Journal of Low Temperature Physics}\ }\textbf {\bibinfo {volume} {147}},\ \bibinfo {pages} {457} (\bibinfo {year} {2007})}\BibitemShut {NoStop}%
\bibitem [{\citenamefont {Sothmann}\ \emph {et~al.}(2014)\citenamefont {Sothmann}, \citenamefont {Weiss}, \citenamefont {Governale},\ and\ \citenamefont {K\"onig}}]{Sothmann2014Dec}%
  \BibitemOpen
  \bibfield  {author} {\bibinfo {author} {\bibfnamefont {B.}~\bibnamefont {Sothmann}}, \bibinfo {author} {\bibfnamefont {S.}~\bibnamefont {Weiss}}, \bibinfo {author} {\bibfnamefont {M.}~\bibnamefont {Governale}},\ and\ \bibinfo {author} {\bibfnamefont {J.}~\bibnamefont {K\"onig}},\ }\bibfield  {title} {\bibinfo {title} {Unconventional superconductivity in double quantum dots},\ }\href {https://doi.org/10.1103/PhysRevB.90.220501} {\bibfield  {journal} {\bibinfo  {journal} {Phys. Rev. B}\ }\textbf {\bibinfo {volume} {90}},\ \bibinfo {pages} {220501} (\bibinfo {year} {2014})}\BibitemShut {NoStop}%
\bibitem [{\citenamefont {Linder}\ and\ \citenamefont {Robinson}(2015{\natexlab{a}})}]{linder2015strong}%
  \BibitemOpen
  \bibfield  {author} {\bibinfo {author} {\bibfnamefont {J.}~\bibnamefont {Linder}}\ and\ \bibinfo {author} {\bibfnamefont {J.~W.}\ \bibnamefont {Robinson}},\ }\bibfield  {title} {\bibinfo {title} {Strong odd-frequency correlations in fully gapped zeeman-split superconductors},\ }\href {https://doi.org/10.1038/srep15483} {\bibfield  {journal} {\bibinfo  {journal} {Scientific reports}\ }\textbf {\bibinfo {volume} {5}},\ \bibinfo {pages} {15483} (\bibinfo {year} {2015}{\natexlab{a}})}\BibitemShut {NoStop}%
\bibitem [{\citenamefont {Burset}\ \emph {et~al.}(2015)\citenamefont {Burset}, \citenamefont {Lu}, \citenamefont {Tkachov}, \citenamefont {Tanaka}, \citenamefont {Hankiewicz},\ and\ \citenamefont {Trauzettel}}]{Burset2015Nov}%
  \BibitemOpen
  \bibfield  {author} {\bibinfo {author} {\bibfnamefont {P.}~\bibnamefont {Burset}}, \bibinfo {author} {\bibfnamefont {B.}~\bibnamefont {Lu}}, \bibinfo {author} {\bibfnamefont {G.}~\bibnamefont {Tkachov}}, \bibinfo {author} {\bibfnamefont {Y.}~\bibnamefont {Tanaka}}, \bibinfo {author} {\bibfnamefont {E.~M.}\ \bibnamefont {Hankiewicz}},\ and\ \bibinfo {author} {\bibfnamefont {B.}~\bibnamefont {Trauzettel}},\ }\bibfield  {title} {\bibinfo {title} {Superconducting proximity effect in three-dimensional topological insulators in the presence of a magnetic field},\ }\href {https://doi.org/10.1103/PhysRevB.92.205424} {\bibfield  {journal} {\bibinfo  {journal} {Phys. Rev. B}\ }\textbf {\bibinfo {volume} {92}},\ \bibinfo {pages} {205424} (\bibinfo {year} {2015})}\BibitemShut {NoStop}%
\bibitem [{\citenamefont {Burset}\ \emph {et~al.}(2016)\citenamefont {Burset}, \citenamefont {Lu}, \citenamefont {Ebisu}, \citenamefont {Asano},\ and\ \citenamefont {Tanaka}}]{Burset2016May}%
  \BibitemOpen
  \bibfield  {author} {\bibinfo {author} {\bibfnamefont {P.}~\bibnamefont {Burset}}, \bibinfo {author} {\bibfnamefont {B.}~\bibnamefont {Lu}}, \bibinfo {author} {\bibfnamefont {H.}~\bibnamefont {Ebisu}}, \bibinfo {author} {\bibfnamefont {Y.}~\bibnamefont {Asano}},\ and\ \bibinfo {author} {\bibfnamefont {Y.}~\bibnamefont {Tanaka}},\ }\bibfield  {title} {\bibinfo {title} {All-electrical generation and control of odd-frequency $s$-wave {Cooper} pairs in double quantum dots},\ }\href {https://doi.org/10.1103/PhysRevB.93.201402} {\bibfield  {journal} {\bibinfo  {journal} {Phys. Rev. B}\ }\textbf {\bibinfo {volume} {93}},\ \bibinfo {pages} {201402} (\bibinfo {year} {2016})}\BibitemShut {NoStop}%
\bibitem [{\citenamefont {Lu}\ \emph {et~al.}(2016)\citenamefont {Lu}, \citenamefont {Burset}, \citenamefont {Tanuma}, \citenamefont {Golubov}, \citenamefont {Asano},\ and\ \citenamefont {Tanaka}}]{Lu2016Jul}%
  \BibitemOpen
  \bibfield  {author} {\bibinfo {author} {\bibfnamefont {B.}~\bibnamefont {Lu}}, \bibinfo {author} {\bibfnamefont {P.}~\bibnamefont {Burset}}, \bibinfo {author} {\bibfnamefont {Y.}~\bibnamefont {Tanuma}}, \bibinfo {author} {\bibfnamefont {A.~A.}\ \bibnamefont {Golubov}}, \bibinfo {author} {\bibfnamefont {Y.}~\bibnamefont {Asano}},\ and\ \bibinfo {author} {\bibfnamefont {Y.}~\bibnamefont {Tanaka}},\ }\bibfield  {title} {\bibinfo {title} {Influence of the impurity scattering on charge transport in unconventional superconductor junctions},\ }\href {https://doi.org/10.1103/PhysRevB.94.014504} {\bibfield  {journal} {\bibinfo  {journal} {Phys. Rev. B}\ }\textbf {\bibinfo {volume} {94}},\ \bibinfo {pages} {014504} (\bibinfo {year} {2016})}\BibitemShut {NoStop}%
\bibitem [{\citenamefont {Burset}\ \emph {et~al.}(2017)\citenamefont {Burset}, \citenamefont {Lu}, \citenamefont {Tamura},\ and\ \citenamefont {Tanaka}}]{Burset2017Jun}%
  \BibitemOpen
  \bibfield  {author} {\bibinfo {author} {\bibfnamefont {P.}~\bibnamefont {Burset}}, \bibinfo {author} {\bibfnamefont {B.}~\bibnamefont {Lu}}, \bibinfo {author} {\bibfnamefont {S.}~\bibnamefont {Tamura}},\ and\ \bibinfo {author} {\bibfnamefont {Y.}~\bibnamefont {Tanaka}},\ }\bibfield  {title} {\bibinfo {title} {Current fluctuations in unconventional superconductor junctions with impurity scattering},\ }\href {https://doi.org/10.1103/PhysRevB.95.224502} {\bibfield  {journal} {\bibinfo  {journal} {Phys. Rev. B}\ }\textbf {\bibinfo {volume} {95}},\ \bibinfo {pages} {224502} (\bibinfo {year} {2017})}\BibitemShut {NoStop}%
\bibitem [{\citenamefont {Diesch}\ \emph {et~al.}(2018)\citenamefont {Diesch}, \citenamefont {Machon}, \citenamefont {Wolz}, \citenamefont {S{\"u}rgers}, \citenamefont {Beckmann}, \citenamefont {Belzig},\ and\ \citenamefont {Scheer}}]{diesch2018}%
  \BibitemOpen
  \bibfield  {author} {\bibinfo {author} {\bibfnamefont {S.}~\bibnamefont {Diesch}}, \bibinfo {author} {\bibfnamefont {P.}~\bibnamefont {Machon}}, \bibinfo {author} {\bibfnamefont {M.}~\bibnamefont {Wolz}}, \bibinfo {author} {\bibfnamefont {C.}~\bibnamefont {S{\"u}rgers}}, \bibinfo {author} {\bibfnamefont {D.}~\bibnamefont {Beckmann}}, \bibinfo {author} {\bibfnamefont {W.}~\bibnamefont {Belzig}},\ and\ \bibinfo {author} {\bibfnamefont {E.}~\bibnamefont {Scheer}},\ }\bibfield  {title} {\bibinfo {title} {Creation of equal-spin triplet superconductivity at the {Al}/{EuS} interface},\ }\href {https://doi.org/10.1038/s41467-018-07597-w} {\bibfield  {journal} {\bibinfo  {journal} {Nature communications}\ }\textbf {\bibinfo {volume} {9}},\ \bibinfo {pages} {1} (\bibinfo {year} {2018})}\BibitemShut {NoStop}%
\bibitem [{\citenamefont {Hwang}\ \emph {et~al.}(2018)\citenamefont {Hwang}, \citenamefont {Burset},\ and\ \citenamefont {Sothmann}}]{Hwang2018Oct}%
  \BibitemOpen
  \bibfield  {author} {\bibinfo {author} {\bibfnamefont {S.-Y.}\ \bibnamefont {Hwang}}, \bibinfo {author} {\bibfnamefont {P.}~\bibnamefont {Burset}},\ and\ \bibinfo {author} {\bibfnamefont {B.}~\bibnamefont {Sothmann}},\ }\bibfield  {title} {\bibinfo {title} {Odd-frequency superconductivity revealed by thermopower},\ }\href {https://doi.org/10.1103/PhysRevB.98.161408} {\bibfield  {journal} {\bibinfo  {journal} {Phys. Rev. B}\ }\textbf {\bibinfo {volume} {98}},\ \bibinfo {pages} {161408} (\bibinfo {year} {2018})}\BibitemShut {NoStop}%
\bibitem [{\citenamefont {Bobkova}\ \emph {et~al.}(2019)\citenamefont {Bobkova}, \citenamefont {Bobkov},\ and\ \citenamefont {Belzig}}]{Bobkova2019}%
  \BibitemOpen
  \bibfield  {author} {\bibinfo {author} {\bibfnamefont {I.~V.}\ \bibnamefont {Bobkova}}, \bibinfo {author} {\bibfnamefont {A.~M.}\ \bibnamefont {Bobkov}},\ and\ \bibinfo {author} {\bibfnamefont {W.}~\bibnamefont {Belzig}},\ }\bibfield  {title} {\bibinfo {title} {Signatures of spin-triplet {Cooper} pairing in the density of states of spin-textured superconductor-ferromagnet bilayers},\ }\href {https://doi.org/10.1088/1367-2630/ab0e20} {\bibfield  {journal} {\bibinfo  {journal} {New Journal of Physics}\ }\textbf {\bibinfo {volume} {21}},\ \bibinfo {pages} {043001} (\bibinfo {year} {2019})}\BibitemShut {NoStop}%
\bibitem [{\citenamefont {Johnsen}\ \emph {et~al.}(2021)\citenamefont {Johnsen}, \citenamefont {Jacobsen},\ and\ \citenamefont {Linder}}]{Johnsen2021Feb}%
  \BibitemOpen
  \bibfield  {author} {\bibinfo {author} {\bibfnamefont {L.~G.}\ \bibnamefont {Johnsen}}, \bibinfo {author} {\bibfnamefont {S.~H.}\ \bibnamefont {Jacobsen}},\ and\ \bibinfo {author} {\bibfnamefont {J.}~\bibnamefont {Linder}},\ }\bibfield  {title} {\bibinfo {title} {Magnetic control of superconducting heterostructures using compensated antiferromagnets},\ }\href {https://doi.org/10.1103/PhysRevB.103.L060505} {\bibfield  {journal} {\bibinfo  {journal} {Phys. Rev. B}\ }\textbf {\bibinfo {volume} {103}},\ \bibinfo {pages} {L060505} (\bibinfo {year} {2021})}\BibitemShut {NoStop}%
\bibitem [{\citenamefont {Hijano}\ \emph {et~al.}(2022)\citenamefont {Hijano}, \citenamefont {Golovach},\ and\ \citenamefont {Bergeret}}]{hijano2022quasiparticle}%
  \BibitemOpen
  \bibfield  {author} {\bibinfo {author} {\bibfnamefont {A.}~\bibnamefont {Hijano}}, \bibinfo {author} {\bibfnamefont {V.~N.}\ \bibnamefont {Golovach}},\ and\ \bibinfo {author} {\bibfnamefont {F.~S.}\ \bibnamefont {Bergeret}},\ }\bibfield  {title} {\bibinfo {title} {Quasiparticle density of states and triplet correlations in superconductor/ferromagnetic-insulator structures across a sharp domain wall},\ }\href {https://doi.org/10.1103/PhysRevB.105.174507} {\bibfield  {journal} {\bibinfo  {journal} {Phys. Rev. B}\ }\textbf {\bibinfo {volume} {105}},\ \bibinfo {pages} {174507} (\bibinfo {year} {2022})}\BibitemShut {NoStop}%
\bibitem [{\citenamefont {Fyhn}\ \emph {et~al.}(2023)\citenamefont {Fyhn}, \citenamefont {Brataas}, \citenamefont {Qaiumzadeh},\ and\ \citenamefont {Linder}}]{Fyhn_2023_PRL}%
  \BibitemOpen
  \bibfield  {author} {\bibinfo {author} {\bibfnamefont {E.~H.}\ \bibnamefont {Fyhn}}, \bibinfo {author} {\bibfnamefont {A.}~\bibnamefont {Brataas}}, \bibinfo {author} {\bibfnamefont {A.}~\bibnamefont {Qaiumzadeh}},\ and\ \bibinfo {author} {\bibfnamefont {J.}~\bibnamefont {Linder}},\ }\bibfield  {title} {\bibinfo {title} {Superconducting proximity effect and long-ranged triplets in dirty metallic antiferromagnets},\ }\href {https://doi.org/10.1103/PhysRevLett.131.076001} {\bibfield  {journal} {\bibinfo  {journal} {Phys. Rev. Lett.}\ }\textbf {\bibinfo {volume} {131}},\ \bibinfo {pages} {076001} (\bibinfo {year} {2023})}\BibitemShut {NoStop}%
\bibitem [{\citenamefont {Kamra}\ \emph {et~al.}(2023)\citenamefont {Kamra}, \citenamefont {Chourasia}, \citenamefont {Bobkov}, \citenamefont {Gordeeva}, \citenamefont {Bobkova},\ and\ \citenamefont {Kamra}}]{Johnsen2023}%
  \BibitemOpen
  \bibfield  {author} {\bibinfo {author} {\bibfnamefont {L.~J.}\ \bibnamefont {Kamra}}, \bibinfo {author} {\bibfnamefont {S.}~\bibnamefont {Chourasia}}, \bibinfo {author} {\bibfnamefont {G.~A.}\ \bibnamefont {Bobkov}}, \bibinfo {author} {\bibfnamefont {V.~M.}\ \bibnamefont {Gordeeva}}, \bibinfo {author} {\bibfnamefont {I.~V.}\ \bibnamefont {Bobkova}},\ and\ \bibinfo {author} {\bibfnamefont {A.}~\bibnamefont {Kamra}},\ }\bibfield  {title} {\bibinfo {title} {Complete ${T}_{c}$ suppression and {N\'eel} triplets mediated exchange in antiferromagnet-superconductor-antiferromagnet trilayers},\ }\href {https://doi.org/10.1103/PhysRevB.108.144506} {\bibfield  {journal} {\bibinfo  {journal} {Phys. Rev. B}\ }\textbf {\bibinfo {volume} {108}},\ \bibinfo {pages} {144506} (\bibinfo {year} {2023})}\BibitemShut {NoStop}%
\bibitem [{\citenamefont {Chourasia}\ \emph {et~al.}(2023)\citenamefont {Chourasia}, \citenamefont {Kamra}, \citenamefont {Bobkova},\ and\ \citenamefont {Kamra}}]{Akash2023}%
  \BibitemOpen
  \bibfield  {author} {\bibinfo {author} {\bibfnamefont {S.}~\bibnamefont {Chourasia}}, \bibinfo {author} {\bibfnamefont {L.~J.}\ \bibnamefont {Kamra}}, \bibinfo {author} {\bibfnamefont {I.~V.}\ \bibnamefont {Bobkova}},\ and\ \bibinfo {author} {\bibfnamefont {A.}~\bibnamefont {Kamra}},\ }\bibfield  {title} {\bibinfo {title} {Generation of spin-triplet {Cooper} pairs via a canted antiferromagnet},\ }\href {https://doi.org/10.1103/PhysRevB.108.064515} {\bibfield  {journal} {\bibinfo  {journal} {Phys. Rev. B}\ }\textbf {\bibinfo {volume} {108}},\ \bibinfo {pages} {064515} (\bibinfo {year} {2023})}\BibitemShut {NoStop}%
\bibitem [{\citenamefont {Li}\ \emph {et~al.}(2024)\citenamefont {Li}, \citenamefont {Huang}, \citenamefont {Wang}, \citenamefont {Li}, \citenamefont {Tao},\ and\ \citenamefont {Fu}}]{Li2024Nov}%
  \BibitemOpen
  \bibfield  {author} {\bibinfo {author} {\bibfnamefont {R.}~\bibnamefont {Li}}, \bibinfo {author} {\bibfnamefont {C.}~\bibnamefont {Huang}}, \bibinfo {author} {\bibfnamefont {D.}~\bibnamefont {Wang}}, \bibinfo {author} {\bibfnamefont {M.}~\bibnamefont {Li}}, \bibinfo {author} {\bibfnamefont {Y.}~\bibnamefont {Tao}},\ and\ \bibinfo {author} {\bibfnamefont {H.}~\bibnamefont {Fu}},\ }\bibfield  {title} {\bibinfo {title} {Perfect spin and/or valley triplet pairing states in an antiferromagnetic-silicene/superconductor hybrid structure},\ }\href {https://doi.org/10.1103/PhysRevB.110.205425} {\bibfield  {journal} {\bibinfo  {journal} {Phys. Rev. B}\ }\textbf {\bibinfo {volume} {110}},\ \bibinfo {pages} {205425} (\bibinfo {year} {2024})}\BibitemShut {NoStop}%
\bibitem [{\citenamefont {Linder}\ and\ \citenamefont {Robinson}(2015{\natexlab{b}})}]{Linder_NatPhys_2015}%
  \BibitemOpen
  \bibfield  {author} {\bibinfo {author} {\bibfnamefont {J.}~\bibnamefont {Linder}}\ and\ \bibinfo {author} {\bibfnamefont {J.~W.~A.}\ \bibnamefont {Robinson}},\ }\bibfield  {title} {\bibinfo {title} {{Superconducting spintronics}},\ }\href {https://doi.org/10.1038/nphys3242} {\bibfield  {journal} {\bibinfo  {journal} {Nat. Phys.}\ }\textbf {\bibinfo {volume} {11}},\ \bibinfo {pages} {307} (\bibinfo {year} {2015}{\natexlab{b}})}\BibitemShut {NoStop}%
\bibitem [{\citenamefont {Eschrig}(2015)}]{Eschrig_RPP_2015}%
  \BibitemOpen
  \bibfield  {author} {\bibinfo {author} {\bibfnamefont {M.}~\bibnamefont {Eschrig}},\ }\bibfield  {title} {\bibinfo {title} {{Spin-polarized supercurrents for spintronics: a review of current progress}},\ }\href {https://doi.org/10.1088/0034-4885/78/10/104501} {\bibfield  {journal} {\bibinfo  {journal} {Rep. Prog. Phys.}\ }\textbf {\bibinfo {volume} {78}},\ \bibinfo {pages} {104501} (\bibinfo {year} {2015})}\BibitemShut {NoStop}%
\bibitem [{\citenamefont {Breunig}\ \emph {et~al.}(2018)\citenamefont {Breunig}, \citenamefont {Burset},\ and\ \citenamefont {Trauzettel}}]{Breunig2018Jan}%
  \BibitemOpen
  \bibfield  {author} {\bibinfo {author} {\bibfnamefont {D.}~\bibnamefont {Breunig}}, \bibinfo {author} {\bibfnamefont {P.}~\bibnamefont {Burset}},\ and\ \bibinfo {author} {\bibfnamefont {B.}~\bibnamefont {Trauzettel}},\ }\bibfield  {title} {\bibinfo {title} {Creation of spin-triplet cooper pairs in the absence of magnetic ordering},\ }\href {https://doi.org/10.1103/PhysRevLett.120.037701} {\bibfield  {journal} {\bibinfo  {journal} {Phys. Rev. Lett.}\ }\textbf {\bibinfo {volume} {120}},\ \bibinfo {pages} {037701} (\bibinfo {year} {2018})}\BibitemShut {NoStop}%
\bibitem [{\citenamefont {Keidel}\ \emph {et~al.}(2020)\citenamefont {Keidel}, \citenamefont {Hwang}, \citenamefont {Trauzettel}, \citenamefont {Sothmann},\ and\ \citenamefont {Burset}}]{Keidel2020Apr}%
  \BibitemOpen
  \bibfield  {author} {\bibinfo {author} {\bibfnamefont {F.}~\bibnamefont {Keidel}}, \bibinfo {author} {\bibfnamefont {S.-Y.}\ \bibnamefont {Hwang}}, \bibinfo {author} {\bibfnamefont {B.}~\bibnamefont {Trauzettel}}, \bibinfo {author} {\bibfnamefont {B.}~\bibnamefont {Sothmann}},\ and\ \bibinfo {author} {\bibfnamefont {P.}~\bibnamefont {Burset}},\ }\bibfield  {title} {\bibinfo {title} {On-demand thermoelectric generation of equal-spin cooper pairs},\ }\href {https://doi.org/10.1103/PhysRevResearch.2.022019} {\bibfield  {journal} {\bibinfo  {journal} {Phys. Rev. Res.}\ }\textbf {\bibinfo {volume} {2}},\ \bibinfo {pages} {022019} (\bibinfo {year} {2020})}\BibitemShut {NoStop}%
\bibitem [{\citenamefont {Mahan}(2013)}]{mahan2013many}%
  \BibitemOpen
  \bibfield  {author} {\bibinfo {author} {\bibfnamefont {G.~D.}\ \bibnamefont {Mahan}},\ }\href {https://doi.org/10.1007/978-1-4757-5714-9} {\emph {\bibinfo {title} {Many-particle physics}}}\ (\bibinfo  {publisher} {Springer Science \& Business Media},\ \bibinfo {year} {2013})\BibitemShut {NoStop}%
\bibitem [{\citenamefont {Zagoskin}(2014)}]{zagoskin}%
  \BibitemOpen
  \bibfield  {author} {\bibinfo {author} {\bibfnamefont {A.}~\bibnamefont {Zagoskin}},\ }\href {https://doi.org/10.1007/978-3-319-07049-0} {\emph {\bibinfo {title} {Quantum Theory of Many-Body Systems: {Techniques} and Applications}}}\ (\bibinfo  {publisher} {Springer},\ \bibinfo {year} {2014})\BibitemShut {NoStop}%
\bibitem [{\citenamefont {San-Jose}(2024)}]{Quantica}%
  \BibitemOpen
  \bibfield  {author} {\bibinfo {author} {\bibfnamefont {P.}~\bibnamefont {San-Jose}},\ }\href {https://doi.org/10.5281/zenodo.4762963} {\bibinfo {title} {pablosanjose/quantica.jl}} (\bibinfo {year} {2024})\BibitemShut {NoStop}%
\bibitem [{\citenamefont {Perrin}\ \emph {et~al.}(2020)\citenamefont {Perrin}, \citenamefont {Santos}, \citenamefont {M{\ifmmode\acute{e}\else\'{e}\fi}nard}, \citenamefont {Brun}, \citenamefont {Cren}, \citenamefont {Civelli},\ and\ \citenamefont {Simon}}]{Perrin_PRL_2020}%
  \BibitemOpen
  \bibfield  {author} {\bibinfo {author} {\bibfnamefont {V.}~\bibnamefont {Perrin}}, \bibinfo {author} {\bibfnamefont {F.~L.~N.}\ \bibnamefont {Santos}}, \bibinfo {author} {\bibfnamefont {G.~C.}\ \bibnamefont {M{\ifmmode\acute{e}\else\'{e}\fi}nard}}, \bibinfo {author} {\bibfnamefont {C.}~\bibnamefont {Brun}}, \bibinfo {author} {\bibfnamefont {T.}~\bibnamefont {Cren}}, \bibinfo {author} {\bibfnamefont {M.}~\bibnamefont {Civelli}},\ and\ \bibinfo {author} {\bibfnamefont {P.}~\bibnamefont {Simon}},\ }\bibfield  {title} {\bibinfo {title} {{Unveiling Odd-Frequency Pairing around a Magnetic Impurity in a Superconductor}},\ }\href {https://doi.org/10.1103/PhysRevLett.125.117003} {\bibfield  {journal} {\bibinfo  {journal} {Phys. Rev. Lett.}\ }\textbf {\bibinfo {volume} {125}},\ \bibinfo {pages} {117003} (\bibinfo {year} {2020})}\BibitemShut {NoStop}%
\bibitem [{\citenamefont {Seoane~Souto}\ \emph {et~al.}(2024)\citenamefont {Seoane~Souto}, \citenamefont {Kuzmanovski}, \citenamefont {Sardinero}, \citenamefont {Burset},\ and\ \citenamefont {Balatsky}}]{SeoaneSouto2024Oct}%
  \BibitemOpen
  \bibfield  {author} {\bibinfo {author} {\bibfnamefont {R.}~\bibnamefont {Seoane~Souto}}, \bibinfo {author} {\bibfnamefont {D.}~\bibnamefont {Kuzmanovski}}, \bibinfo {author} {\bibfnamefont {I.}~\bibnamefont {Sardinero}}, \bibinfo {author} {\bibfnamefont {P.}~\bibnamefont {Burset}},\ and\ \bibinfo {author} {\bibfnamefont {A.~V.}\ \bibnamefont {Balatsky}},\ }\bibfield  {title} {\bibinfo {title} {P-wave pairing near a spin-split {Josephson} junction},\ }\href {https://doi.org/10.1007/s10909-024-03176-0} {\bibfield  {journal} {\bibinfo  {journal} {J. Low Temp. Phys.}\ }\textbf {\bibinfo {volume} {217}},\ \bibinfo {pages} {106} (\bibinfo {year} {2024})}\BibitemShut {NoStop}%
\bibitem [{\citenamefont {Ren}\ \emph {et~al.}(2019)\citenamefont {Ren}, \citenamefont {Pientka}, \citenamefont {Hart}, \citenamefont {Pierce}, \citenamefont {Kosowsky}, \citenamefont {Lunczer}, \citenamefont {Schlereth}, \citenamefont {Scharf}, \citenamefont {Hankiewicz}, \citenamefont {Molenkamp}, \citenamefont {Halperin},\ and\ \citenamefont {Yacoby}}]{Ren_Nature2019}%
  \BibitemOpen
  \bibfield  {author} {\bibinfo {author} {\bibfnamefont {H.}~\bibnamefont {Ren}}, \bibinfo {author} {\bibfnamefont {F.}~\bibnamefont {Pientka}}, \bibinfo {author} {\bibfnamefont {S.}~\bibnamefont {Hart}}, \bibinfo {author} {\bibfnamefont {A.~T.}\ \bibnamefont {Pierce}}, \bibinfo {author} {\bibfnamefont {M.}~\bibnamefont {Kosowsky}}, \bibinfo {author} {\bibfnamefont {L.}~\bibnamefont {Lunczer}}, \bibinfo {author} {\bibfnamefont {R.}~\bibnamefont {Schlereth}}, \bibinfo {author} {\bibfnamefont {B.}~\bibnamefont {Scharf}}, \bibinfo {author} {\bibfnamefont {E.~M.}\ \bibnamefont {Hankiewicz}}, \bibinfo {author} {\bibfnamefont {L.~W.}\ \bibnamefont {Molenkamp}}, \bibinfo {author} {\bibfnamefont {B.~I.}\ \bibnamefont {Halperin}},\ and\ \bibinfo {author} {\bibfnamefont {A.}~\bibnamefont {Yacoby}},\ }\bibfield  {title} {\bibinfo {title} {Topological superconductivity in a phase-controlled {Josephson} junction},\ }\href {https://doi.org/10.1038/s41586-019-1148-9} {\bibfield  {journal} {\bibinfo  {journal} {Nature}\
  }\textbf {\bibinfo {volume} {569}},\ \bibinfo {pages} {93} (\bibinfo {year} {2019})}\BibitemShut {NoStop}%
\bibitem [{\citenamefont {Ikegaya}\ \emph {et~al.}(2020)\citenamefont {Ikegaya}, \citenamefont {Tamura}, \citenamefont {Manske},\ and\ \citenamefont {Tanaka}}]{Ikegaya2020Oct}%
  \BibitemOpen
  \bibfield  {author} {\bibinfo {author} {\bibfnamefont {S.}~\bibnamefont {Ikegaya}}, \bibinfo {author} {\bibfnamefont {S.}~\bibnamefont {Tamura}}, \bibinfo {author} {\bibfnamefont {D.}~\bibnamefont {Manske}},\ and\ \bibinfo {author} {\bibfnamefont {Y.}~\bibnamefont {Tanaka}},\ }\bibfield  {title} {\bibinfo {title} {Anomalous proximity effect of planar topological {Josephson} junctions},\ }\href {https://doi.org/10.1103/PhysRevB.102.140505} {\bibfield  {journal} {\bibinfo  {journal} {Phys. Rev. B}\ }\textbf {\bibinfo {volume} {102}},\ \bibinfo {pages} {140505} (\bibinfo {year} {2020})}\BibitemShut {NoStop}%
\bibitem [{\citenamefont {Lesser}\ and\ \citenamefont {Oreg}(2022{\natexlab{b}})}]{Lesser_2022}%
  \BibitemOpen
  \bibfield  {author} {\bibinfo {author} {\bibfnamefont {O.}~\bibnamefont {Lesser}}\ and\ \bibinfo {author} {\bibfnamefont {Y.}~\bibnamefont {Oreg}},\ }\bibfield  {title} {\bibinfo {title} {Majorana zero modes induced by superconducting phase bias},\ }\href {https://doi.org/10.1088/1361-6463/ac4a37} {\bibfield  {journal} {\bibinfo  {journal} {Journal of Physics D: Applied Physics}\ }\textbf {\bibinfo {volume} {55}},\ \bibinfo {pages} {164001} (\bibinfo {year} {2022}{\natexlab{b}})}\BibitemShut {NoStop}%
\bibitem [{\citenamefont {Oshima}\ \emph {et~al.}(2022)\citenamefont {Oshima}, \citenamefont {Ikegaya}, \citenamefont {Schnyder},\ and\ \citenamefont {Tanaka}}]{Oshima2022May}%
  \BibitemOpen
  \bibfield  {author} {\bibinfo {author} {\bibfnamefont {D.}~\bibnamefont {Oshima}}, \bibinfo {author} {\bibfnamefont {S.}~\bibnamefont {Ikegaya}}, \bibinfo {author} {\bibfnamefont {A.~P.}\ \bibnamefont {Schnyder}},\ and\ \bibinfo {author} {\bibfnamefont {Y.}~\bibnamefont {Tanaka}},\ }\bibfield  {title} {\bibinfo {title} {Flat-band {Majorana} bound states in topological {Josephson} junctions},\ }\href {https://doi.org/10.1103/PhysRevResearch.4.L022051} {\bibfield  {journal} {\bibinfo  {journal} {Phys. Rev. Res.}\ }\textbf {\bibinfo {volume} {4}},\ \bibinfo {pages} {L022051} (\bibinfo {year} {2022})}\BibitemShut {NoStop}%
\end{thebibliography}%

\end{document}